\pdfoutput=1

\documentclass[11pt,twoside,a4paper,cmspaper,final,collab]{cms-tdr}
\def\svnVersion{386450}\def\cmsCernNoTag{CERN-EP/2016-207}\def\cmsCernDate{\today}\def\cmsMessage{Published in the Journal of High Energy Physics as \href{http://dx.doi.org/10.1007/JHEP02(2017)028}{\doi{10.1007/JHEP02(2017)028.}}}

\begin{document}\cmsNoteHeader{TOP-14-007}

\hyphenation{had-ron-i-za-tion}
\hyphenation{cal-or-i-me-ter}
\hyphenation{de-vices}
\RCS$Revision: 386436 $
\RCS$HeadURL: svn+ssh://svn.cern.ch/reps/tdr2/papers/TOP-14-007/trunk/TOP-14-007.tex $
\RCS$Id: TOP-14-007.tex 386436 2017-02-10 18:18:52Z doudko $
\renewcommand{\labelenumi}{\theenumi}
\newcommand{\zjets}{\ensuremath{\PZ/\PGg^*\text{+jets}}\xspace}
\newcommand{\wjets}{\ensuremath{\PW\text{+jets}}\xspace}
\newcommand{\VPtmiss}{\ensuremath{\vec {P}_{\mathrm{T}}^{\text{miss}}}\xspace}
\newcommand{\MtW}{\ensuremath{m_{\mathrm T}(\PW)}\xspace}
\newlength\cmsFigWidth
\ifthenelse{\boolean{cms@external}}{\setlength\cmsFigWidth{0.85\columnwidth}}{\setlength\cmsFigWidth{0.4\textwidth}}
\ifthenelse{\boolean{cms@external}}{\providecommand{\cmsLeft}{top}}{\providecommand{\cmsLeft}{left}}
\ifthenelse{\boolean{cms@external}}{\providecommand{\cmsRight}{bottom}}{\providecommand{\cmsRight}{right}}
\cmsNoteHeader{TOP-14-007}
\title{\texorpdfstring{Search for anomalous Wtb couplings and flavour-changing neutral
currents in $t$-channel single top quark production in pp collisions at $\sqrt{s}=7$ and 8\TeV}{Search for anomalous Wtb couplings and flavour-changing neutral currents in t-channel single top quark production from pp collisions at sqrt(s) = 7 and 8 TeV}}

\date{\today}
\abstract{
Single top quark events produced in the $t$ channel are used to set limits on anomalous Wtb couplings and to search for top quark flavour-changing neutral current (FCNC) interactions. The data taken with the CMS detector at the LHC in proton-proton collisions at $\sqrt{s}=7$ and 8\TeV correspond to integrated luminosities of 5.0 and 19.7\fbinv, respectively. The analysis is performed using events with one muon and two or three jets. A Bayesian neural network technique is used to discriminate between the signal and backgrounds, which are observed to be consistent with the standard model prediction. The 95\% confidence level (CL) exclusion limits on anomalous right-handed vector, and left- and right-handed tensor Wtb couplings are measured to be $\abs{f_\mathrm{V}^\mathrm{R}}< 0.16$, $\abs{f_\mathrm{T}^\mathrm{L}}<0.057$, and $-0.049<f_{\rm T}^{\rm R}<0.048$, respectively. For the FCNC couplings $\kappa_\mathrm{tug}$ and $\kappa_\mathrm{tcg}$, the 95\% CL upper limits on coupling strengths are
$\abs{\kappa_\mathrm{tug}}/\Lambda < 4.1 \times 10^{-3}\TeV^{-1}$ and
$\abs{\kappa_\mathrm{tcg}} /\Lambda < 1.8 \times 10^{-2}\TeV^{-1}$,
where $\Lambda$ is the scale for new physics, and correspond to upper limits on the branching fractions of
$2.0\times10^{-5}$ and
$4.1\times10^{-4}$  for the decays
$\PQt\to \PQu\Pg$ and
$\PQt\to \PQc\Pg$, respectively.
}
\hypersetup{%
pdfauthor={CMS Collaboration},%
pdftitle={Search for anomalous Wtb couplings and flavour-changing neutral currents in t-channel single top quark production in pp collisions at sqrt(s) = 7 and 8 TeV},%
pdfsubject={CMS},%
pdfkeywords={CMS, physics, software, computing}}
\maketitle

\section{Introduction}
\label{sec:introduction}
Single top quark ($\PQt$) production provides ways to investigate aspects of top quark physics that cannot be studied with \ttbar events~\cite{Beneke:2000hk}.
The theory of electroweak interactions predicts three mechanisms for producing single top quarks in hadron-hadron collisions.
At leading order (LO), these are classified according to the virtuality of the W boson propagation in $t$-channel, $s$-channel, or associated  tW production~\cite{Willenbrock:1986cr}.
Single top quark production in all channels is directly related to the squared modulus of the Cabibbo--Kobayashi--Maskawa  matrix element $V_{\rm tb}$. As a consequence, it provides a direct measurement of this quantity and thereby a check of the standard model (SM).
The single top quark topology also opens a window for searches of anomalous Wtb couplings relative to the SM, where the interaction vertex of the top quark with the bottom quark ($\PQb$) and the W boson (Wtb vertex) has a V--A axial-vector structure.
Flavour-changing neutral currents (FCNC) are absent at lowest order in the SM, and are significantly suppressed through the Glashow--Iliopoulos--Maiani mechanism~\cite{Glashow:1970gm} at higher orders.
Various rare decays of  $\PK$, $\PD$, and $\PB$ mesons, as well as the oscillations in $\PKz\PAKz$, $\PDz\PADz$, and $\PBz\PABz$
systems, strongly constrain FCNC interactions involving the first two generations and the b quark~\cite{Agashe:2014kda}.
The V--A structure of the charged current with light quarks is well established~\cite{Agashe:2014kda}.
However, FCNC involving the top quark, as well as the structure of the Wtb vertex, are significantly less constrained.
In the SM, the FCNC couplings of the top quark are predicted to be very small and not detectable at current experimental sensitivity.
However, they can be significantly enhanced in various SM extensions, such as supersymmetry~\cite{Eilam:1990zc,Atwood:1995ud,Yang:1997dk}, and models with multiple Higgs boson doublets~\cite{Grossman:1994jb,Pich:2009sp,Keus:2013hya}, extra quarks~\cite{DiazCruz:1989ub,Arhrib:2006pm,Branco:2013tda}, or a composite top quark \cite{Georgi:1994ha}.
New vertices with top quarks  are predicted, in particular, in models with light composite Higgs bosons \cite{Giudice:2007fh,Konig:2014iqa}, extra-dimension models with warped geometry \cite{Agashe:2014jca}, or holographic structures \cite{Contino:2003ve}.
Such possibilities can be encoded in an effective field theory through higher-dimensional gauge-invariant operators \cite{AguilarSaavedra:2008zc,Willenbrock:2014bja}.
Direct limits on top quark FCNC parameters have been established by the CDF~\cite{Aaltonen:2008qr}, D0~\cite{Abazov:2010qk}, and ATLAS~\cite{Aad:2015gea} Collaborations.
There are two complementary strategies to search for FCNC in single top quark production.
A search can be performed in the $s$ channel for resonance production through the fusion of a gluon (g) with an up (u) or charm (c) quark, as was the case in analyses by the CDF and ATLAS Collaborations.
However, as pointed out by the D0 Collaboration, the $s$-channel production rate is proportional to the square of the FCNC coupling parameter and is therefore expected to be small~\cite{Abazov:2010qk}.
On the other hand, the $t$-channel cross section and its corresponding kinematic properties have been measured accurately at the LHC \cite{Khachatryan:2014iya,Khachatryan:2015dzz,Aad:2014fwa}, with an important feature being that the $t$-channel signature contains a light-quark jet produced in association with the single top quark.
This light-quark jet can be used to search for deviations from the SM prediction caused by FCNC in the top quark sector.
This strategy was applied by the D0 Collaboration~\cite{Abazov:2010qk}, as well as in our analysis.
Models that have contributions from FCNC in the production of single top quarks can have sizable deviations relative to SM predictions.
Processes with FCNC vertices in the decay of the top quark are negligible.
In contrast, the modelling of Wtb couplings can involve anomalous Wtb interactions in both the production and the decay, because both are significantly affected by anomalous contributions.
All these features are explicitly taken into account in the {\sc CompHEP} Monte Carlo (MC) generator~\cite{Boos:2004kh}.
In this paper, we present a search by the CMS experiment at the CERN LHC for anomalous Wtb couplings and FCNC interactions of the top quark through the u or c quarks and a gluon (tug or tcg vertices), by selecting muons arising from  W boson decay (including through a $\tau$ lepton) from the top quarks in $\mathrm{muon{+}jets}$ events.
Separation of signal and background is achieved through a Bayesian neural network (BNN) technique~\cite{FBMBook,Bhat:2005hq}, performed using the Flexible Bayesian modelling package~\cite{FBMPackage}.
Limits on Wtb and top quark FCNC anomalous couplings are obtained from the distribution in the BNN discriminants.

\section{The CMS detector}
The central feature of the CMS apparatus is a superconducting solenoid of 6\unit{m} internal diameter, providing a magnetic field of 3.8\unit{T}. Within the solenoid volume are a silicon pixel and strip tracker, a lead tungstate crystal electromagnetic calorimeter (ECAL), and a brass and scintillator hadron calorimeter (HCAL), each composed of a barrel and two endcap sections. Forward calorimeters extend the pseudorapidity $\eta$~\cite{JINST} coverage provided by the barrel and endcap detectors. Muons are measured in gas-ionization detectors embedded in the steel flux-return yoke outside the solenoid.
The first level of the CMS trigger system, composed of custom hardware processors, uses information from the calorimeters and muon detectors to select the most interesting events in a fixed time interval of less than 4\mus. The high-level trigger processor farm further decreases the event rate from around 100\unit{kHz} to less than 1\unit{kHz}, before data storage. A more detailed description of the CMS detector, together with a definition of the coordinate system used and the relevant kinematic variables, can be found in Ref.~\cite{JINST}.
The particle-flow event algorithm~\cite{CMS-PAS-PFT-09-001,CMS-PAS-PFT-10-001} reconstructs and identifies each individual particle with an optimized combination of information from the various elements of the CMS detector. The energy of photons is directly obtained from the ECAL measurement, corrected for zero-suppression effects. The energy of electrons is determined from a combination of the electron momentum at the primary interaction vertex as determined by the tracker, the energy of the corresponding ECAL cluster, and the energy sum of all bremsstrahlung photons spatially compatible with originating from the electron track. The energy of muons is obtained from the curvature of the corresponding track. The energy of charged hadrons is determined from a combination of their momentum measured in the tracker and the matching ECAL and HCAL energy deposits, corrected for zero-suppression effects and for the response function of the calorimeters to hadronic showers. Finally, the energy of neutral hadrons is obtained from the corresponding corrected ECAL and HCAL energy.
Jets are reconstructed offline from particle-flow candidates clustered by the anti-\kt algorithm \cite{Cacciari:2008gp,Cacciari:2011ma} with a size parameter of 0.5.
Jet momentum is determined as the vectorial sum of all particle momenta in the jet, and is found from simulation to be within 5 to 10\% of the true momentum over the whole transverse momentum (\pt) spectrum and detector acceptance. An offset correction is applied to jet energies to take into account the contribution from additional proton-proton interactions within the same or nearby bunch crossing (pileup). Jet energy corrections are derived from simulation, and are confirmed with in situ measurements of the energy balance in dijet and $\mathrm{photon{+}jet}$ events. Additional selection criteria are applied to each event to remove spurious jet-like features originating from isolated noise patterns in certain HCAL regions.
The missing transverse momentum vector \ptvecmiss is defined as the projection on the plane perpendicular to the beams of the negative vector sum of the momenta of all reconstructed particles in an event. Its magnitude is referred to as \MET~\cite{CMS-PAS-PFT-09-001}.

\section{Data and simulated events}
\label{sec:datasets}
The analysis is performed using proton-proton collisions recorded with the CMS detector in 2011 and 2012  at centre-of-mass energies of 7 and 8\TeV, respectively, and corresponding to integrated luminosities of 5.0 and 19.7\fbinv.
The $t$-channel production of a single top quark is modelled using the  {\sc CompHEP}~4.5 package~\cite{Boos:2004kh}, supplemented by an additional matching method used to simulate an effective next-to-leading-order (NLO) approach~\cite{Boos:2006af}.
The NLO cross sections used for $t$-channel single top production are $\sigma(7\TeV)=64.6^{+2.6}_{-1.9}\unit{pb}$~\cite{Kidonakis:tch} and $\sigma(8\TeV)=84.7^{+3.8}_{-3.2}\unit{pb}$~\cite{Hathor1,Hathor2}.
The \POWHEG~1.0 NLO MC generator~\cite{Alioli:2010xd} provides an alternative model to estimate the sensitivity of the analysis to the modelling of the signal.
Contributions from anomalous operators are added to the {\sc CompHEP} simulation for both the production and decay of top quarks.
This takes into account the width of the top quark, spin correlations between the production and decay, and the b quark mass in the anomalous and SM contributions.
The LO \MADGRAPH~5.1~\cite{Alwall:2011uj} generator is used to simulate the main background processes: top quark pair production with total cross sections of $\sigma(7\TeV)=172.0^{+6.5}_{-7.6}\unit{pb}$~\cite{Czakon:2013goa} and $\sigma(8\TeV)=253^{+13}_{-14}\unit{pb}$~\cite{Czakon:ttbar}, and W boson production with total cross sections of $\sigma(7\TeV)=31.3\pm1.6\unit{nb}$ and $\sigma(8\TeV)=36.7\pm 1.3\unit{nb}$~\cite{Gavin:2010az}, for processes with up to 3 and 4 additional jets in the matrix element calculations, respectively.
The subdominant backgrounds from Drell--Yan in association with jets (\zjets) production, corresponding to $\sigma(7\TeV)=5.0\pm 0.3\unit{nb}$ and $\sigma(8\TeV)=4.3\pm 0.2\unit{nb}$~\cite{Gavin:2010az}, and from WW, WZ, and ZZ (dibosons) production, corresponding to $\sigma(7\TeV)=67.1\pm1.7\unit{pb}$ and $\sigma(8\TeV)=73.8\pm1.9\unit{pb}$~\cite{Campbell:2010ff} are modelled using LO \PYTHIA~6.426~\cite{Sjostrand:2006za}.
The contribution from multijet events, with one of the jets misidentified as a lepton, is estimated using a mutually exclusive data sample. The details are given in the next section.
Single top quark production in the $s$ channel with $\sigma(7\TeV)=4.6^{+0.2}_{-0.2}\unit{pb}$, $\sigma(8\TeV)=5.5\pm0.2\unit{pb}$, and in the tW channel with $\sigma(7\TeV)=15.7\pm1.2\unit{pb}$, $\sigma(8\TeV)=22.2\pm1.5\unit{pb}$ ~\cite{Kidonakis:twsch_new} are modelled using the \POWHEG generator. The \PYTHIA~6.4 program is also used to simulate parton showers for the hard processes calculated in the {\sc CompHEP}, \MADGRAPH, and \POWHEG generators. The PDF4LHC recipe~\cite{Alekhin:2011sk} is used to reweight all simulated events to the central value of CT10 PDF~\cite{Gao:2013xoa}. The Z2Star~\cite{Chatrchyan:2013gfi,Khachatryan:2015pea} set of parameters is used to simulate the underlying-events.
Because of the importance of the \wjets background and the significant difference in the kinematic distributions, the following contributions are considered separately in the analysis: W boson produced together with a pair of b or c quarks ($\mathrm{W{+}Q\overline Q}$); W boson produced in association with a c quark ($\mathrm{W{+}c}$); W boson events that do not contain heavy quarks ($\mathrm{W{+}light}$); and events associated with underlying events (UE) that contain heavy quarks originating from the initial parton interaction ($\mathrm{W{+}QX}$).
Different nuisance parameters for the normalization scale factors are used for these components of the complete \wjets \MADGRAPH simulation.
Simulated events are reweighted to reproduce the observed particle multiplicity from pileup.
Small differences between the data and simulation in trigger efficiency~\cite{Chatrchyan:2012xi, CMS-DP-2013-009}, lepton identification and isolation~\cite{Chatrchyan:2012xi, CMS-DP-2013-009}, and b tagging~\cite{BTag2011} are corrected via scale factors, which are generally close to unity.

\section{Event selection}
\label{sec:selection}
\begin{figure}[!h]
\centering
\includegraphics[width=0.49\textwidth]{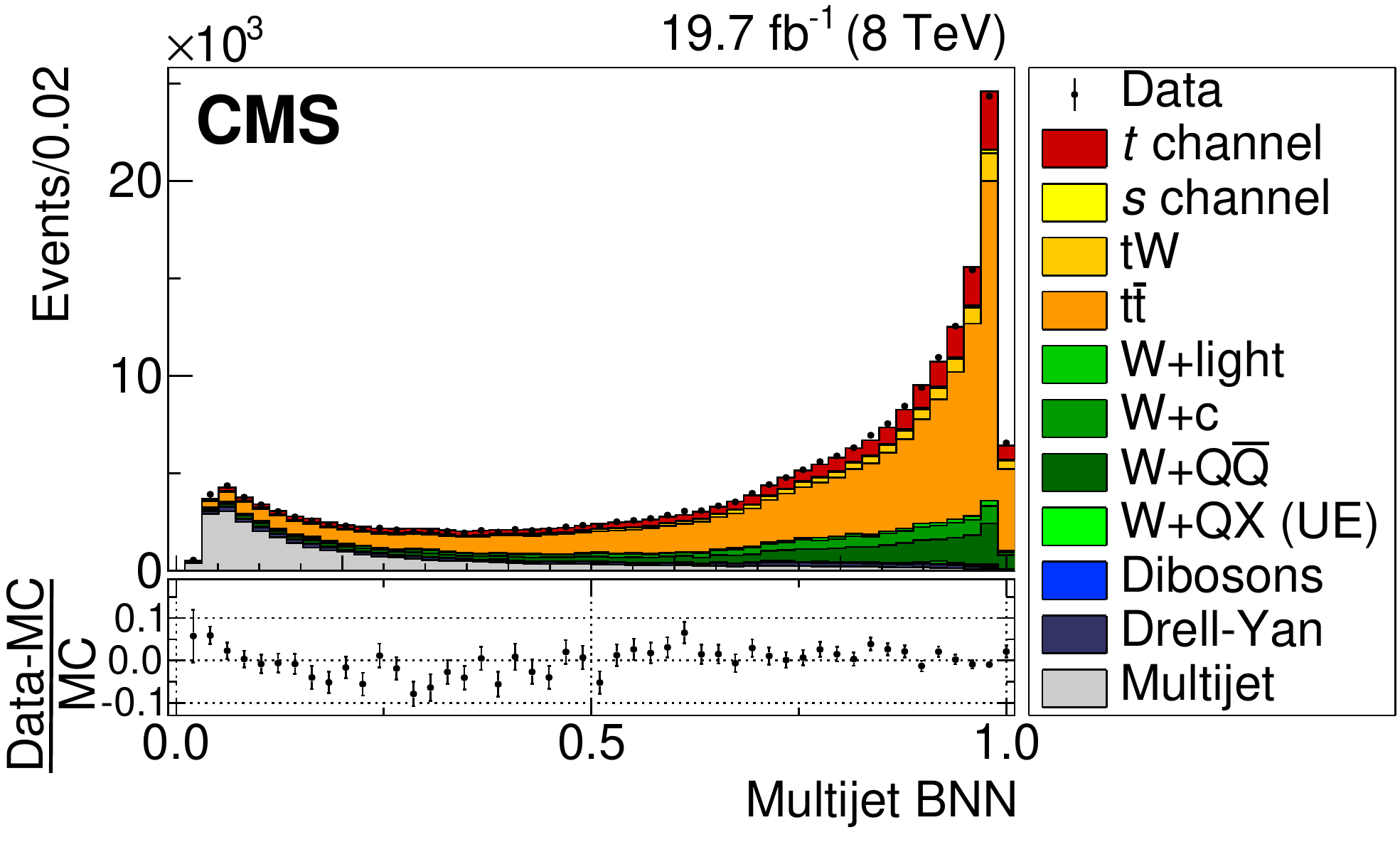}
\includegraphics[width=0.49\textwidth]{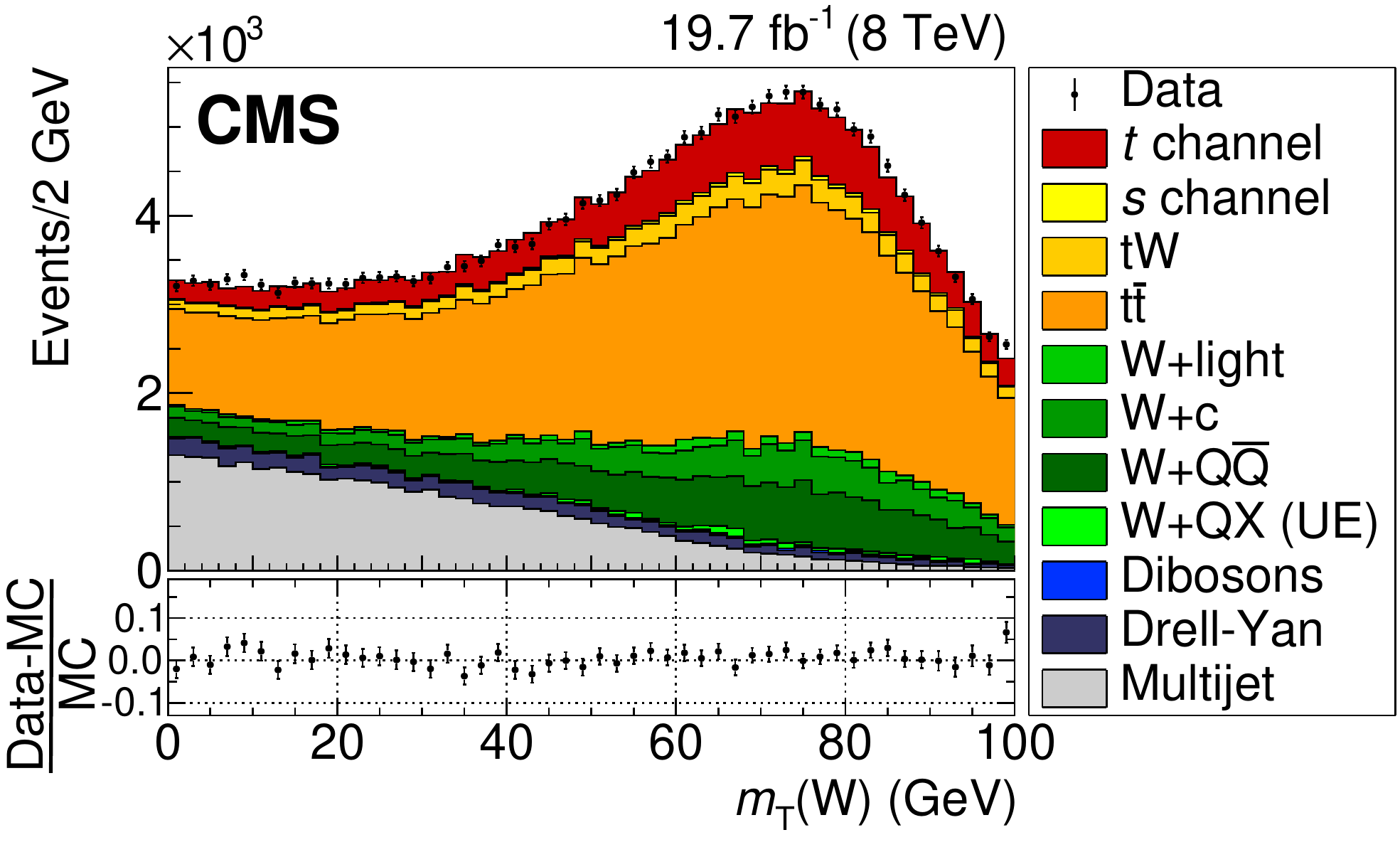}
\caption{The distributions of the multijet BNN discriminant used for the QCD multijet background rejection (left) and the reconstructed transverse W boson mass (right) from data (points) and the predicted backgrounds from simulation (filled histograms) for $\sqrt{s}=8\TeV$.
The lower part of each plot shows the relative difference between the data and the total predicted background.
The vertical bars represent the statistical uncertainties.}
\label{fig:QCD normalization}
\end{figure}

The following signature is used to identify $t$-channel single top quark production candidates: exactly one isolated muon~\cite{Chatrchyan:2012xi}, one light-flavour jet in the forward region (defined below); one \cPqb-tagged jet~\cite{BTag2011} from the b quark originating from the decay of the top quark, and an associated ``soft'' \cPqb \ jet.
The ``soft'' \cPqb \ jet is likely to fail either the \pt or $\eta$ threshold (given below).
The presence of a neutrino in the decay of the W boson leads to a significant amount of \MET, which is used to enhance the signal.
The analysis is performed using  data  collected with a trigger requiring at least one muon in each event.
To accommodate the increasing instantaneous luminosity delivered by the LHC in 2011, different triggers were used for various data-taking periods, with the muon $\pt$ threshold ranging from 20 to 27\GeV.
A single trigger with muon threshold $\pt > 24\GeV$ was used in 2012.
The selected events are required to have:
\renewcommand{\theenumi}{(\roman{enumi})}
\begin{enumerate}
\item  at least one primary vertex reconstructed from at least four tracks, and located within 24\unit{cm} in the longitudinal direction and 2\unit{cm} in the radial direction from the centre of the detector;
\item  only one isolated ($I_\text{rel}^\mu < 0.12$) muon~\cite{Chatrchyan:2012xi} with $\pt > 20\ (27)$\GeV according to the variation of the trigger \pt threshold at $\sqrt{s} = 7$ and $\pt > 26$\GeV at $\sqrt{s} = 8$  \TeV, and $\abs{\eta} < 2.1$, originating from the primary vertex, where the relative isolation parameter of the muon, $I_\text{rel}^\mu$, is defined as the sum of the energy deposited by long-lived charged hadrons, neutral hadrons, and photons in a cone with radius $\Delta R = \sqrt{\smash[b]{(\Delta \eta^2 + \Delta \phi^2)}} = 0.4$, divided by the muon \pt, where $\Delta\eta$ and $\Delta\phi$ are the differences in pseudorapidity and azimuthal angle (in radians), respectively, between the muon and the other particle's directions.
Events with additional muons or electrons are rejected using a looser quality requirement of  $\pt >10$\GeV for muons and 15\GeV for electrons,  $\abs{\eta} < 2.5$, and having $I_\text{rel}^\mu < 0.2$ and $I_\text{rel}^{\Pe} < 0.15$, where the electron relative isolation parameter $I_\text{rel}^{\Pe}$ is measured similarly to that for a muon;
\item  two or three jets with $\pt > 30\GeV$ and $\abs{\eta} < 4.7$, and, at $\sqrt{s}=8\TeV$, the highest-\pt jet ($\rm j_1$) is required to satisfy $\pt(\rm j_1) > 40\GeV$.
For events with 3 jets we require the second-highest-\pt jet ($\rm j_2$) to have $\pt(\rm j_2) > 40\GeV$;
\item  at least one b-tagged jet and at least one jet that fails the combined secondary vertex algorithm tight b tagging working-point requirement~\cite{BTag2011}.
A tight b tagging selection corresponds to an efficiency of ${\approx} 50\%$ for jets originating from true b quarks and a mistagging rate of ${\approx} 0.1\%$ for other jets in the signal simulation.
\end{enumerate}
Control regions containing events with 2 or 3 jets and no b-tagged jet, and events with 4 jets, 2 of which are b-tagged, are used to validate the modelling of the \wjets and \ttbar backgrounds, respectively.
The multijet events contribute background when there is a muon from the semileptonic decay of a b or c quark, or a light charged hadron is misidentified as a muon.
These background muons candidates are usually surrounded by hadrons.
This feature is exploited to define a control region by demanding exactly one muon with an inverted isolation criteria for hadronic activity of ${0.35 < I_\text{rel}^\mu < 1}$.
The jets falling inside the cone of a size $\Delta R = 0.5$ around the selected muon are removed and the remaining jets are subject to the criteria that define the signal.
To suppress the multijet background, we use a dedicated Bayesian neural network (multijet BNN), with the following input variables, sensitive to multijet production: the transverse mass $\MtW = \sqrt{\smash[b]{2\pt(\mu)\MET(1-\cos[\Delta \phi(\mu,\ptvecmiss)])}}$ of the reconstructed W boson, the azimuthal angle $\Delta \phi(\mu,\ptvecmiss)$ between the muon direction and \ptvecmiss, the quantity \MET, and the muon $\pt$.
The same set of variables is used for both the $\sqrt{s}=7$ and 8\TeV data sets, but because of the different selection criteria, different BNNs are trained for each set.
In Fig.~\ref{fig:QCD normalization}, data-to-simulation comparisons are shown for the multijet BNN discriminant and the \MtW distributions for the $\sqrt{s}=8\TeV$ data.
The predictions for the multijet BNN discriminant and \MtW agree with the data.
The  normalization of the multijet background is taken from a fit to the multijet BNN distribution, and all other processes involving a W boson are normalized to their theoretical cross sections.
To reduce the multijet background, the multijet BNN discriminant is required to have a value greater than 0.7.
Using the value of the discriminant rather than a selection on \MtW increases the signal efficiency by 10\%, with a similar background rejection.
This requirement rejects about 90\% of the multijet background, while rejecting only about 20\% of the signal, as determined from simulation.
The observed and predicted event yields before and after the multijet background suppression are listed in  Table~\ref{tab:QCD cut flow}.
\begin{table}[hbt]
\center{
\def\arraystretch{1.25}
\topcaption{The predicted and observed events yields before and after the multijet BNN selection for the two data sets. The uncertainties include the estimation of the scale and parton distribution function uncertainties.}
\label{tab:QCD cut flow}
\begin{tabular}{c|cc|cc}
& \multicolumn{2}{c|}{ $\sqrt{s}=7\TeV$ } & \multicolumn{2}{c}{ $\sqrt{s}=8\TeV$} \\ \hline
{Process} & {Basic selection}         & {Multijet BNN $>$ 0.7} & {Basic selection}         & {Multijet BNN $>$ 0.7} \\ \hline
$t$ channel        & \phantom{0}5\,580$^{+220}_{-160}$             &  \phantom{0}4\,560$^{+180}_{-130}$     & 21\,900$^{+980}_{-840}$\phantom{$^{0}$}            & 14\,800$^{+660}_{-560}$\phantom{$^{0}$}    \\
$s$ channel        & \phantom{0}373$^{+16}_{-14}$                & \phantom{0}301$^{+13}_{-12}$        & 1\,307$\pm$47                     & \phantom{0\,}865$\pm$31               \\
tW                 & \phantom{0}2\,080$\pm$160                     & \phantom{0}1\,760$\pm$130             & \phantom{0}9\,220$\pm$620                   & \phantom{0}6\,620$\pm$450             \\
\ttbar           & 20\,450$^{+770}_{-900}$\phantom{$^{0}$}            & 17\,360$^{+660}_{-770}$\phantom{$^{0}$}    & 101\,100$^{+5100}_{-6100}$\phantom{0}         & 72\,200$^{+3\,600}_{-4\,300}$  \\
\wjets           & 16\,100$\pm$800                    & 12\,700$\pm$630            & 36\,100$^{+1200}_{-1200}$          & 23\,700$\pm$800    \\
Dibosons          & \phantom{0,}380$\pm$10                       & \phantom{0,}300$\pm$8\phantom{0}                & \phantom{0\,}780$\pm$20                       & \phantom{0\,}537$\pm$14               \\
Drell--Yan        & 1\,520$\pm$80                      & \phantom{0\,}660$\pm$40               & \phantom{0}5\,960$\pm$320                     & \phantom{0}2\,060$\pm$110             \\
Multijets              & \phantom{0}7\,340$^{+3\,700}_{-3\,400}$           & \phantom{0\,}740$^{+380}_{-350}$      & 30\,200$^{+6\,000}_{-6\,300}$         & 2\,630$^{+520}_{-550}$     \\
\hline
Total       & \phantom{\,}53\,800$^{+3\,900}_{-3\,700}$          & \phantom{\,}38\,380$^{+1\,000}_{-1\,100}$  & 206\,650$^{+8\,100}_{-8\,900}$\phantom{0}         & 123\,400$^{+3\,800}_{-4\,500}$\phantom{0} \\
\hline
Data             & 56\,145                            & 40\,681                    & 222\,242                           & 135\,071                   \\
\end{tabular}
}
\end{table}

\section{Signal extraction with Bayesian neural networks}
\label{sec:analysis}
\begin{figure}[b]
\centering
\includegraphics[width=0.49\textwidth]{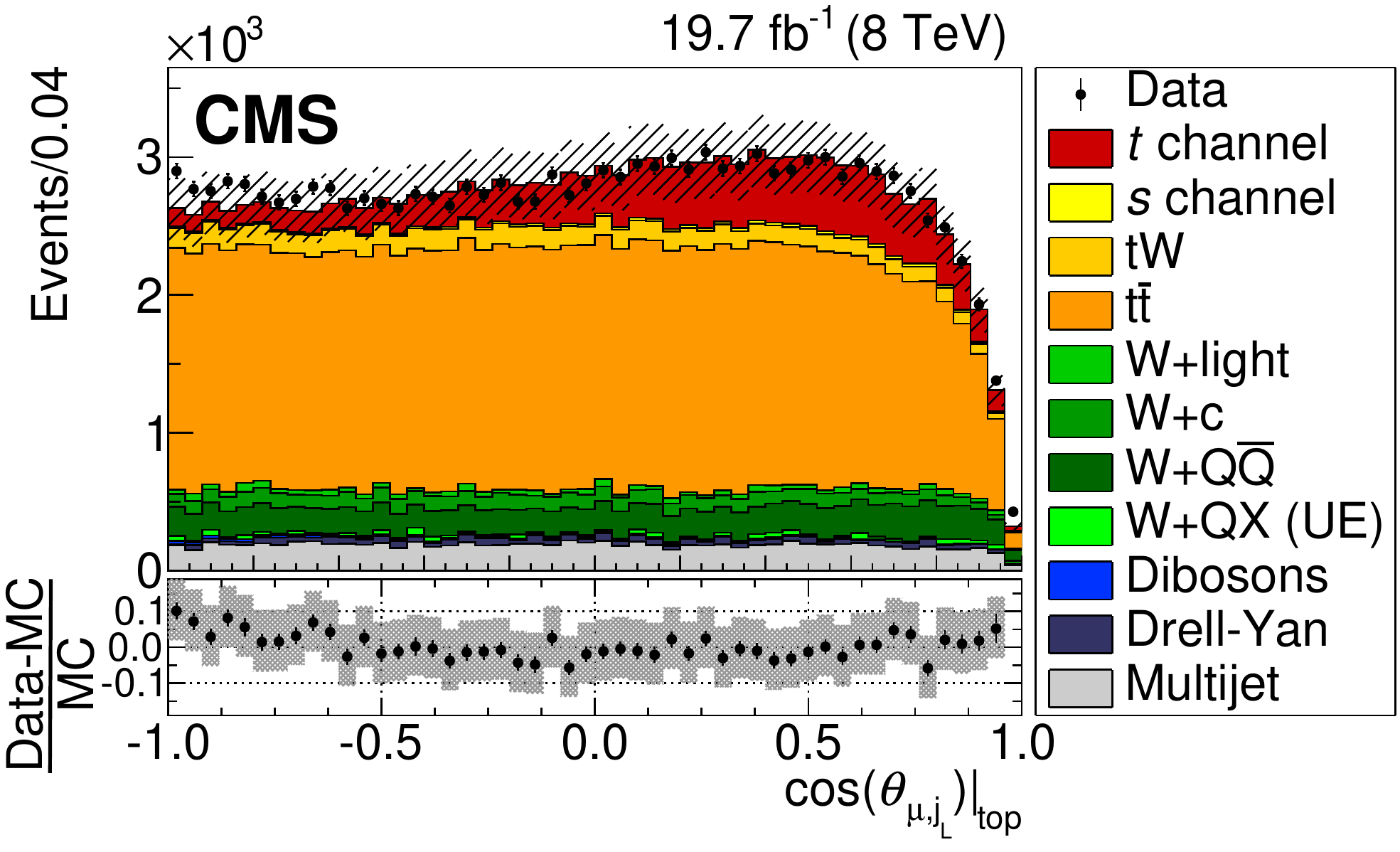}
\includegraphics[width=0.49\textwidth]{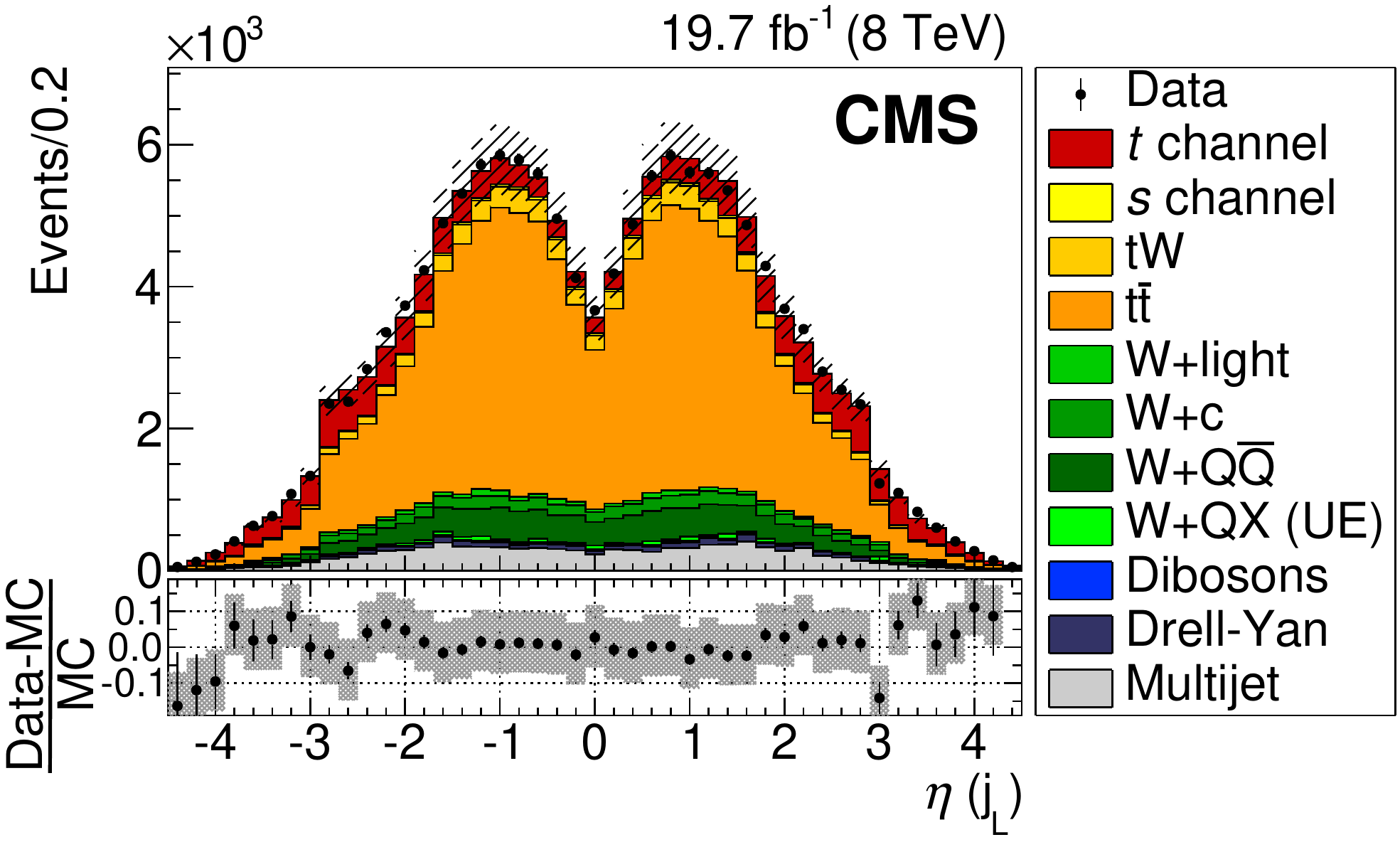}
\includegraphics[width=0.49\textwidth]{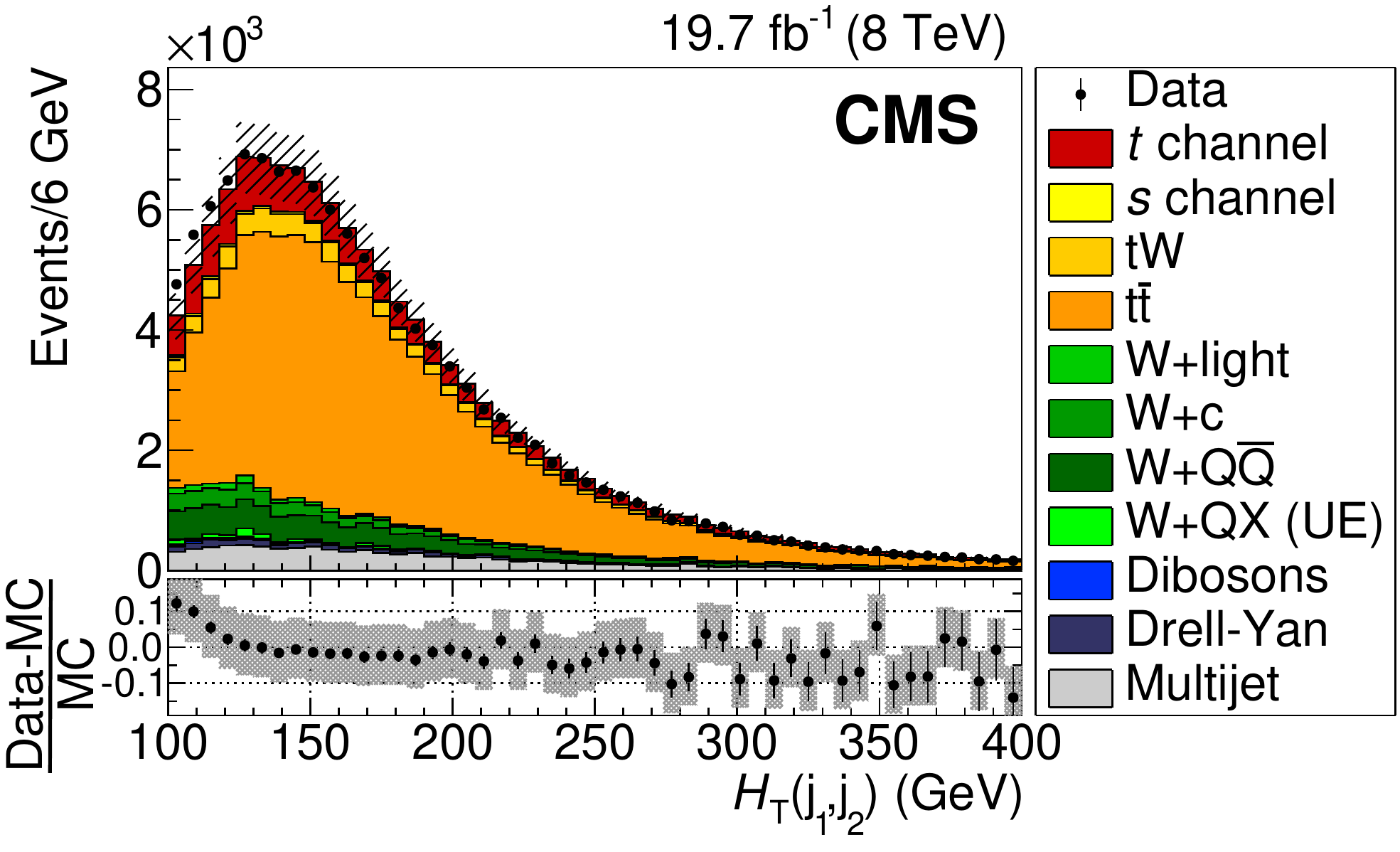}
\includegraphics[width=0.49\textwidth]{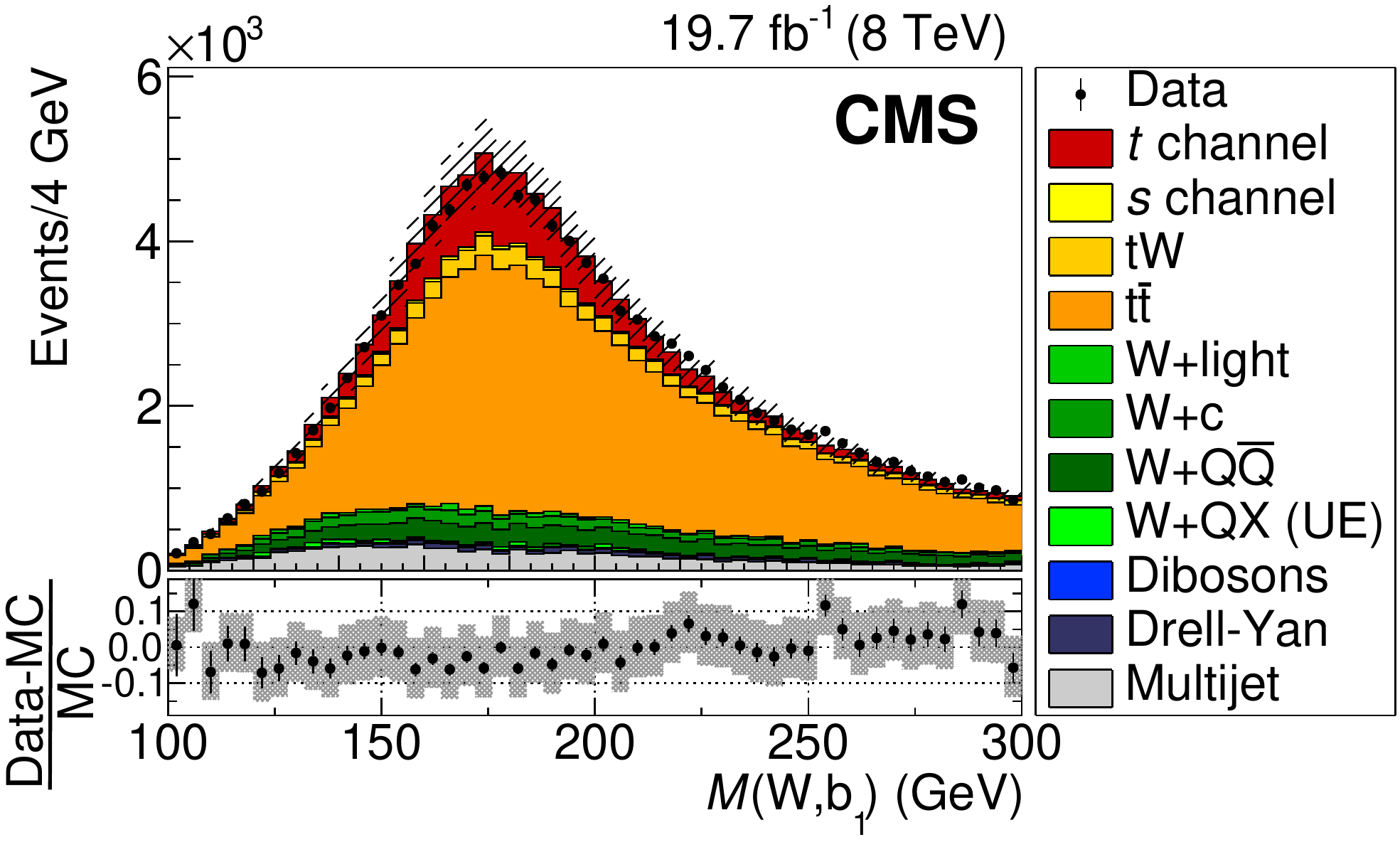}
\caption{Comparison of experimental with simulated data of the BNNs input variables $\cos(\theta_{\mu,\rm {\rm j_L}})|_\text{top}$, $\eta(\rm j_L)$, $H_{\rm T}(\rm j_1, j_2)$, and $M(\rm W,\rm {b_1})$. The variables are described in Table~\ref{tab:BNN input vars}.
The lower part of each plot shows the relative difference between the data and the total predicted background.
The hatched band corresponds to the total simulation uncertainty.
The vertical bars represent the statistical uncertainties.
Plots are for the $\sqrt{s}=$ 8\TeV data set.
}
\label{fig:BNN input vars}
\end{figure}

Events that pass the initial selection and the multijet BNN discriminant requirement are considered in the final analysis, which requires the training of the BNN (SM BNN) to distinguish the $t$-channel single top quark production process from other SM processes.
The $s$- and tW-channels, \ttbar, \wjets, diboson, and Drell--Yan processes with their relative normalizations are treated as a combined background for the training of the SM BNN.
The SM BNN discriminant is used to remove the SM backgrounds in the search for an anomalous structure at the Wtb vertex.
Three additional Wtb BNNs are used to separate the individual contributions of right-handed vector ($f_{\rm V}^{\rm R}$), left-handed ($f_{\rm T}^{\rm L}$) and right-handed ($f_{\rm T}^{\rm R}$) tensor couplings from the left-handed vector coupling  ($f_{\rm V}^{\rm L}$) expected in the SM. The physical meanings of these couplings are discussed in Section~\ref{sec:awtb}.
The FCNC processes with anomalous tcg and tug vertices are assumed to be completely independent of the SM contribution.
In addition tcg BNN and tug BNN are trained to distinguish the corresponding contributions from the SM contribution.
The kinematic properties of the potential tcg and tug contributions are slightly different owing to the different initial states and the discussion of the couplings appears in Section~\ref{sec:fcnc}.
The input variables used by each BNN are summarised in Table~\ref{tab:BNN input vars}.
Their choice is based on the difference in the structure of the Feynman diagrams contributing to the signal and background processes.
Distributions of four representative variables for data and simulated events are shown in Fig.~\ref{fig:BNN input vars}.
Several variables in the analysis require full kinematic reconstruction of the top quark and W boson candidates.
For the kinematic reconstruction of the top quark, the W boson mass constraint is applied to extract the component of the neutrino momentum along the beam direction ($p_z$).
This leads to a quadratic equation in $p_z$.
For two real solutions of the equation, the smaller value of $p_z$ is used as the solution.
For events with complex solutions, the imaginary components are eliminated by modifying \MET such that $ m_{\rm T}(\rm W) = M_{\rm W}$~\cite{Agashe:2014kda}.
\newcommand\T{\rule{0pt}{2.6ex}}       
\newcommand\B{\rule[-1.2ex]{0pt}{0pt}} 

\begin{figure}[htb]
\centering
\includegraphics[width=0.49\textwidth]{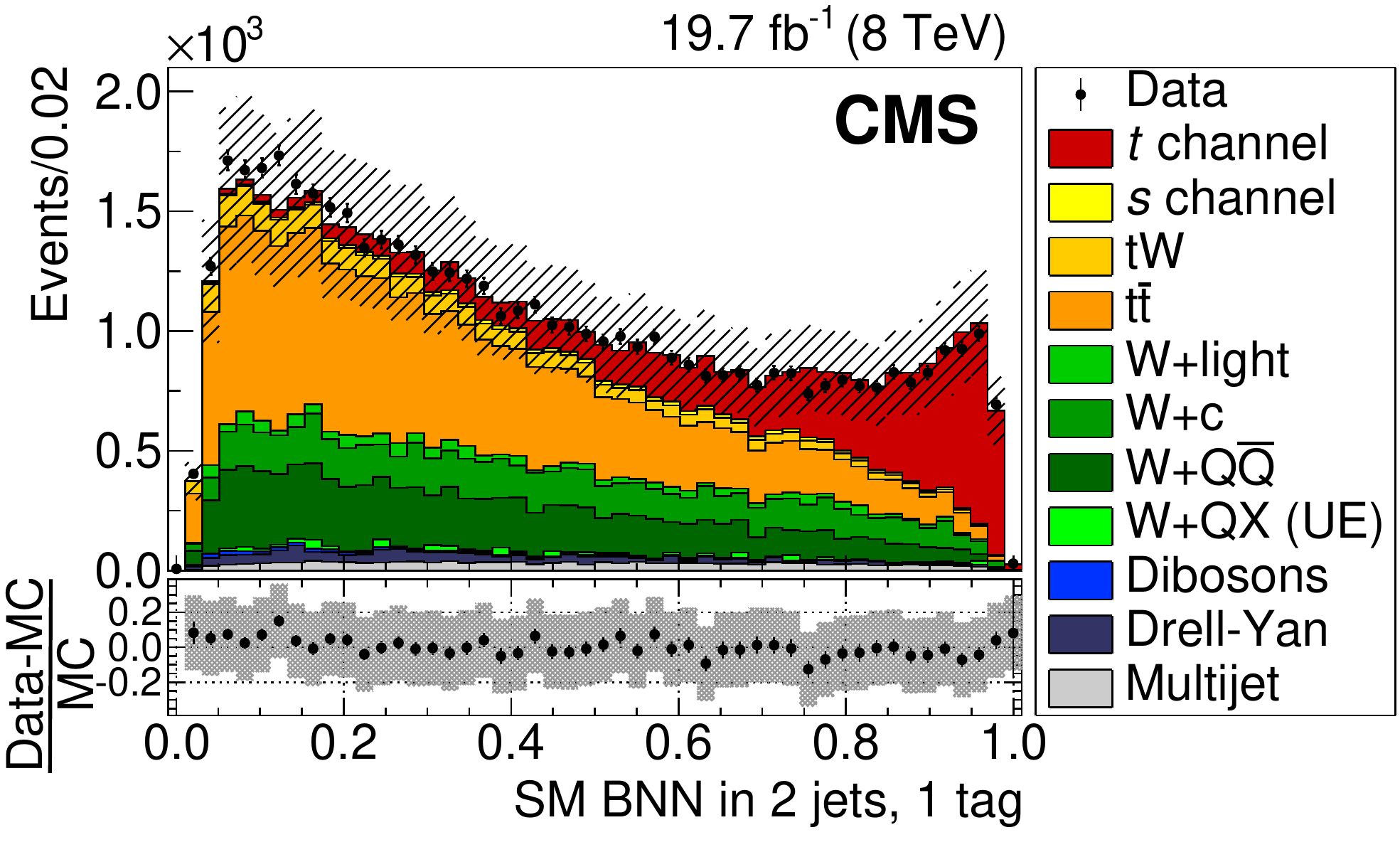}\\
\includegraphics[width=0.49\textwidth]{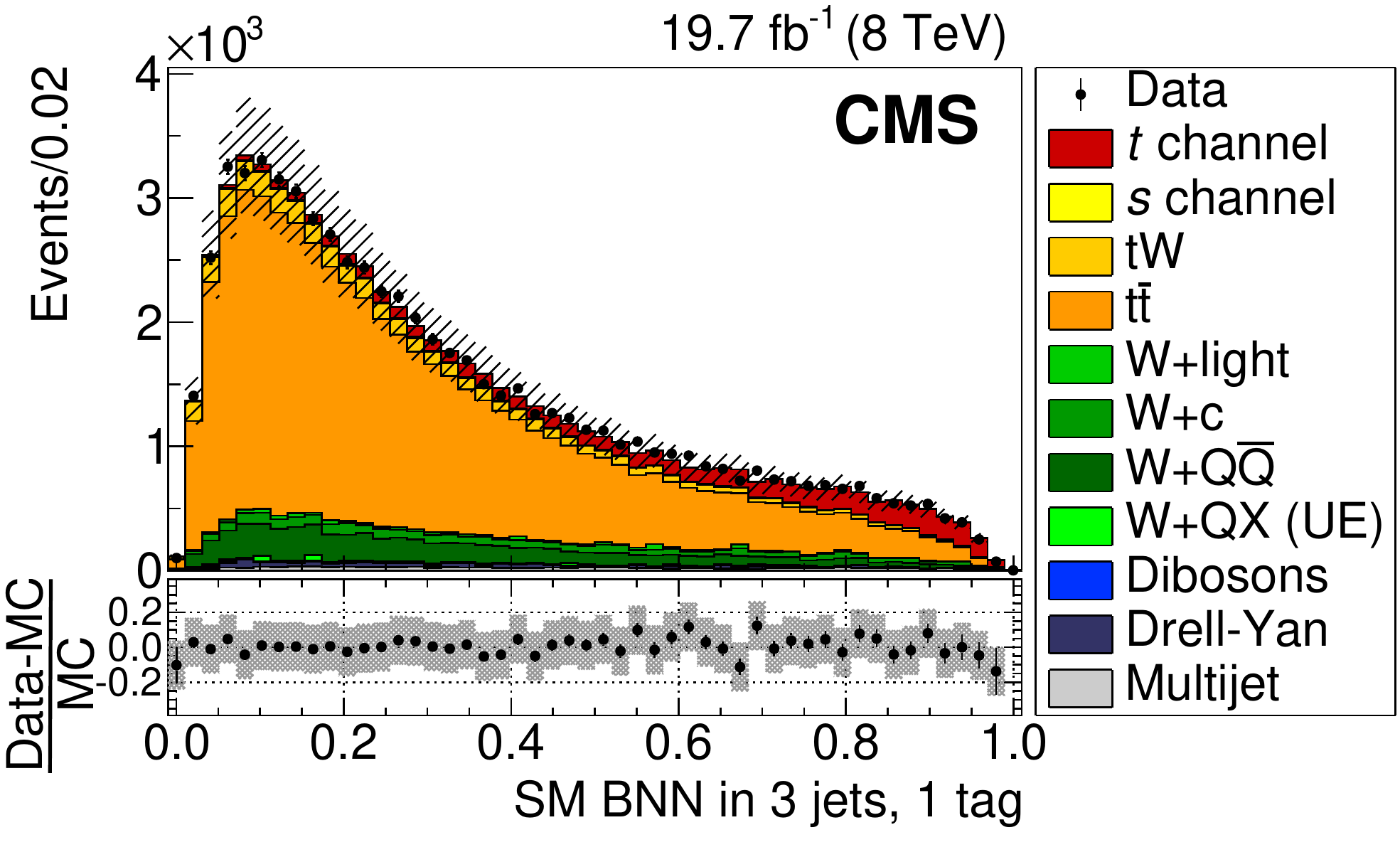}
\includegraphics[width=0.49\textwidth]{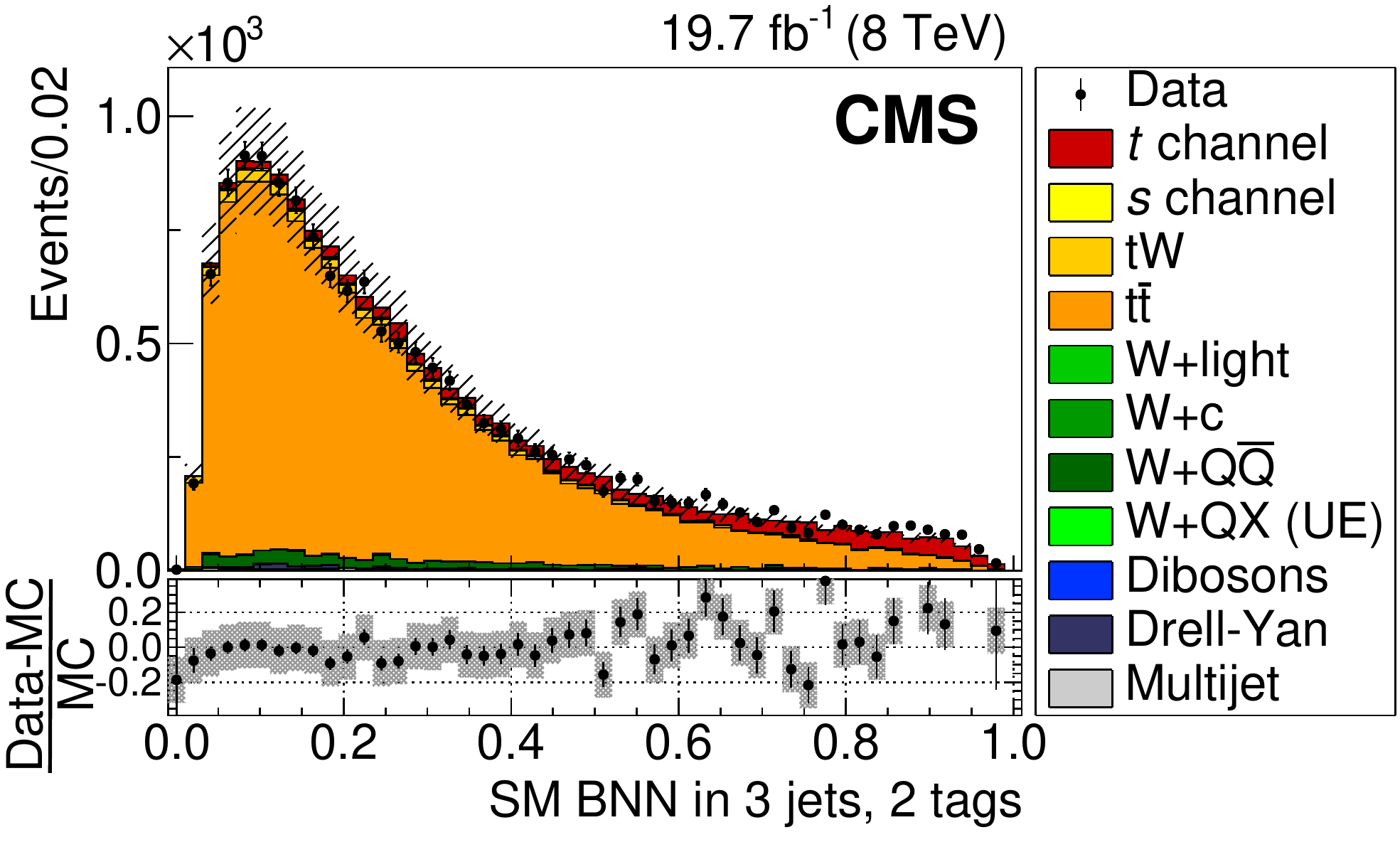}
\includegraphics[width=0.49\textwidth]{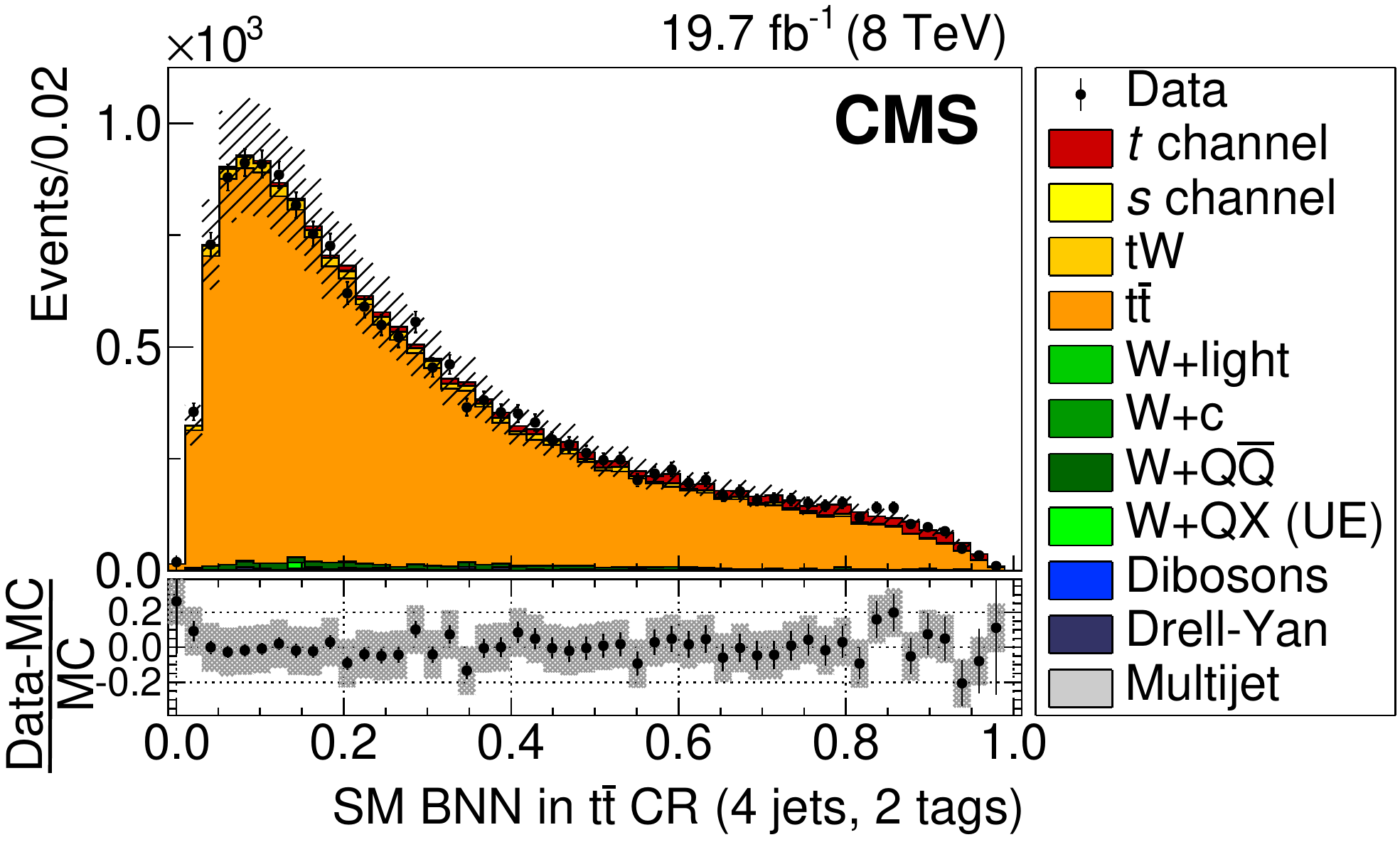}
\includegraphics[width=0.49\textwidth]{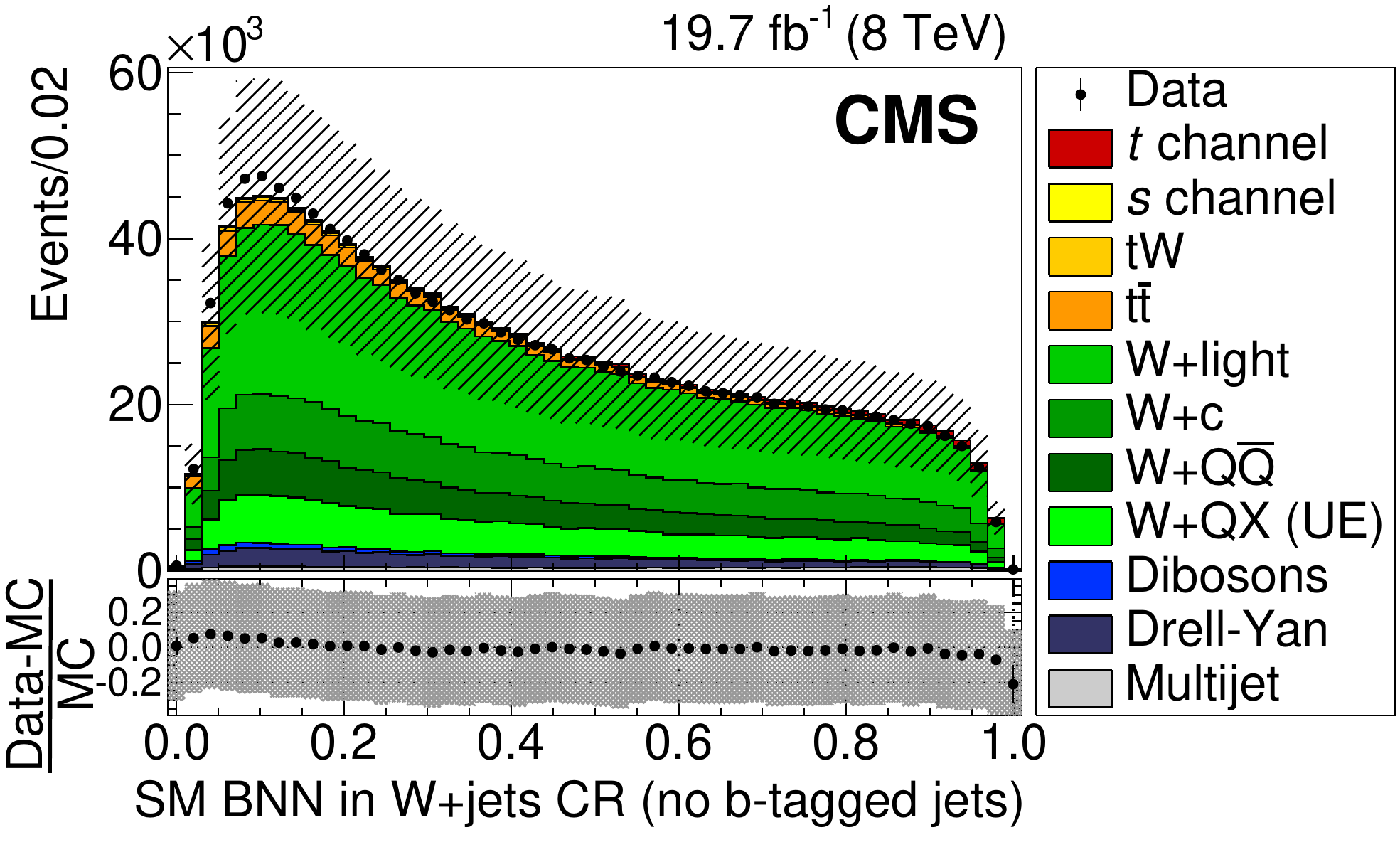}
\caption{Comparison of $\sqrt{s} = 8\TeV$ data and simulation using the SM BNN discriminant in three separate signal regions of two jets with one b-tagged (2 jets, 1 tag) (upper), three jets with one of them b-tagged (3 jets, 1 tag) (middle left), and three jets with two of them b-tagged (3 jets, 2 tags) (middle right), and in \ttbar (4 jets, 2 tags) (lower left) and \wjets (no b-tagged jets) (lower right) background control regions (CR).
The lower part of each plot shows the relative difference between the data and the total predicted background.
The hatched band corresponds to the total simulation uncertainty.
The vertical bars represent the statistical uncertainties.
}
\label{fig:BNNs discr}
\end{figure}

\begin{table}[ph]
\begin{center}
\topcaption{Input variables for the BNNs used in the analysis. The symbol $\times$ represents the variables used for each particular BNN.
The number 7 or 8 marks the variables used in just the $\sqrt{s}=$ 7 or 8\TeV  analysis.
The symbol "tug" marks the variables used just in the training of the tug FCNC BNN.
The notations "leading" and "next-to-leading" refer to the highest-\pt and second-highest-\pt jet, respectively. The notation "best" jet is used for the jet that gives a reconstructed mass of the top quark closest to the value of 172.5\GeV, which is used in the MC simulation.}
\label{tab:BNN input vars}
\small
\def\arraystretch{1.1}
\resizebox{\textwidth}{!}{
\begin{tabular}{p{0.16\linewidth}| p{0.53\linewidth}| c | c | c | c | c }
Variable & \multicolumn{1}{c|}{Description} & SM & $f_{\rm V}^{\rm L}f_{\rm V}^{\rm R}$ & $f_{\rm V}^{\rm L}f_{\rm T}^{\rm L}$ & $f_{\rm V}^{\rm L}f_{\rm T}^{\rm R}$ & FCNC \B \\
\hline
$\pt(\rm b_{\rm 1})$ & \begin{tabular}{l} \pt of the leading b jet \\ (the b-tagged jet with the highest \pt)  \end{tabular} &  $\times$ &   & & &  \\
\hline
$\pt(\rm b_{\rm 2})$& \begin{tabular}{l} \pt of the next-to-leading b jet \end{tabular} &  7 &   &  &  &     \\
\hline
$\pt(\rm j_1)$ & \begin{tabular}{l} \pt of the leading jet  \end{tabular} &  &   & $\times$  & $\times$ &  $\times$     \\
\hline
$\pt(\rm j_1,j_2)$ & \begin{tabular}{l}vector sum of the \pt of the leading \\and the next-to-leading jet  \end{tabular} &  $\times$ &   & $\times$  & &  \\
\hline
$\pt(\sum_{\small \, i \neq i_{\rm best}} \vec{p_{\rm T}}({\rm j}_i))$ & \begin{tabular}{l} vector sum of the \pt of all jets without the best jet  \end{tabular}  &  7 &   &   & &   \B \\
\hline
$\pt(\rm j_L)$ & \begin{tabular}{l} \pt of the light-flavour jet \\(untagged jet with the highest value of $\abs{\eta}$) \end{tabular} &  $\times$ &   & $\times$  & $\times$ & $\times$    \\
\hline
$\pt(\mu)$ & \begin{tabular}{l} \pt of the muon \end{tabular} &  7 & $\times$  & $\times$  & &     \\
\hline
$\pt(\rm W,b_{\rm 1})$ & \begin{tabular}{l} \pt of the W boson and the leading b jet \end{tabular} &  $\times$ &   & $\times$  & $\times$ & $\times$    \\
\hline
$H_{\rm T}(\rm j_1,j_2)$ & \begin{tabular}{l} scalar sum of the \pt of the leading \\and the next-to-leading jet \end{tabular} & $\times$  &   & $\times$  & $\times$ & $\times$     \\
\hline
\MET & \begin{tabular}{l} missing transverse energy  \end{tabular} &   & $\times$  &   & &  \B  \\
\hline
$\eta(\mu)$& \begin{tabular}{l} $\eta$ of the muon \end{tabular} &  $\times$ &   &   & &     \\
\hline
$\eta(\rm j_L)$& \begin{tabular}{l} $\eta$ of the light-flavour jet  \end{tabular} &  $\times$ &   & $\times$  & & $\times$    \\
\hline
$M(\rm j_1,j_2)$ & \begin{tabular}{l} invariant mass of the leading \\and the next-to-leading jets \end{tabular} &  $\times$ &   &  $\times$ & $\times$ & $\times$   \\
\hline
$M(\sum_{\small \, i \neq i_{\rm best}} ({\rm j}_i))$ & \begin{tabular}{l} invariant mass of all jets without the best one  \end{tabular} &  7 &   &   & & \B    \\
\hline
$M(\rm jW)$ & \begin{tabular}{l} invariant mass of the W boson and all jets  \end{tabular} &  $\times$ &   &   &  $\times$ &  $\times$   \\
\hline
$M(\rm W,b_{\rm 1})$ & \begin{tabular}{l} invariant mass of the W boson \\and the leading b jet  \end{tabular} &  $\times$ &   &   & &   \\
\hline
 $\Delta R(\mu,{\rm b_1})$ & \begin{tabular}{l} $\sqrt{\smash[b]{(\eta(\mu)-\eta({\rm b_1}))^2+(\phi(\mu)-\phi({\rm b_1}))^2}}$  \end{tabular} &   &   &   & 8 &  \T\B     \\
\hline
 $\Delta R(\mu,{\rm j_1})$ & \begin{tabular}{l} $\sqrt{\smash[b]{ \, ( \eta(\mu)-\eta({\rm j_1}))^2 \, + \, (\phi(\mu)-\phi({\rm j_1}))^2}}$ \end{tabular} &   &   &   & 7 &    \T\B  \\
\hline
$\Delta \phi(\mu,\MET)$ & \begin{tabular}{l} azimuthal angle between the muon and \ptvecmiss \end{tabular}&   &   & $\times$  & $\times$ &  \B \\
\hline
$\Delta \phi(\mu,\rm W)$ & \begin{tabular}{l} azimuthal angle between the muon \\and the W boson \end{tabular} &   &   & 8  & &      \\
\hline
$\cos(\theta_{\mu,\rm j_L})|_\text{top}$& \begin{tabular}{l} cosine of the angle between the muon \\and the light-flavour jet in the top quark rest frame, \\for top quark reconstructed with the leading b jet~\cite{Mahlon:1996pn} \end{tabular} &  $\times$ & $\times$  &   & 7 & $\times$    \\
\hline
$\cos(\theta_{\mu,\rm W})|_{\rm W}$ & \begin{tabular}{l} cosine of the angle between \\the muon momentum in the W boson rest frame \\and the direction of the W boson boost vector~\cite{Savedra:cos} \end{tabular} &  $\times$  & $\times$  &  $\times$ & &   \\
\hline
$\cos(\theta_{\rm W,j_L})|_\text{top}$ & \begin{tabular}{l} cosine of the angle between the W boson \\and the light-flavour jet \\in the top quark rest frame~\cite{Savedra:cos} \end{tabular} &  8  & $\times$  &   & &      \\
\hline
$Q(\mu)$ & \begin{tabular}{l} charge of the muon \end{tabular} &   &   &   & & tug    \\
\hline
Planarity & \begin{tabular}{l} measure of the flatness of the event \\using the smallest eigenvalue \\of the normalized momentum tensor~\cite{Barger:1987nn} \end{tabular} & 8  &   &   & &      \\
\hline
SM BNN & \begin{tabular}{l} SM BNN discriminant \end{tabular} &  &   &   & & $\times$      \\
\end{tabular}
}
\end{center}
\end{table}
The data-to-simulation comparisons shown in Fig.~\ref{fig:BNNs discr} demonstrate good agreement in the control regions enriched in top quark pair events (4 jets with 2 b tags) and \wjets (no b-tagged jets), as well as in the signal regions, as discussed in Section~\ref{sec:selection}.
In Fig.~\ref{fig:BNNs discr}, the simulated events are normalized to the results obtained in the fit to the data.

\section{Systematic uncertainties and statistical analysis}
\label{sec:systematics}
\begin{figure}[!h!tb]
\centering
\includegraphics[width=0.49\textwidth]{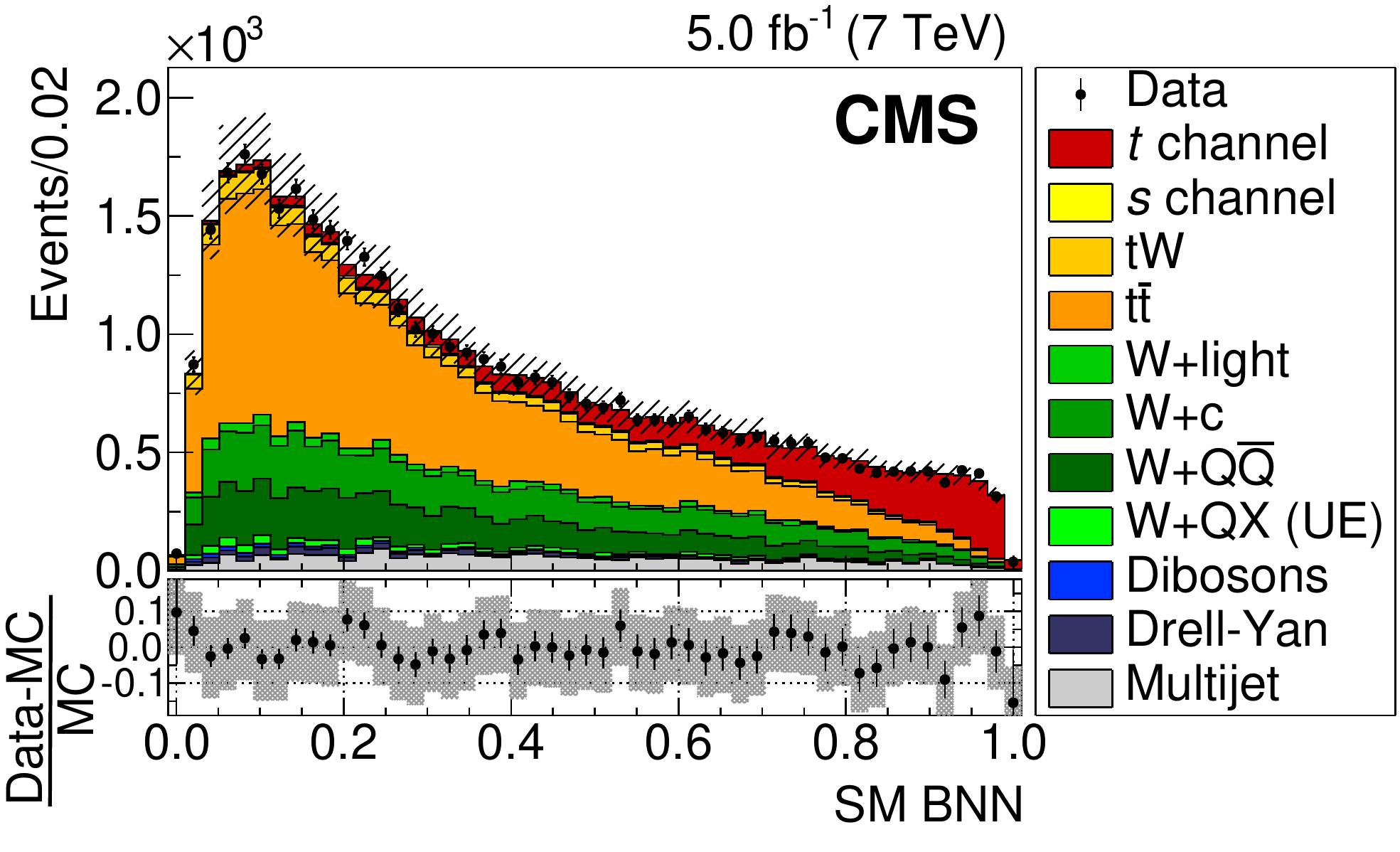}
\includegraphics[width=0.49\textwidth]{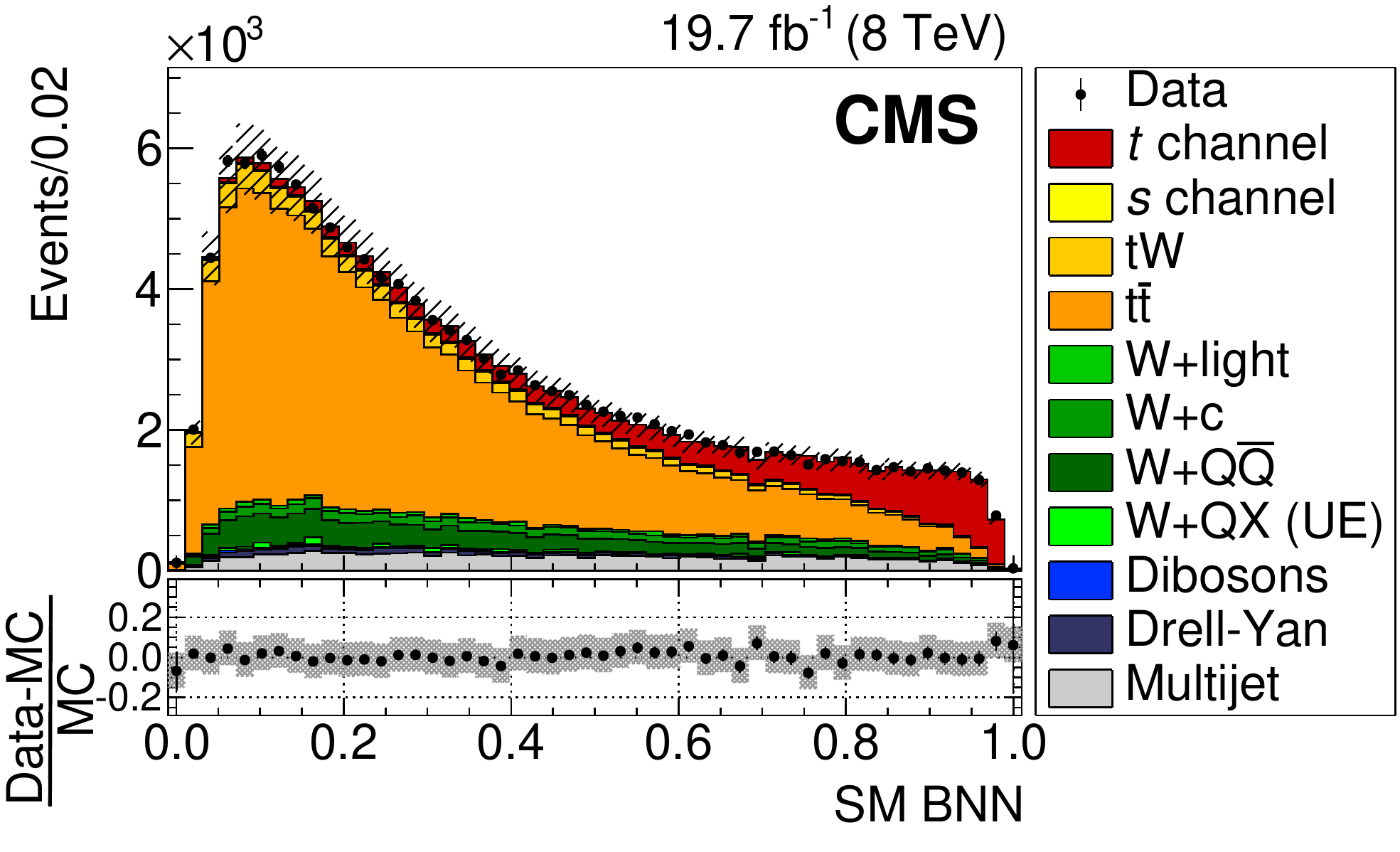}
\caption{The SM BNN discriminant distributions after the statistical analysis and evaluation of all the uncertainties.
The lower part of each plot shows the relative difference between the data and the total predicted background.
The hatched band corresponds to the total simulation uncertainty.
The vertical bars represent the statistical uncertainties.
The left (right) plot corresponds to $\sqrt{s}=7$ (8)\TeV.}
\label{fig:data/mc theta2}
\end{figure}

The analysis extracts the parameters of single top quark production and any signs of beyond the SM behaviour based on the BNN discriminant distributions.
It follows the same methodology for estimating the uncertainties as used in previously CMS measurements of single top quark production~\cite{Chatrchyan:2011vp,Chatrchyan:2012ep}.
Bayesian inference is used to derive the posterior probability.
A signal strength $\vec\mu_s$ is the central value of the posterior probability distribution $p(\vec\mu_s|d)$ with a certain data set $d$. This posterior probability can be obtained as the integral
\begin{linenomath}
\begin{equation}
p(\vec\mu_s|d) = \int p(d|\vec\mu_s, \vec\mu_b, \vec\theta) \frac{\pi(\vec\mu_s)\pi(\vec\mu_b)\pi(\vec\theta)}{\pi(d)} {\rm d} {\vec\mu_b} {\rm d}\, {\vec\theta},
\label{stat_integral}
\end{equation}
\end{linenomath}
where $\vec\mu_b$ are the background yields, $\vec\theta$ are additional nuisance parameters, which are the systematic uncertainties of the analysis, $\pi(\vec\mu_s)$, $\pi(\vec\mu_b)$, and $\pi(\vec\theta)$ are the prior probabilities of the corresponding parameters, $\pi(d)$ is a normalization factor, and $p(d|\vec\mu_s,\vec\mu_b,\vec\theta)$ is the probability to obtain a given $d$ with given $\vec\mu_s$, $\vec\mu_b$, and $\vec\theta$.
Uncertainties considered in the analysis are discussed next.
For the variation of the background normalization, scale parameters are introduced in the statistical model, and the corresponding variations of these parameters are the same as for the SM measurement in Ref.~\cite{Chatrchyan:2012ep}.
All background processes and their normalizations are treated as being statistically independent.
To estimate the uncertainty in the multijet distributions, two different isolation criteria are used ($0.3 < I_\text{rel}^\mu <0.5$ and $0.5 < I_\text{rel}^\mu < 1$).
Also, a comparison is made between data and events generated with the \PYTHIA 6.4 simulation.
The impact of the changes in the multijet template are well within the range of $-50\%$ to $+100\%$, and this is included as a prior uncertainty in the statistical model.
To estimate the uncertainties in the detector-related jet and \MET corrections, the four-momenta of all reconstructed jets in simulated events are scaled simultaneously in accordance with $\pt$- and $\eta$-dependent jet energy correction (JEC) uncertainties~\cite{JEC2010}.
These changes are also propagated to \MET.
The effect of the 10\% uncertainty in \MET coming from unclustered energy deposits in the
calorimeters that are not associated with jets is estimated after subtracting all the
jet and lepton energies from the \MET calculation.
Parameters in the procedure to correct the jet energy resolution (JER) are varied within their uncertainties, and the procedure is repeated for all jets in the simulation~\cite{JER2010,JEC2010}.
The variations coming from the uncertainty in the b quark tagging efficiency and mistagging rate of jets are propagated to the simulated events~\cite{BTag2011}.
The uncertainties for c quark jets are assumed to be twice as large as for b quark jets.
The scale factors for tagging b and c quark jets are treated as fully correlated, whereas the mistagging scale factors are varied independently.
The integrated luminosity in the $\sqrt{s}$ = 7 and 8\TeV data-taking periods is measured with a relative uncertainty of 2.2\%~\cite{Lumi7TeV} and 2.6\%~\cite{Lumi8TeV}, respectively.
In the combined fits, all experimental uncertainties, including these from the integrated luminosity, are treated as uncorrelated between the data sets.
The uncertainty in the pileup modelling is estimated by using different multiplicity distributions obtained by changing the minimum-bias cross section by $\pm5\%$~\cite{Chatrchyan:2012nj}.
Trigger scale factors, muon identification, and muon isolation uncertainties are introduced in the statistical model as additional factors, Gaussian-distributed parameters with a mean of 1 and widths of 0.2\%, 0.5\%, and 0.2\%, respectively.
The uncertainties from additional hard-parton radiation and the matching of the samples with different jet multiplicity are evaluated by doubling or halving the threshold for the \MADGRAPH jet-matching procedure for the top quark pair and \wjets production, using dedicated \MADGRAPH samples generated with such shifts in the parameters~\cite{Alwall:2008qv}.
The renormalization and factorization scale uncertainties are estimated using MC simulated samples generated by doubling or halving the renormalization and factorization scales for the signal and the main background processes.
Uncertainties in the parton distribution functions (PDF) are evaluated with the CT10 PDF set according to the PDF4LHC formulae for Hessian-based sets.
We follow this recommendation and reweight the simulated events to obtain the uncertainty, which is about 5\% on average.
The uncertainty from the choice of the event generator to model the signal is estimated using pseudo-experiments.
These pseudo-experiments are used to fit simulated events, obtained from the  {\sc CompHEP} signal sample, and from the \POWHEG signal sample.
Half of the difference between these two measurements is taken as the uncertainty (5\%).
Previous CMS studies~\cite{Chatrchyan:2012saa,Khachatryan:2015oqa} of top quark pair production showed a softer $\pt$ distribution of the top quark in the data than predicted by the NLO simulation.
A correction for the simulation of \ttbar production background is applied.
The small effect of this reweighting procedure (0.8\%) is taken into account as an uncertainty.
The uncertainty owing to the finite size of the simulated samples is taken into account through the Barlow--Beeston method~\cite{Barlow:1993dm}.
The BNN discriminant distributions can be affected by different types of systematic uncertainties.
Some of these only impact the overall normalization, while others change the shape of the distribution.
Both types of systematic uncertainties are included in the statistical model through additional nuisance parameters.
Systematic uncertainties related to the modelling of JEC, JER, b tagging and mistagging rates, \MET, and pileup, are included as nuisance parameters in the fit.
The variations in these quantities leads to a total uncertainty of about 6\%.
Other systematic uncertainties, i.e. those related to the signal model, renormalization and factorization scales, matching of partons to final jets, and choice of PDF, are handled through the pseudo-experiments to determine the difference between the varied and the nominal result.
The total uncertainty from these sources is about 8\%.
We include uncertainties in the statistical model by following the same approach as described in previous CMS measurements of the single top quark $t$-channel cross section~\cite{Chatrchyan:2012ep,Chatrchyan:2011vp,Khachatryan:2014iya}.
The SM BNN discriminant distribution after the statistical analysis and evaluation of all the uncertainties are shown in Fig.~\ref{fig:data/mc theta2} for the two data sets.
As the $\sqrt{s} = 7$ and 8\TeV data sets have similar selection criteria, reweighting, and uncertainties and the physics is expected to also be similar, the data sets are combined by performing a joint fit.
The previously described systematic uncertainties and methods of statistical analysis are used in the combination.
In the statistical model, the experimental uncertainties are treated as uncorrelated between the data sets.
The theoretical uncertainties (from the choice of generator, scales, and PDF) are treated as fully correlated between the data sets.
The sensitivity of the separate $\sqrt{s} = 7$ and 8\TeV analyses and their combination is limited by their corresponding systematic uncertainties.
Therefore, the combined statistical model does not necessarily provide the tightest exclusion limits.
In order to validate the analysis strategy and the statistical treatment of the uncertainties, we measure the cross sections in the SM $t$ channel, and find
values and uncertainties in agreement with previous measurements~\cite{Chatrchyan:2012ep,Chatrchyan:2011vp} and with the prediction of the SM.

\section{Search for anomalous contributions to the Wtb vertex}
\label{sec:awtb}
\subsection{Modelling the structure of the anomalous Wtb vertex}
\label{subsec:awtb_method}
The $t$-channel single top quark production is sensitive to possible deviations from the SM prediction for the Wtb vertex.
The most general, lowest-dimension, CP-conserving Lagrangian for the Wtb vertex has the following form~\cite{Buchmuller:1985jz,Kane:1991bg}:
\begin{linenomath}
\begin{equation}
\mathfrak{L}=\frac{g}{\sqrt{2}}\bar{\rm b}\gamma^{\mu}\left( f_{\rm V}^{\rm L}P_{\rm L} + f_{\rm V}^{\rm R}P_{\rm R}\right){\rm tW}_{\mu}^{-} - \frac{g}{\sqrt{2}}\bar{\rm b}\frac{\sigma^{\mu\nu} \partial_{\nu} {\rm W}_{\mu}^{-}}{M_{\rm W}}\left( f_{\rm T}^{\rm L}P_{\rm L} + f_{\rm T}^{\rm R}P_{\rm R}\right)\rm t + \rm {h.c.},
\label{lagrangian_wtb}
\end{equation}
\end{linenomath}
where $P_{\rm L,R}=(1\mp\gamma_{5})/2$, $\sigma_{\mu\nu}=i(\gamma_{\mu}\gamma_{\nu} -\gamma_{\nu}\gamma_{\mu})/2$, $g$ is the coupling constant of the weak interaction, the form factor $f_{\rm V}^{\rm L}$ ($f_{\rm V}^{\rm R}$) represents the left-handed (right-handed) vector coupling, and $f_{\rm T}^{\rm L}$ ($f_{\rm T}^{\rm R}$) represents the left-handed (right-handed) tensor coupling.
The SM has the following set of coupling values: $f_{\rm V}^{\rm L} = V_{\rm tb},f_{\rm V}^{\rm R} = f_{\rm T}^{\rm L} = f_{\rm T}^{\rm R} = 0$.
The same analysis scheme proposed in Refs.~\cite{Abazov:2008sz,Abazov:2011pm} is used to look for possible deviations from the SM, by postulating the presence of a left-handed vector coupling.
Two of the four couplings are considered simultaneously in two-dimensional scenarios: $(f_{\rm V}^{\rm L}$, $f_{\rm V}^{\rm R})$ and $(f_{\rm V}^{\rm L}$, $f_{\rm T}^{\rm L})$, where the couplings are allowed to vary from 0 to ${+}\infty$, and $(f_{\rm V}^{\rm L}$, $f_{\rm T}^{\rm R})$ with variation bounds from $\mbox{-}\infty$ to ${+}\infty$.
Then, considering three couplings simultaneously leads to the three-dimensional scenarios ($f_{\rm V}^{\rm L}$, $f_{\rm T}^{\rm L}$, $f_{\rm T}^{\rm R}$) and ($f_{\rm V}^{\rm L}$, $f_{\rm V}^{\rm R}$, $f_{\rm T}^{\rm R}$).
In these scenarios, the couplings have the same variation range of (0; ${+}\infty$) for $f_{\rm V}^{\rm R}$ and $f_{\rm T}^{\rm L}$, and ($\mbox{-}\infty$; ${+}\infty$) for $f_{\rm V}^{\rm L}$ and $f_{\rm T}^{\rm R}$.
In the presence of anomalous Wtb couplings in both the production and decay of the top quark, the kinematic and angular distributions are significantly affected relative to their SM expectations.
It is therefore important to correctly model the kinematics of such processes.
Following the method of Ref.~\cite{Boos:2016zmp}, the event samples with left-handed (SM) interactions and a purely right-handed vector (left-handed tensor) interactions are generated to model the $(f_{\rm V}^{\rm L}$, $f_{\rm V}^{\rm R})$ ($(f_{\rm V}^{\rm L}$, $f_{\rm T}^{\rm L})$) scenario.
Simulated event samples with the left-handed interaction in the production and the right-handed vector (left-handed tensor) interaction in the decay of the top quark, and vice versa, are also generated.
The scenarios with $f_{\rm T}^{\rm R}$ couplings are more complicated because of the presence of cross terms, such as  $(f_{\rm V}^{\rm L} \cdot f_{\rm T}^{\rm R})$, in the squared matrix element describing the single top quark production process. Special event samples are generated for such scenarios.
Owing to the presence of the cross terms with odd power of $f_{\rm V}^{\rm L}$ and $f_{\rm T}^{\rm R}$ couplings, the analysis is sensitive to negative values of these couplings. The details of the simulation approach are provided in Ref.~\cite{Boos:2016zmp}.
All signal samples are simulated at NLO precision following Ref.~\cite{Boos:2006af}.
\subsection{Exclusion limits on anomalous couplings }
\label{subsec:awtb_limits}
\begin{figure}[!h!]
\begin{minipage}[t]{\linewidth}
\begin{center}
\includegraphics[width=0.49\textwidth]{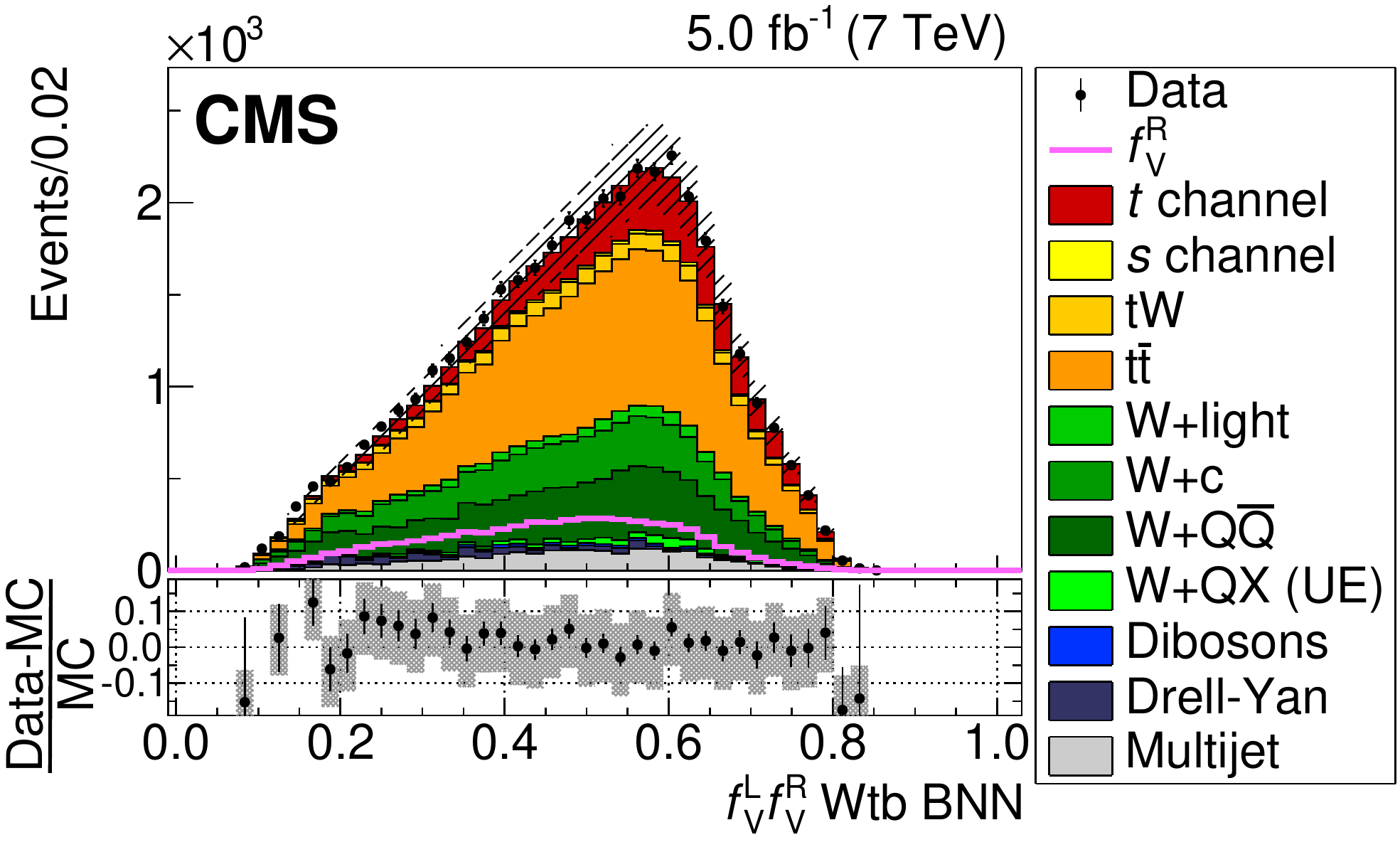}
\includegraphics[width=0.49\textwidth]{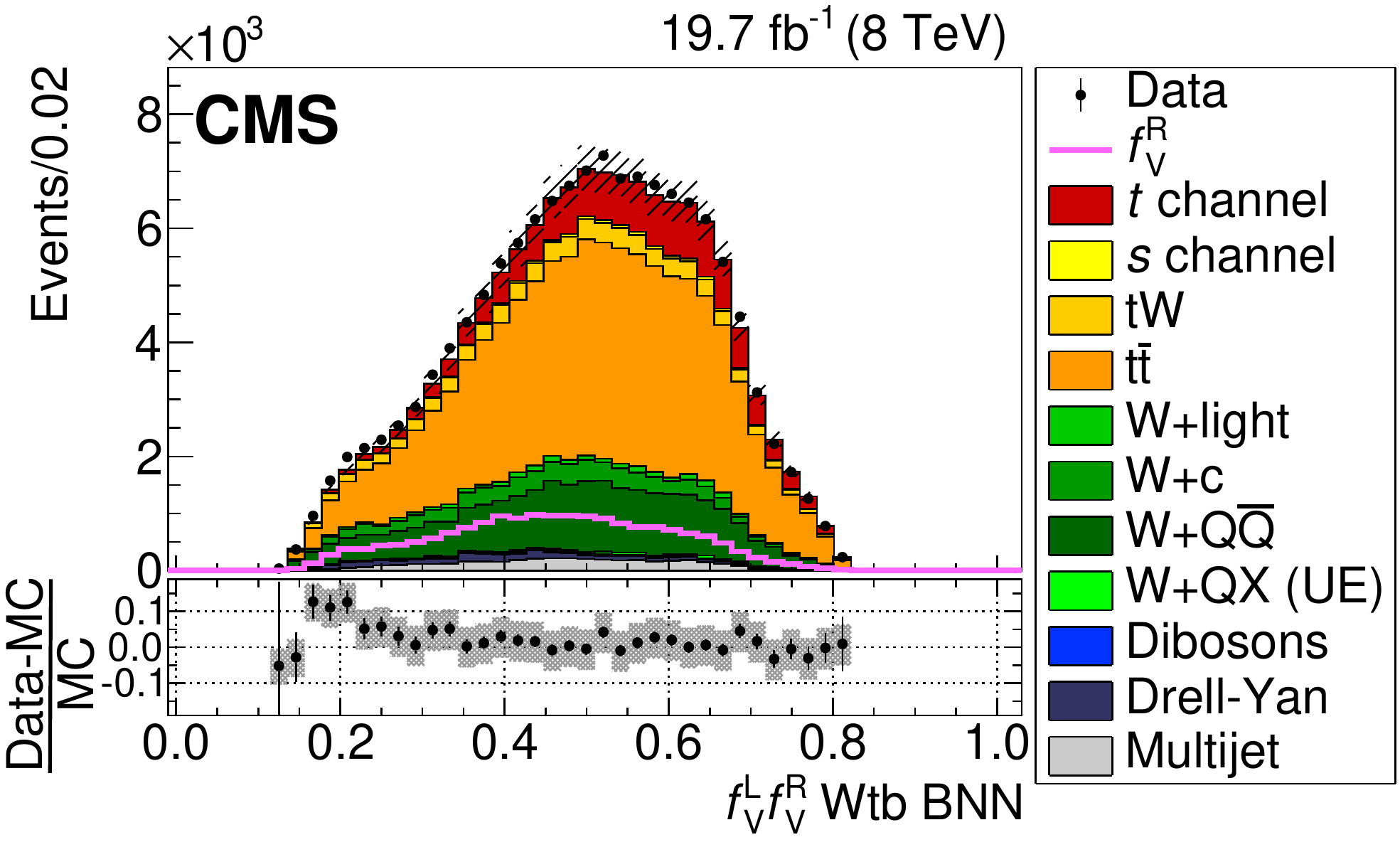}
\includegraphics[width=0.49\textwidth]{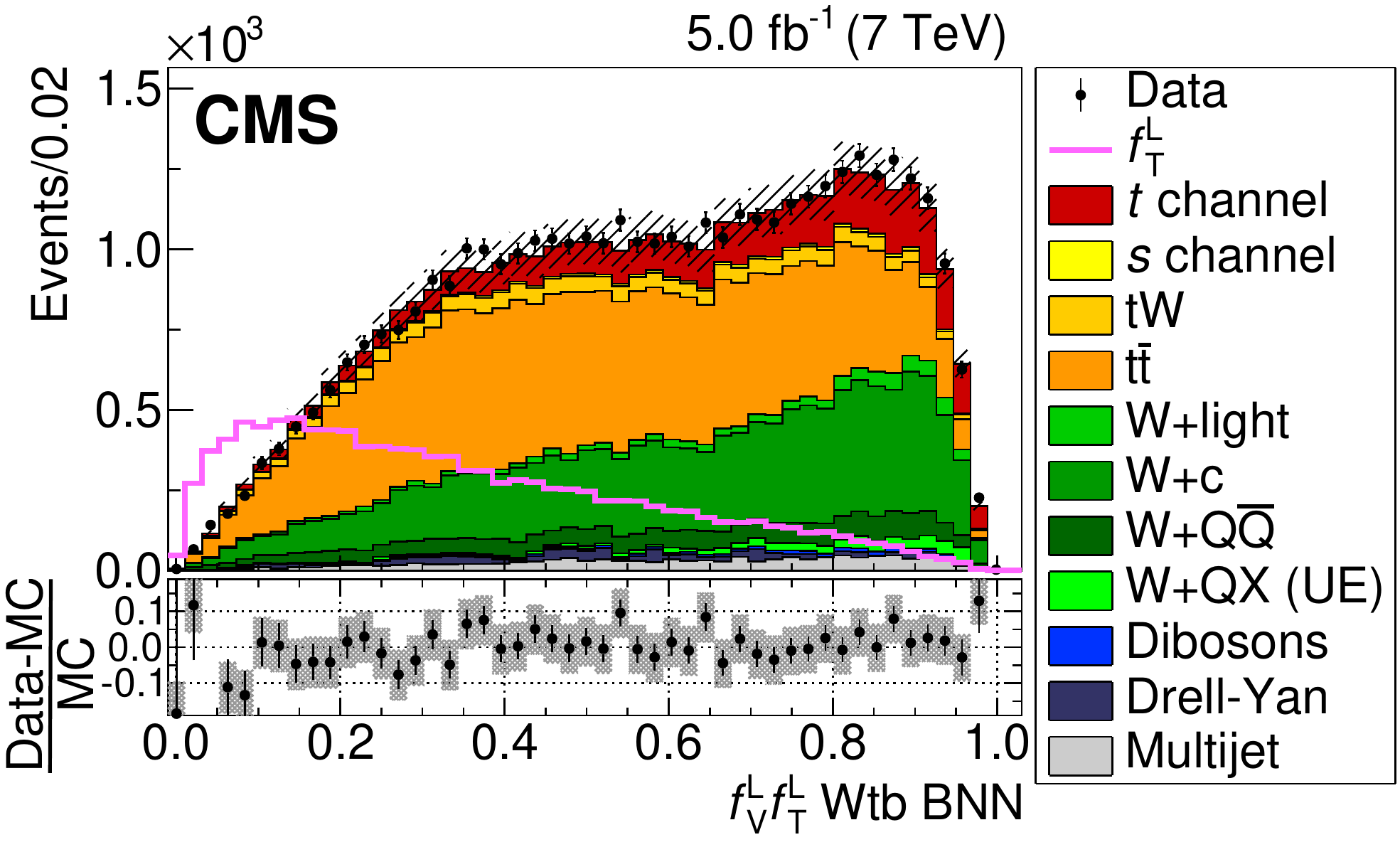}
\includegraphics[width=0.49\textwidth]{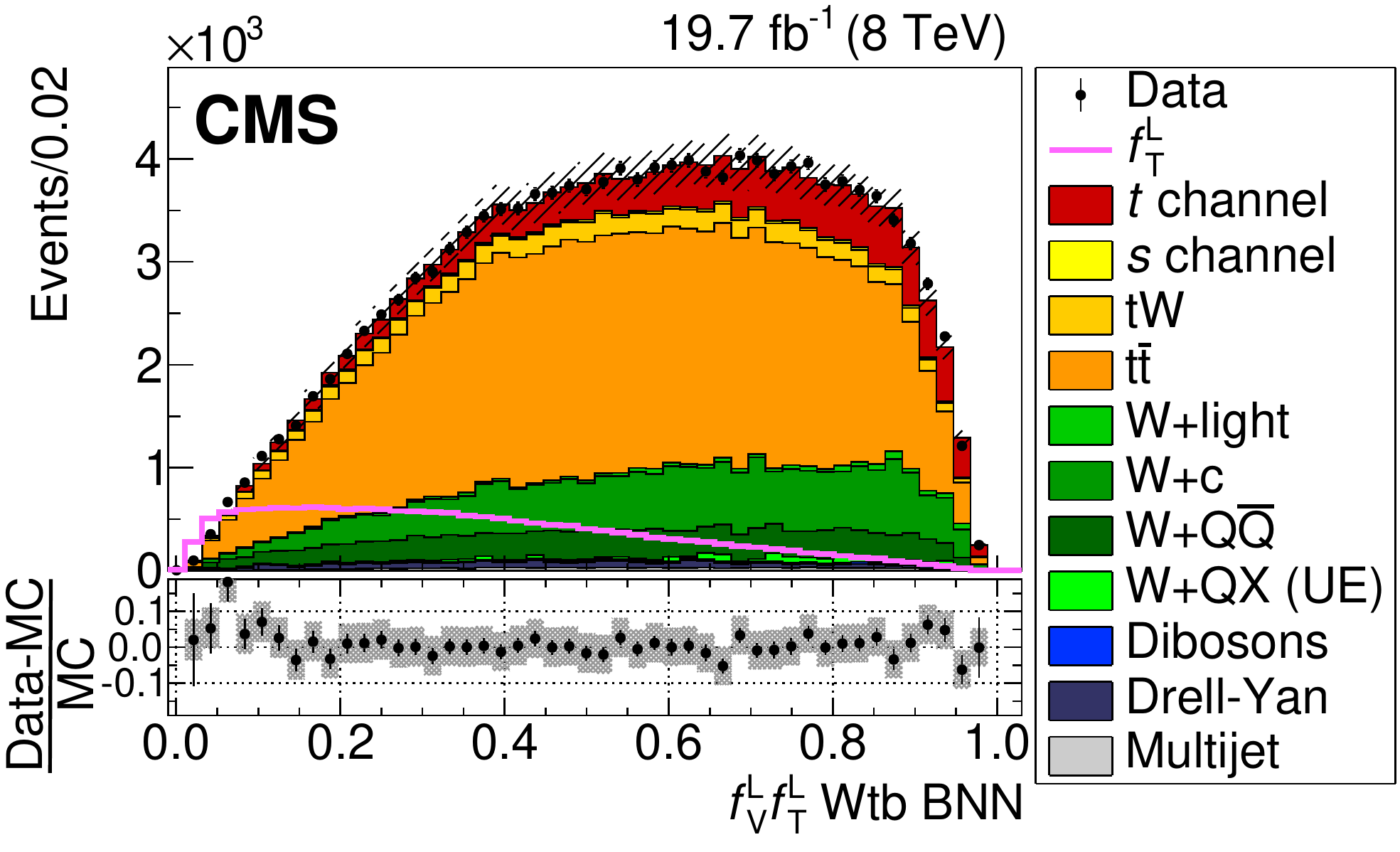}
\includegraphics[width=0.49\textwidth]{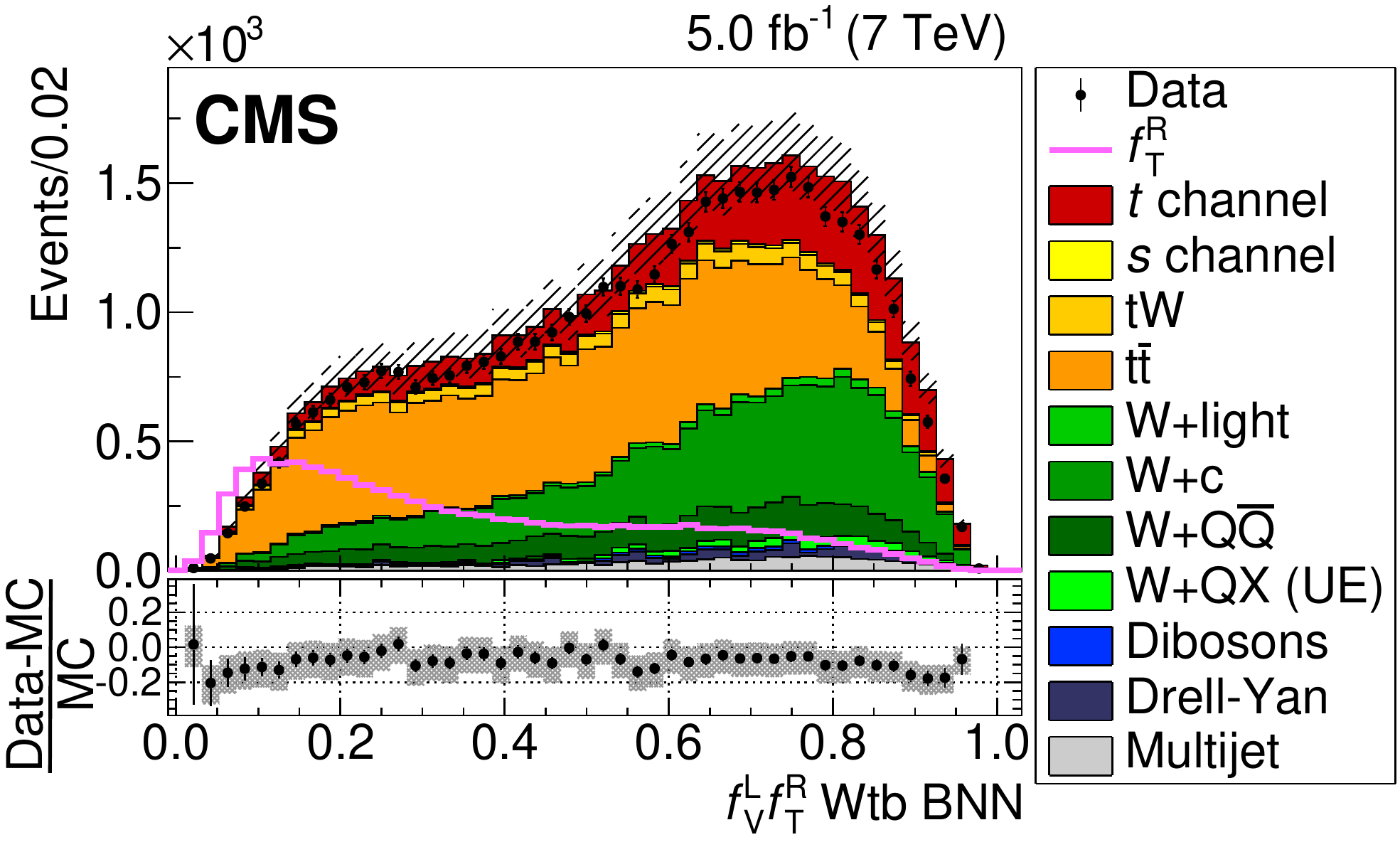}
\includegraphics[width=0.49\textwidth]{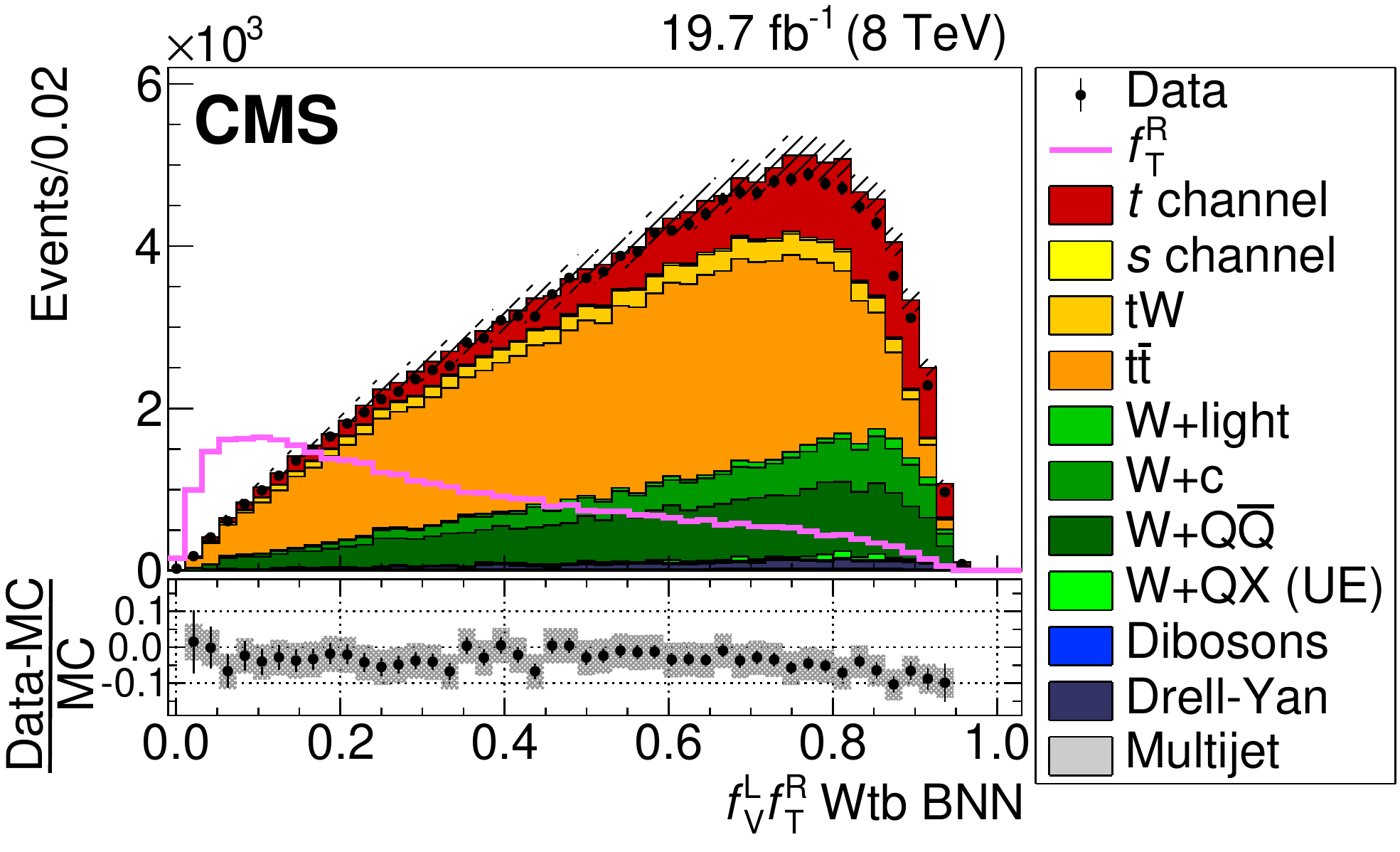}
\end{center}
\end{minipage}
\caption{Distributions of the Wtb BNN discriminants from data (points) and simulation (filled histograms) for the scenarios $(f_{\rm V}^{\rm L}$, $f_{\rm V}^{\rm R})$ (top), $(f_{\rm V}^{\rm L}$, $f_{\rm T}^{\rm L})$ (middle), and $(f_{\rm V}^{\rm L}$, $f_{\rm T}^{\rm R})$ (bottom).
The plots on the left (right) correspond to $\sqrt{s}=7$ (8)\TeV.
The Wtb BNNs are trained to separate SM left-handed interactions from one of the anomalous interactions.
In each plot, the expected distribution with the corresponding anomalous coupling set to 1.0 is shown by the solid curve.
The lower part of each plot shows the relative difference between the data and the total predicted background.
The hatched band corresponds to the total simulation uncertainty.
The vertical bars represent the statistical uncertainties.
}
\label{fig:BNN aWtb LVRV discr}
\end{figure}

\begin{figure}[!h!]
\begin{minipage}[t]{\linewidth}
\begin{center}
\includegraphics[width=0.49\textwidth]{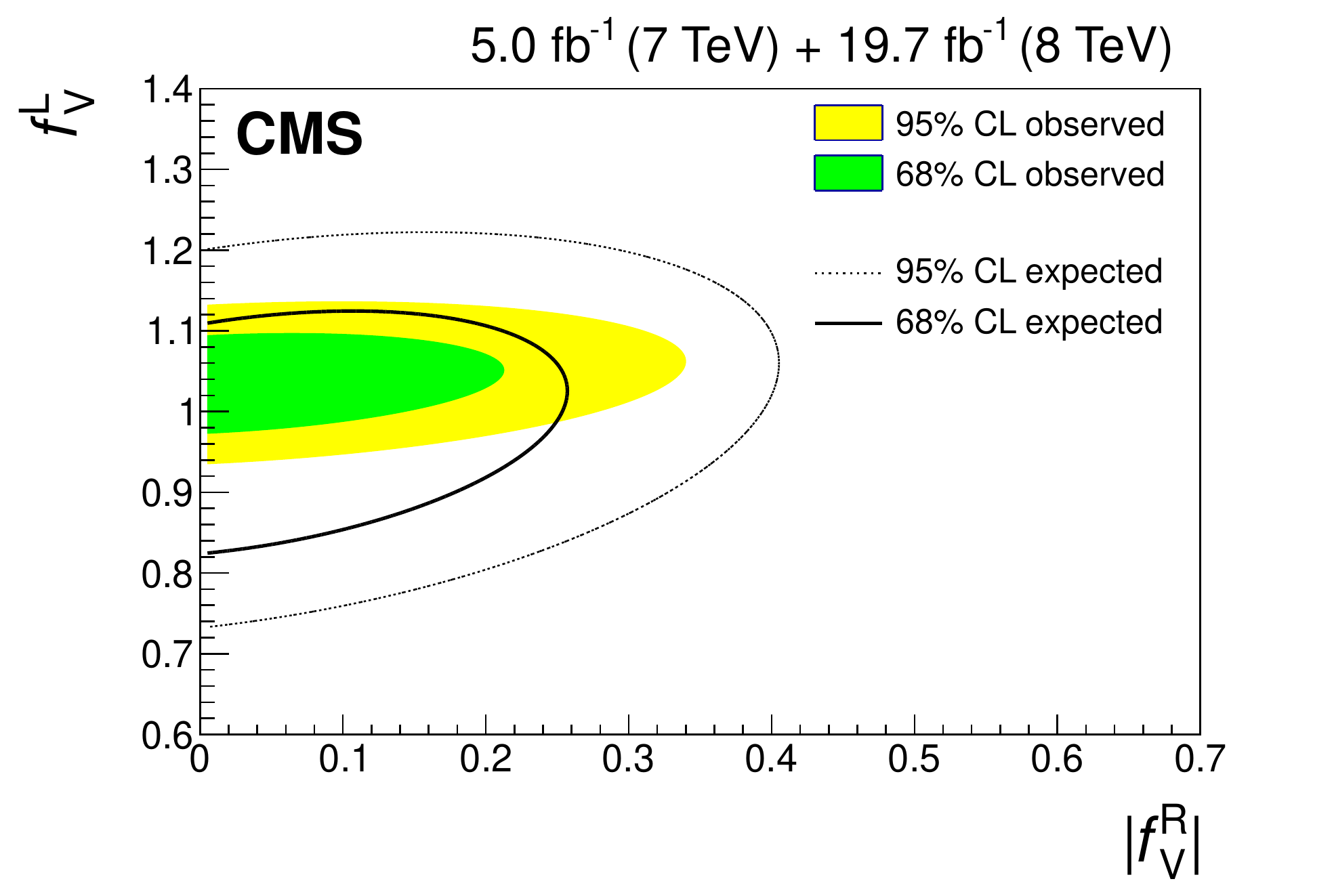}
\includegraphics[width=0.49\textwidth]{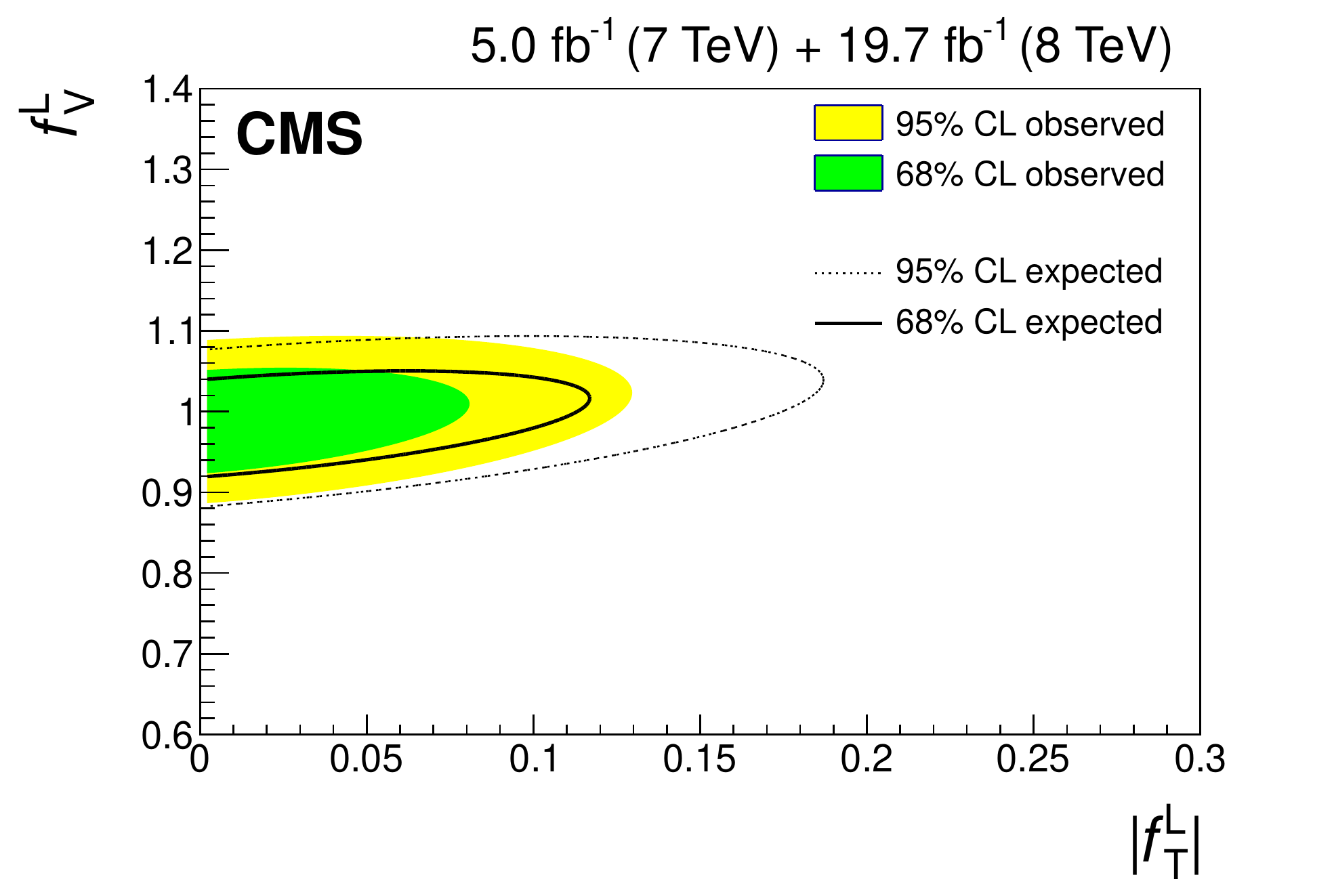}
\includegraphics[width=0.49\textwidth]{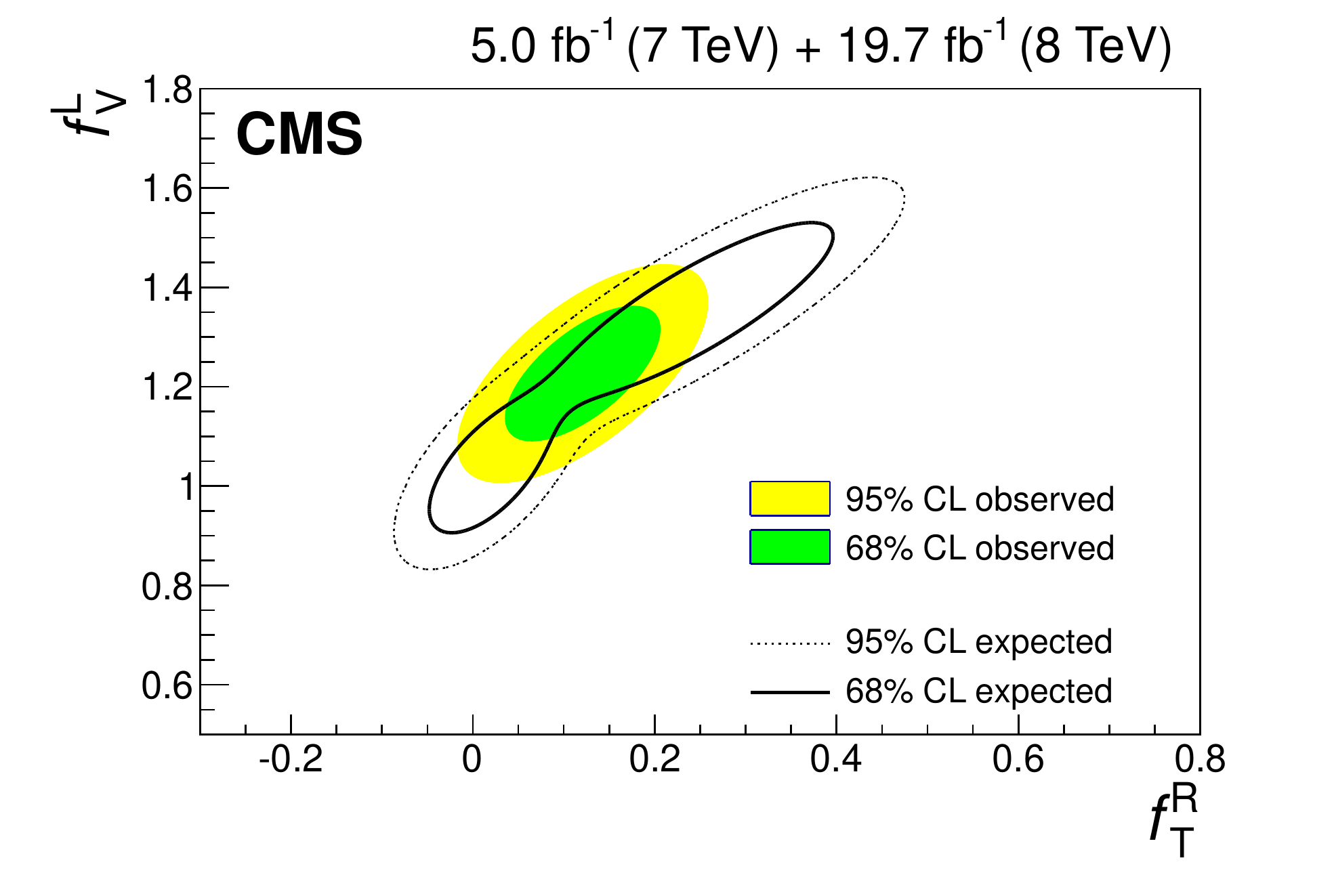}
\end{center}
\end{minipage}
\caption{Combined $\sqrt{s}=7$ and $8\TeV$ observed and expected exclusion limits in the two-dimensional planes $(f_{\rm V}^{\rm L}$, $\lvert f_{\rm V}^{\rm R}\rvert)$ (top-left), $(f_{\rm V}^{\rm L}$, $\lvert f_{\rm T}^{\rm L}\rvert)$ (top-right), and $(f_{\rm V}^{\rm L}$, $f_{\rm T}^{\rm R})$ (bottom) at 68\% and 95\% CL.}
\label{fig:2d-comb}
\end{figure}

\begin{figure}[!h!]
\begin{minipage}[t]{\linewidth}
\begin{center}
\includegraphics[width=0.49\textwidth]{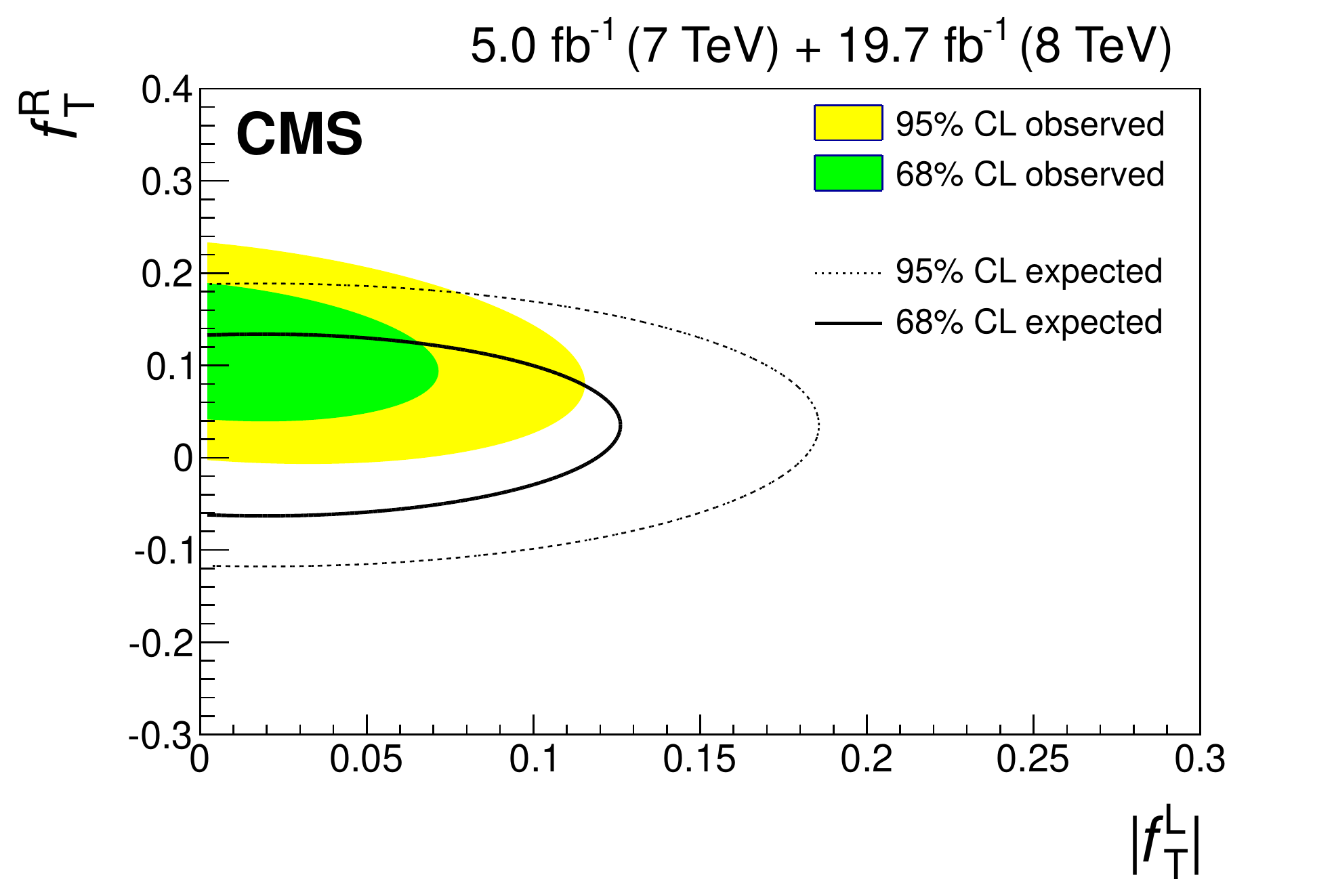}
\includegraphics[width=0.49\textwidth]{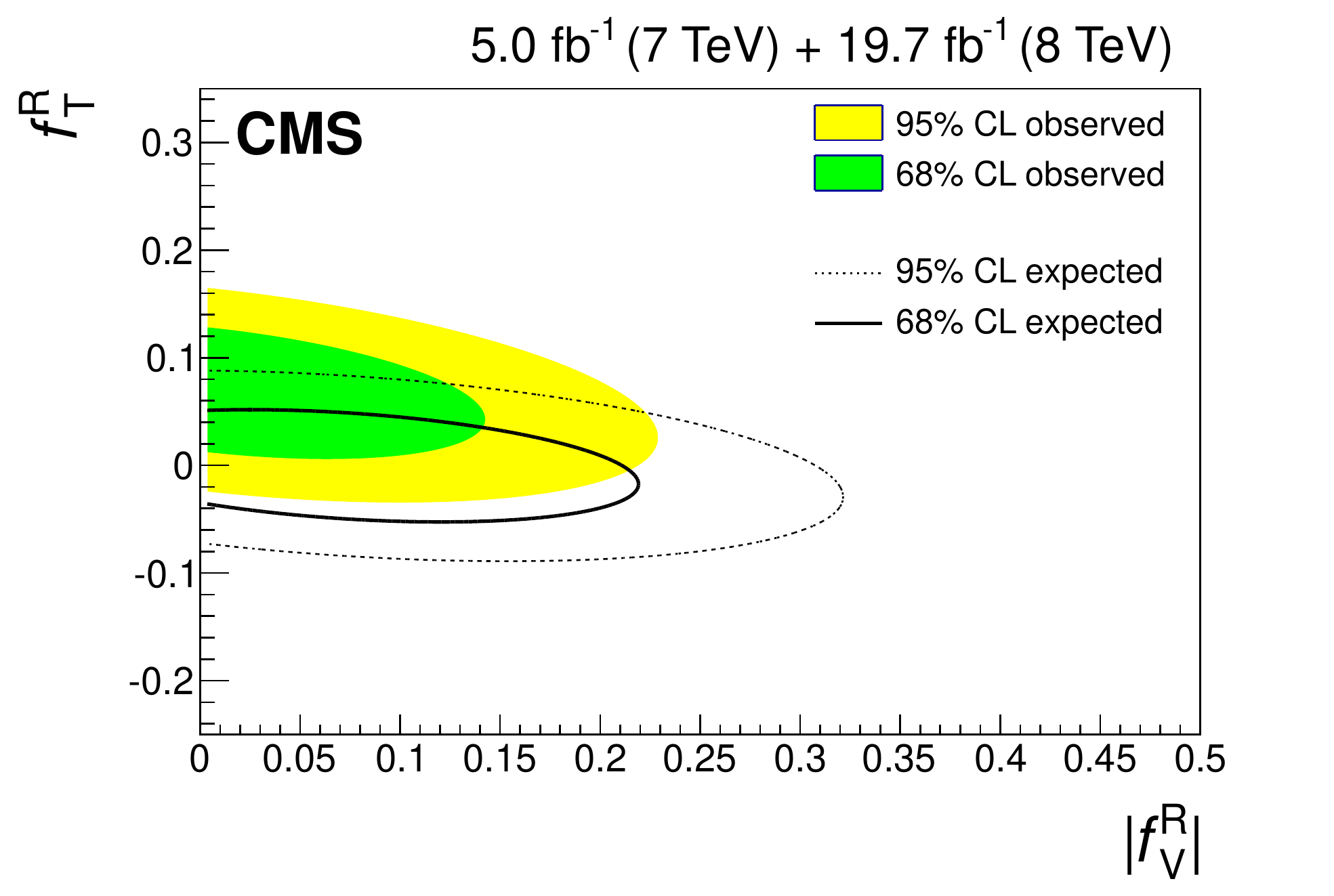}
\end{center}
\end{minipage}
\caption{ Combined $\sqrt{s}=7$ and $8\TeV$ results from the three-dimensional variation of the couplings of $f_{\rm V}^{\rm L}$, $f_{\rm T}^{\rm L}$, $f_{\rm T}^{\rm R}$ (left), and $f_{\rm V}^{\rm L}$, $f_{\rm V}^{\rm R}$, $f_{\rm T}^{\rm R}$ (right)  in the form of observed and expected exclusion limits at 68\% and 95\% CL in the two-dimension planes $(\lvert f_{\rm T}^{\rm L}\rvert$, $f_{\rm T}^{\rm R})$ (left) and  $(\lvert f_{\rm V}^{\rm R}\rvert $, $f_{\rm T}^{\rm R})$ (right).}
\label{fig:3d-comb}
\end{figure}

Following the strategy described in Section~\ref{sec:analysis}, in addition to the SM BNN, the anomalous Wtb BNNs are trained to distinguish possible right-handed vector or left-/right-handed tensor structures from the SM left-handed vector structure in the $t$-channel single top quark events.
The set of variables chosen for the different Wtb BNNs are listed in Table~\ref{tab:BNN input vars}.
The first two-dimension\-al scenario considers a possible mixture of  $f_{\rm V}^{\rm L}$ and (anomalous) $f_{\rm V}^{\rm R}$ couplings.
The corresponding Wtb BNN ($f_{\rm V}^{\rm L}$, $f_{\rm V}^{\rm R}$) is trained to distinguish the contribution of these two couplings.
For the $(f_{\rm V}^{\rm L}$, $f_{\rm T}^{\rm L})$ scenario, another  Wtb BNN is trained to separate the left-handed vector interacting single top quark SM events from events with a left-handed tensor operator in the Wtb vertex.
For the third scenario, $(f_{\rm V}^{\rm L}$, $f_{\rm T}^{\rm R})$, the last Wtb BNN is trained to separate  left-handed-vector-interacting single top quark SM events from events with a right-handed-tensor operator in the Wtb vertex.
Figure~\ref{fig:BNN aWtb LVRV discr} shows the comparison between the data and simulation for the outputs of the Wtb BNN ($f_{\rm V}^{\rm L}$, $f_{\rm V}^{\rm R}$), Wtb BNN $(f_{\rm V}^{\rm L}$, $f_{\rm T}^{\rm L})$, and Wtb BNN $(f_{\rm V}^{\rm L}$, $f_{\rm T}^{\rm R})$.
The SM BNN and one of the Wtb BNN discriminants are used as inputs in the simultaneous fit of the two BNN discriminants.
One-dimensional constraints on the anomalous parameters are obtained by integrating over the other anomalous parameter in the corresponding scenario.
The results of the fits are presented in the form of two-dimensional contours at 68\% and 95\% CL exclusion limits, and as given in Table~\ref{tab:1D_wtb_limits}, as one-dimensional constraints in different scenarios.
Both the one- and two-dimension limits are measured for the individual data sets and their combination.
The combined observed and expected two-dimensional contours in the $(f_{\rm V}^{\rm L}$, $f_{\rm V}^{\rm R})$, $(f_{\rm V}^{\rm L}$, $f_{\rm T}^{\rm L})$, and $(f_{\rm V}^{\rm L}$, $f_{\rm T}^{\rm R})$ spaces are shown in Fig.~\ref{fig:2d-comb}.
\begin{table}[!h!]
\begin{center}
\small
\def\arraystretch{1.35}
\topcaption{One-dimensional exclusion limits obtained in different two- and three-dimensional fit scenarios.
The first column shows the couplings allowed to vary in the fit, with the remaining couplings set to the SM values.
The observed (expected) 95\% CL limits for each of the two data sets and their combination are given in the following columns.}
\label{tab:1D_wtb_limits}
\begin{tabular}{c|c|c|c|cc}
Scenario & $f_{\rm V}^{\rm L} >$ & $ \lvert f_{\rm V}^{\rm R}\rvert <$ & $ \lvert f_{\rm T}^{\rm L}\rvert <$  & \multicolumn{2}{c}{$< f_{\rm T}^{\rm R} <$}  \\  \hline
\multicolumn{6}{l}{ $\sqrt{s}=7\TeV$ } \\ \hline
$(f_{\rm V}^{\rm L}$, $f_{\rm V}^{\rm R})$ & 0.96 (0.91) & 0.29 (0.37) &  &  &        \\
$(f_{\rm V}^{\rm L}$, $f_{\rm T}^{\rm L})$ & 0.88 (0.89) &  & 0.11 (0.16) &  &        \\
$(f_{\rm V}^{\rm L}$, $f_{\rm T}^{\rm R})$ & 0.94 (0.91) &  &  & --0.077 (--0.067) & 0.046 (0.053)       \\
$(f_{\rm V}^{\rm L}$, $f_{\rm T}^{\rm L}$, $f_{\rm T}^{\rm R})$ & 0.95 (0.91) &  & 0.16 (0.22) & --0.074 (--0.065) & 0.037 (0.055)       \\
$(f_{\rm V}^{\rm L}$, $f_{\rm V}^{\rm R}$, $f_{\rm T}^{\rm R})$ & 0.94 (0.89) & 0.24 (0.29) &  & --0.087 (--0.076) & 0.040 (0.064)       \\
\hline
\multicolumn{6}{l}{ $\sqrt{s}=8\TeV$ } \\ \hline
$(f_{\rm V}^{\rm L}$, $f_{\rm V}^{\rm R})$ & 0.96 (0.92) & 0.24 (0.29) &  &  &        \\
$(f_{\rm V}^{\rm L}$, $f_{\rm T}^{\rm L})$ & 0.91 (0.92) &  & 0.15 (0.18) &  &        \\
$(f_{\rm V}^{\rm L}$, $f_{\rm T}^{\rm R})$ & 0.92 (0.92) &  &  & --0.041 (--0.050) & 0.060(0.048)       \\
$(f_{\rm V}^{\rm L}$, $f_{\rm T}^{\rm L}$, $f_{\rm T}^{\rm R})$ & 0.93 (0.94) &  & 0.070(0.12) & --0.049 (--0.067) & 0.080 (0.066)       \\
$(f_{\rm V}^{\rm L}$, $f_{\rm V}^{\rm R}$, $f_{\rm T}^{\rm R})$ & 0.95 (0.95) & 0.18 (0.20) &  & --0.035 (--0.044) & 0.043 (0.032)       \\  \hline
\multicolumn{6}{l}{ $\sqrt{s}=7$ and $8\TeV$ } \\ \hline
$(f_{\rm V}^{\rm L}$, $f_{\rm V}^{\rm R})$ & 0.97 (0.92) & 0.28 (0.31) &  &  &        \\
$(f_{\rm V}^{\rm L}$, $f_{\rm T}^{\rm L})$ & 0.92 (0.92) &  & 0.10 (0.14) &  &        \\
$(f_{\rm V}^{\rm L}$, $f_{\rm T}^{\rm R})$ & 0.94 (0.93) &  &  & --0.046 (--0.050) & 0.046 (0.041)       \\
$(f_{\rm V}^{\rm L}$, $f_{\rm T}^{\rm L}$, $f_{\rm T}^{\rm R})$ & 0.98 (0.97) &  & 0.057 (0.10) & --0.049 (--0.051) & 0.048 (0.046)       \\
$(f_{\rm V}^{\rm L}$, $f_{\rm V}^{\rm R}$, $f_{\rm T}^{\rm R})$ & 0.98 (0.97) & 0.16 (0.22) &  & --0.049 (--0.049) & 0.039 (0.037)       \\
\end{tabular}
\end{center}
\end{table}
As the interference terms between $f_{\rm T}^{\rm L}$ and $f_{\rm T}^{\rm R}$ or  $f_{\rm V}^{\rm R}$ and $f_{\rm T}^{\rm R}$ couplings are negligible~\cite{Willenbrock:2014bja}, it is possible to consider three-dimensional scenarios with simultaneous variation of $f_{\rm V}^{\rm L}$, $f_{\rm T}^{\rm L}$, $f_{\rm T}^{\rm R}$ or $f_{\rm V}^{\rm L}$, $f_{\rm V}^{\rm R}$, $f_{\rm T}^{\rm R}$ couplings.
The three-dimensional statistical analysis is performed using the SM BNN, Wtb BNN ($f_{\rm V}^{\rm L}$, $f_{\rm T}^{\rm R}$), and either the Wtb BNN ($f_{\rm V}^{\rm L}$, $f_{\rm T}^{\rm L}$) or Wtb BNN ($f_{\rm V}^{\rm L}$, $f_{\rm V}^{\rm R}$) discriminants to obtain the excluded regions at 68\% and 95\% CL for $f_{\rm T}^{\rm L}$ and $f_{\rm T}^{\rm R}$, again by integrating over the other anomalous couplings.
The combined $\sqrt{s}=7$ and $8\TeV$ results in the three-dimensional simultaneous fit of $f_{\rm V}^{\rm L}$, $f_{\rm T}^{\rm L}$, and $f_{\rm T}^{\rm R}$ couplings are presented in Fig.~\ref{fig:3d-comb} (left) in the form of observed and expected 68\% and 95\% exclusion contours on the $(f_{\rm T}^{\rm L}$, $f_{\rm T}^{\rm R})$ couplings.
The corresponding results for the $f_{\rm V}^{\rm L}$, $f_{\rm V}^{\rm R}$, and $f_{\rm T}^{\rm R}$ couplings are shown in Fig.~\ref{fig:3d-comb} (right) as two-dimensional exclusion limits in the $(f_{\rm V}^{\rm R}$, $f_{\rm T}^{\rm R})$ plane.
The measured exclusion limits from the three-dimensional fits with the combined data sets are $f_{\rm V}^{\rm L} > 0.98$, $\lvert f_{\rm V}^{\rm R}\rvert < 0.16$, and $\lvert f_{\rm T}^{\rm L}\rvert < 0.057$. For $f_{\rm T}^{\rm R}$ we take the more-conservative limits from the three-dimensional fits of $-0.049<f_{\rm T}^{\rm R}<0.048$ as our measurement.
These limits are much more restrictive than those obtained by the D0 Collaboration in a direct search~\cite{Abazov:2011pm}, and agree well with the recent results obtained by the ATLAS~\cite{Aad:2012ky} and CMS~\cite{Chatrchyan:2013jna,Khachatryan:2014vma} experiments from measurements of the W boson helicity fractions.

\section{Search for tcg and tug FCNC interactions}
\label{sec:fcnc}
\subsection{Theoretical introduction}
\begin{figure}[!h]
\centering
\includegraphics[width=0.99\textwidth]{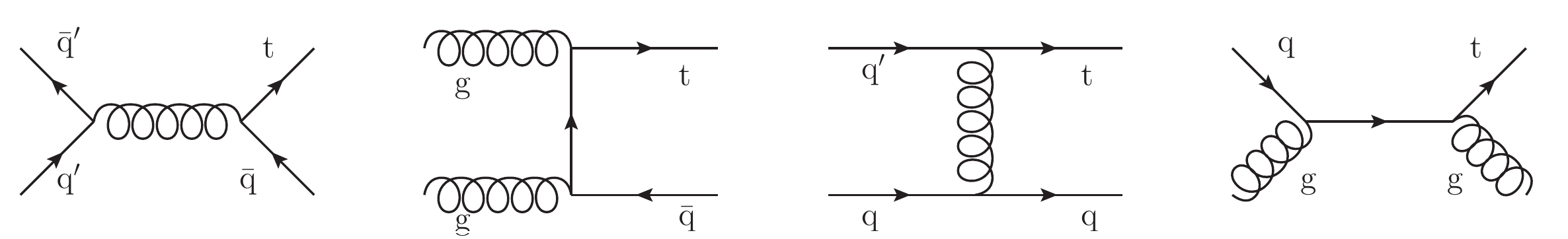}
\caption {Representative Feynman diagrams for the FCNC processes.} {\label{fcnc_diags} }
\end{figure}

The FCNC tcg and tug interactions can be written in a model-independent form with the following effective Lagrangian~\cite{Beneke:2000hk}:
\begin{linenomath}
\begin{equation}
\label{fcnc_lagrangian}
\mathfrak{L}=\frac{\kappa_{\rm tqg}}{\Lambda} g_s \overline{\rm q} \sigma^{ \mu\nu} \frac{\lambda^{\rm a}}{2}
{\rm t} G^{\rm a}_{ \mu\nu},
\end{equation}
\end{linenomath}
where $\Lambda$ is the scale of new physics $({\approx} 1\TeV)$, q refers to either the u or c quarks, $\kappa_{\rm tqg}$ defines the strength of the FCNC interactions in the tug or tcg vertices,
$\lambda^{\rm a}/2$ are the generators of the SU(3) colour gauge group, $g_s$ is the coupling constant of the strong interaction, and $G^{\rm a}_{\mu\nu}$ is a gluon field strength tensor.
The Lagrangian is assumed to be symmetric with respect to the left and right projectors.
Single top quark production through FCNC interactions contains 48 subprocesses for both the tug and tcg channels, and the cross section is proportional to $(\kappa_{\rm tqg}/\Lambda)^2$.
Representative Feynman diagrams for the FCNC processes are shown in Fig.~\ref{fcnc_diags}.
Since the influence of the FCNC parameters on the total top quark width is negligible for the allowed region of FCNC parameters, the SM value for the top quark width is used in this analysis.
The \textsc{CompHEP} generator is used to simulate of the signal tug and tcg processes.
The FCNC samples are normalized to the NLO cross sections using a K factor of 1.6 for higher-order QCD corrections~\cite{Liu:2005dp}.
\subsection{Exclusion limits on tug and tcg anomalous couplings}
\begin{figure}[!h]
\centering
\includegraphics[width=0.49\textwidth]{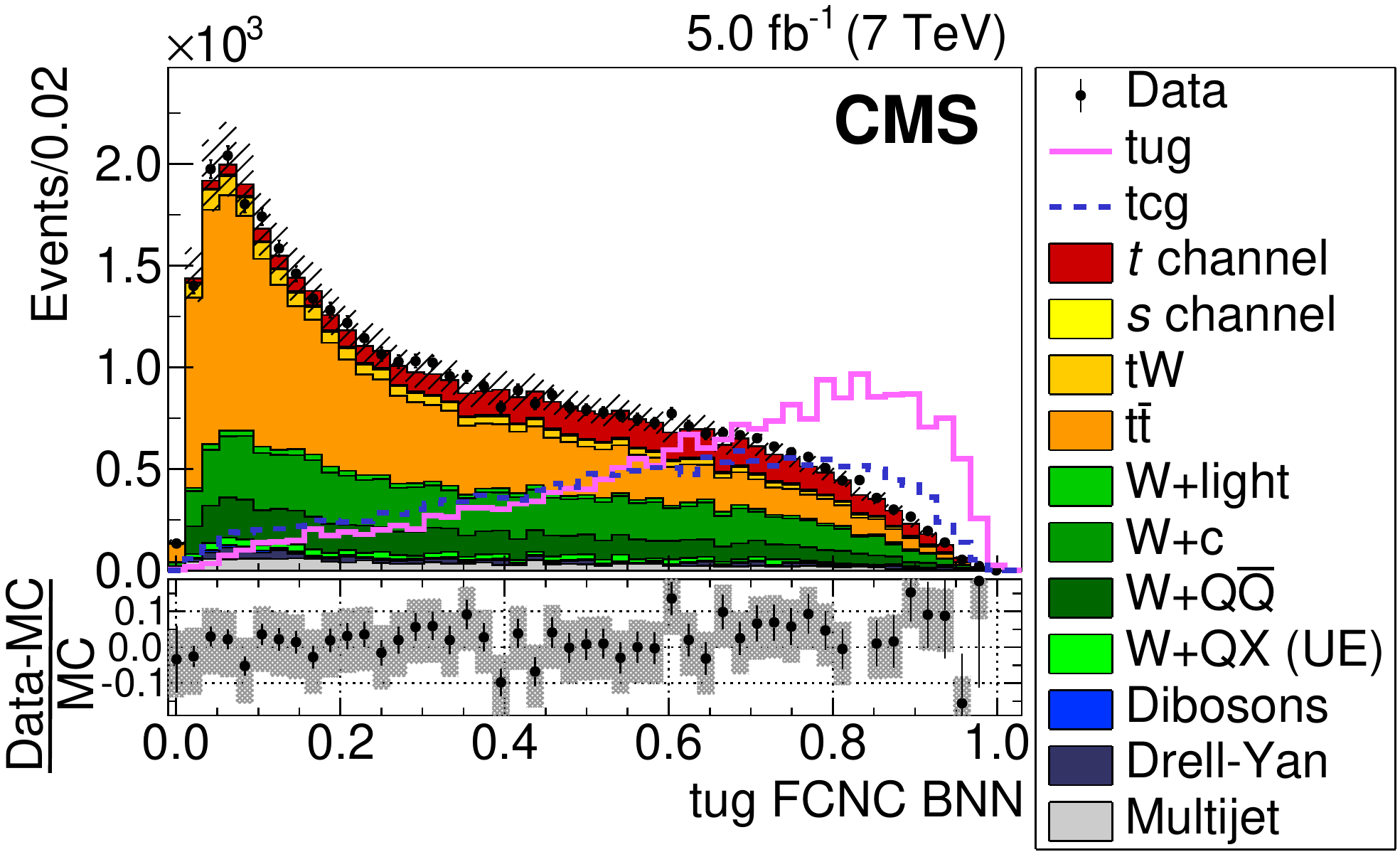}
\includegraphics[width=0.49\textwidth]{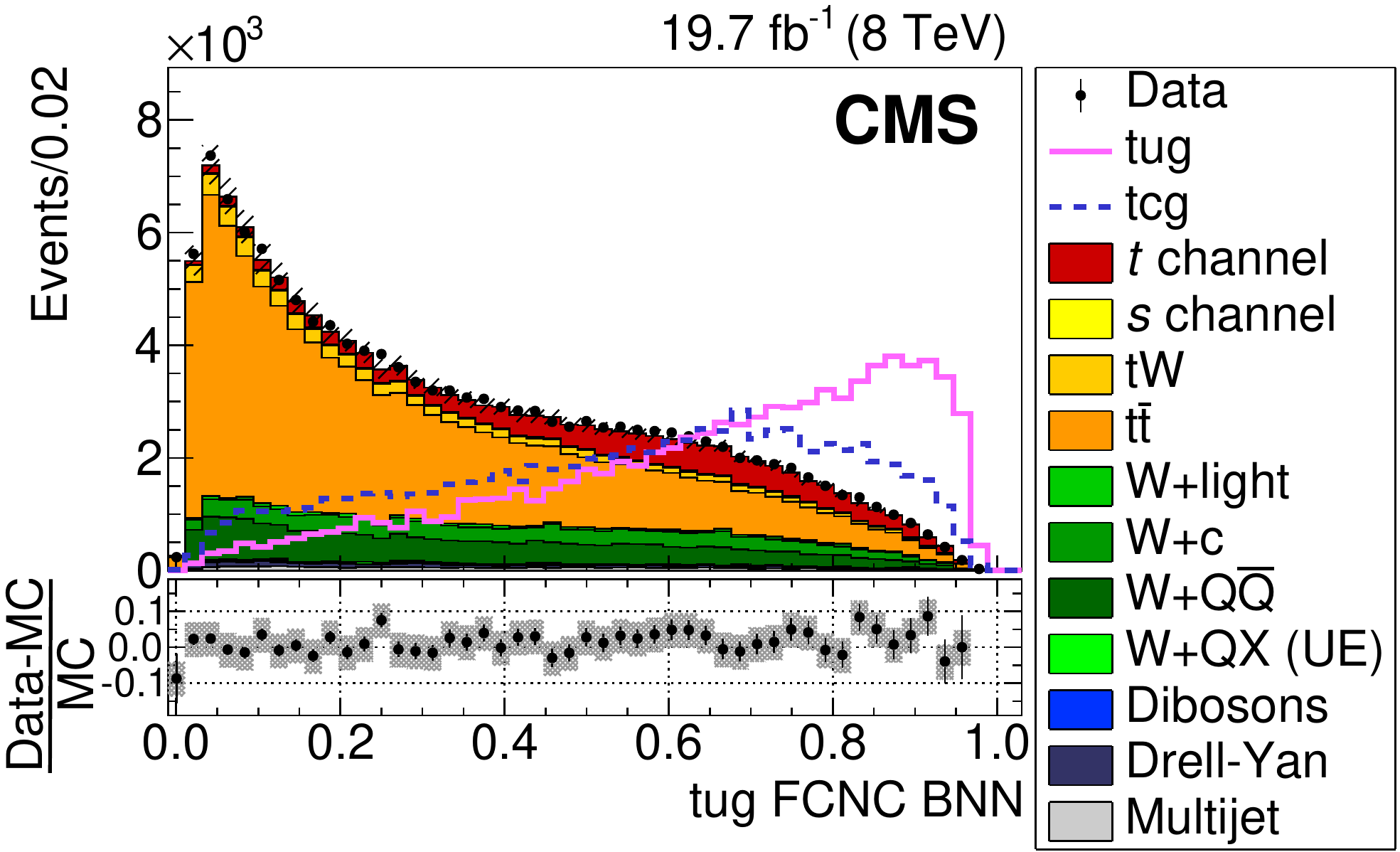}
\includegraphics[width=0.49\textwidth]{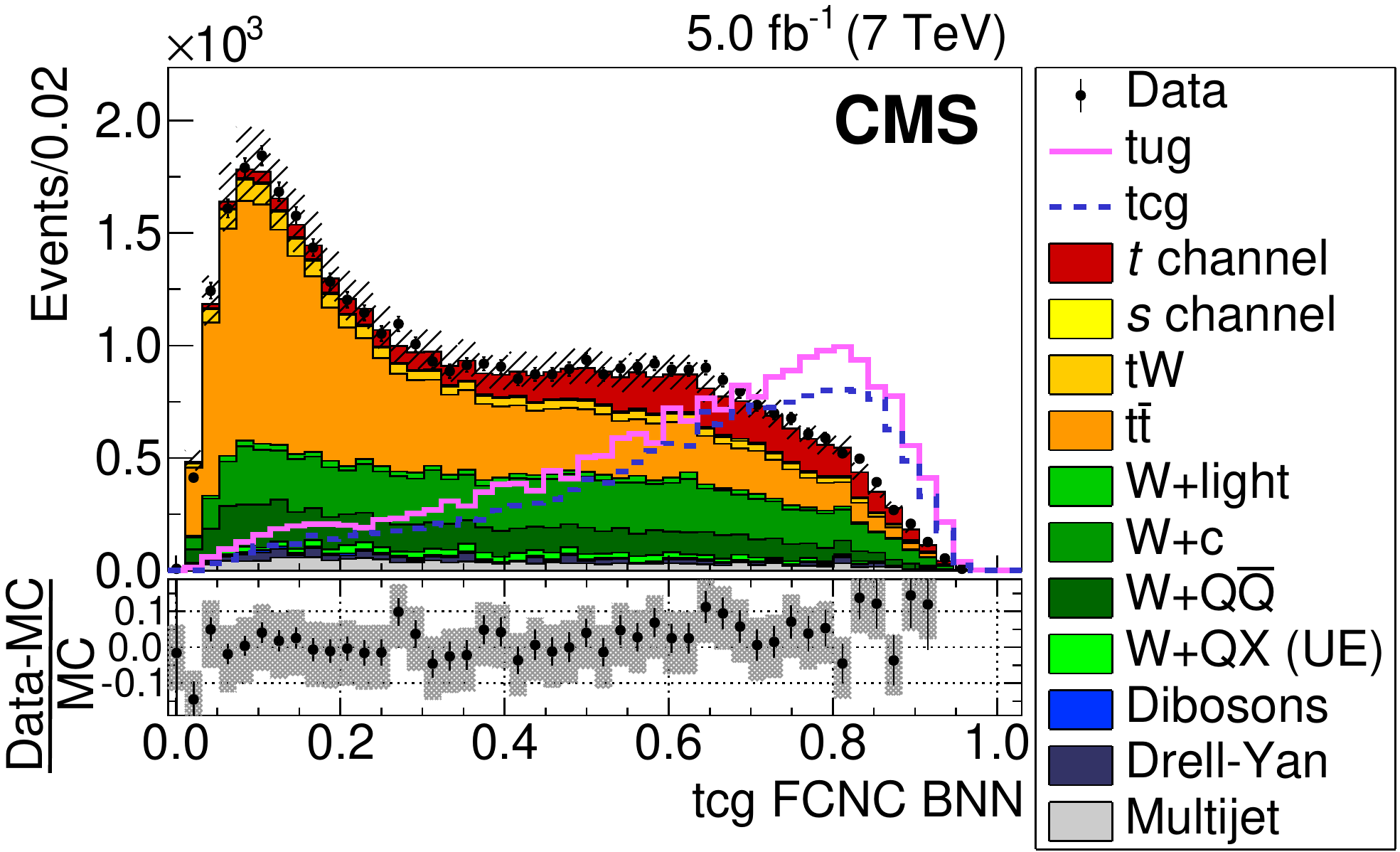}
\includegraphics[width=0.49\textwidth]{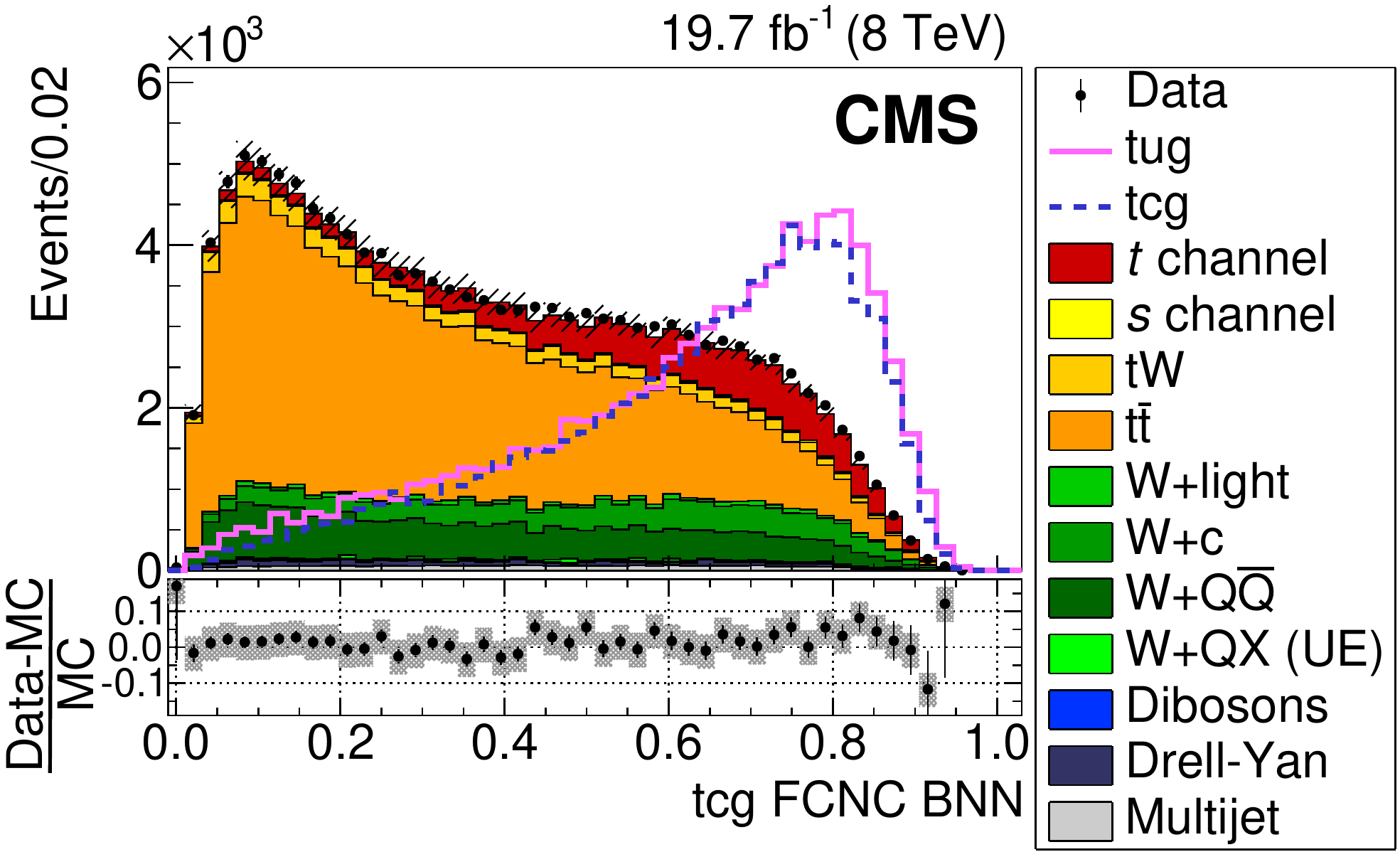}
\caption{The FCNC BNN discriminant distributions when the BNN is trained to distinguish $\rm t\to ug$ (upper) or $\rm t\to cg$ (lower) processes as signal from the SM processes as background.
The results from data are shown as points and the predicted distributions from the background simulations by the filled histograms.
The plots on the left (right) correspond to the $\sqrt{s} = 7$ (8)\TeV data.
The solid and dashed lines give the expected distributions for $\rm t\to ug$ and $\rm t\to cg$, respectively, assuming a coupling of $\lvert\kappa_\mathrm{tug}\rvert/\Lambda = 0.04\ (0.06)$ and $\lvert\kappa_\mathrm{tcg}\rvert/\Lambda = 0.08\ (0.12)\TeV^{-1}$ on the left (right) plots.
The lower part of each plot shows the relative difference between the data and the total predicted background.
The hatched band corresponds to the total simulation uncertainty.
The vertical bars represent the statistical uncertainties.
}
\label{fcnc:datamc_tug_tcg}
\end{figure}

\begin{figure}[!h!]
\centering
\includegraphics[width=0.6\textwidth]{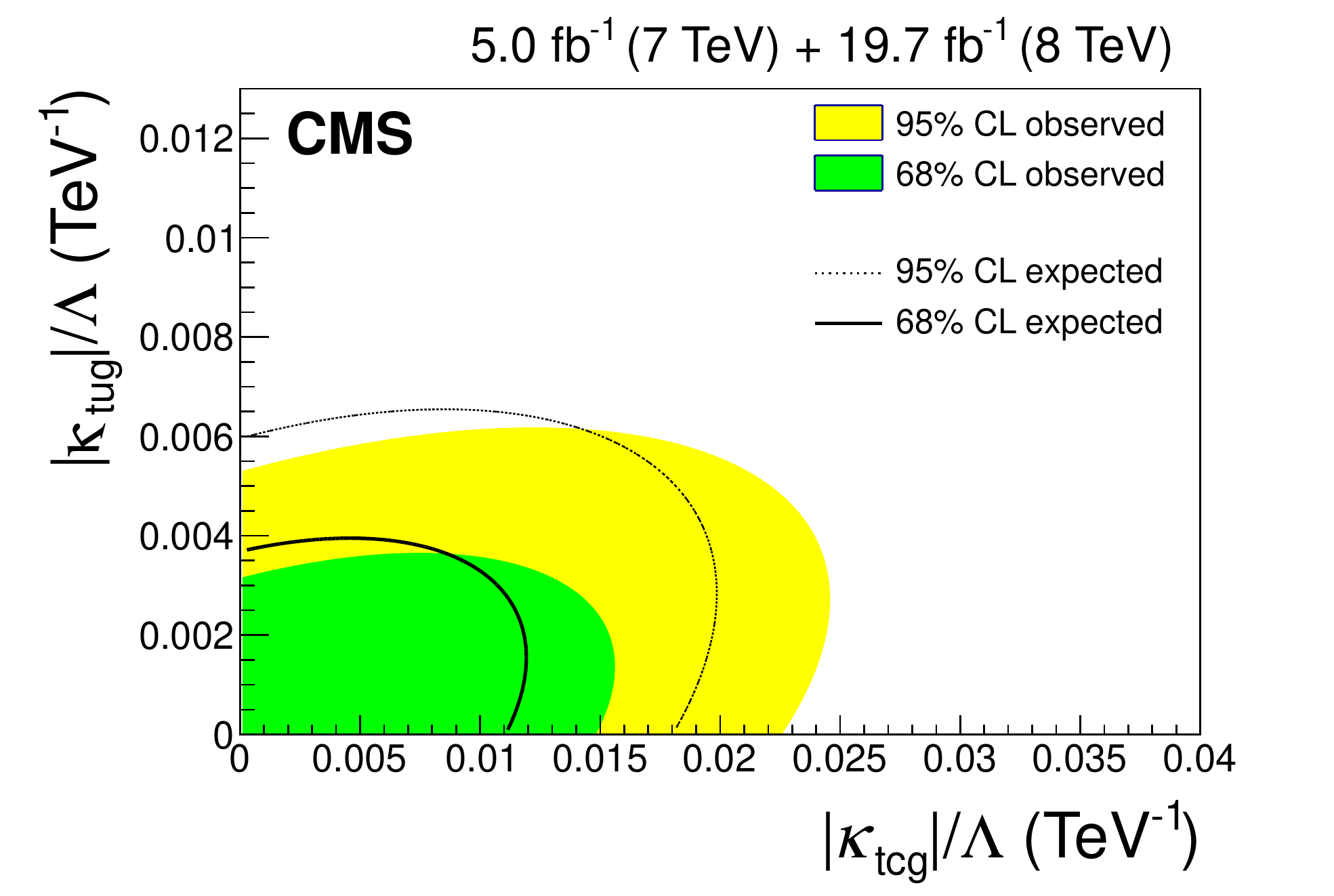}
\caption{Combined $\sqrt{s}$ = 7 and 8\TeV observed and expected limits for the 68\% and 95\% CL on the
$\lvert\kappa_\mathrm{tug}\rvert/\Lambda$ and $\lvert\kappa_\mathrm{tcg}\rvert/\Lambda$ couplings.
}
\label{fig:fcnc_kukc_2d_contours-comb}
\end{figure}

FCNC processes are kinematically different from any SM processes, therefore, it is reasonable to train a new BNN to discriminate between FCNC production as the signal and the SM background, including the $t$-channel single top quark production.
Owing to the possible presence of a FCNC tug or tcg signal, two BNNs are trained to distinguish each of the couplings.
The variable choices for these BNNs, shown in Table~\ref{tab:BNN input vars}, are motivated by analysis of the Feynman diagrams of the FCNC and SM processes.  The comparison of the neural network output for the data and model is shown in Fig.~\ref{fcnc:datamc_tug_tcg}.
Output histograms from the tug and tcg FCNC BNN discriminants for the SM backgrounds are used as input to the analysis.
The posterior probability distributions of $\lvert\kappa_\mathrm{tug}\rvert/\Lambda$ and $\lvert\kappa_\mathrm{tcg}\rvert/\Lambda$ are obtained by fitting the histograms.
The combined  $\sqrt{s}$ = 7 and 8\TeV observed and expected exclusion limits at 68\% and 95\% CL on the anomalous FCNC parameters in the form of two-dimensional contours are shown in Fig.~\ref{fig:fcnc_kukc_2d_contours-comb}.
The two-dimensional contours reflect the possible simultaneous presence of the two FCNC parameters.
Individual exclusion limits on $\lvert\kappa_\mathrm{tug}\rvert/\Lambda$ are obtained by integrating over $\lvert\kappa_\mathrm{tcg}\rvert/\Lambda$ and vice versa.
These individual limits can be used to calculate the upper limits on the branching fractions $\mathcal{B}(\rm t~\to~ug)$ and $\mathcal{B}(\rm t~\to~cg)$~\cite{Zhang:2008yn}.
The observed and expected exclusion limits at 95\% CL on the FCNC couplings and the corresponding branching fractions are given in Table~\ref{tab:1D_fcnc_limits}.
These limits are significantly better than those obtained by the D0~\cite{Abazov:2010qk} and CDF~\cite{Aaltonen:2008qr} experiments, and in previous CMS results, and are comparable to recent ATLAS measurements~\cite{Aad:2015gea}.
\begin{table}[!h]
\centering
\small
\def\arraystretch{1.4}
\topcaption{Observed (expected) upper limits at 95\% CL for the FCNC couplings and branching fractions obtained using the $\sqrt{s} = 7$ and 8\TeV data, and their combination.
}
\label{tab:1D_fcnc_limits}
\begin{tabular}{c|c|c|c|c}
$\sqrt{s}$   & $\lvert\kappa_\mathrm{tug}\rvert/\Lambda\,\mathrm{(TeV^{ -1})}$ & $\mathcal{B}(\rm t~\to~ug)$         & $\vert\kappa_\mathrm{tcg}\rvert/\Lambda \,\mathrm{(TeV^{ -1})} $ & $\mathcal{B}(\rm t~\to~cg)$ \\
\hline
7\TeV       & 14 (13) $\times 10^{-3}$                             &                    24 (21)$\times10^{-5}$&                  2.9 (2.4) $\times 10^{-2}$           &                    10.1 (6.9)$\times10^{-4}$ \\
8\TeV       & 5.1  (5.9)  $\times 10^{-3}$                             &                    3.1 (4.2)$\times10^{-5}$  &                  2.2 (2.0) $\times 10^{-2}$           &                    5.6 (4.8)$\times10^{-4}$ \\
7 and 8\TeV & {4.1  (4.8)  $\times 10^{-3}$}          & { 2.0 (2.8)$\times10^{-5}$}& {1.8 (1.5) $\times 10^{-2}$}          & { 4.1 (2.8)$\times10^{-4}$} \\
\end{tabular}
\end{table}

\section{Summary}
\label{sec:results}
A direct search for model-independent anomalous operators in the Wtb vertex and FCNC couplings has been performed using single top quark $t$-channel production in data collected by the CMS experiment in pp collisions at $\sqrt{s}$ = 7 and 8\TeV.
Different possible anomalous contributions are investigated.
The observed event rates are consistent with the SM prediction, and exclusion limits are extracted at 95\% CL.
The combined limits in three-dimensional scenarios on possible Wtb anomalous couplings are $f_{\rm V}^{\rm L} > 0.98$ for the left-handed vector coupling, $\lvert f_{\rm V}^{\rm R}\rvert < 0.16$ for the right-handed vector coupling, $\lvert f_{\rm T}^{\rm L}\rvert <0.057$ for the left-handed tensor coupling, and $-0.049<f_{\rm T}^{\rm R}<0.048$ for the right-handed tensor coupling.
For FCNC couplings of the gluon to top and up quarks (tug) or top and charm quarks (tcg), the 95\% CL exclusion limits on the coupling strengths are $\lvert\kappa_\mathrm{tug}\rvert /\Lambda < 4.1 \times 10^{-3}\,\mathrm{TeV^{ -1}}$ and  $\lvert\kappa_\mathrm{tcg}\rvert /\Lambda < 1.8 \times 10^{-2}\,\mathrm{TeV^{ -1}}$ or, in terms of branching fractions, $\mathcal{B}(\rm t\to ug) < 2.0\times 10^{-5}$ and $\mathcal{B}(\rm t \to cg) < 4.1\times10^{-4}$.

\begin{acknowledgments}

\hyphenation{Bundes-ministerium Forschungs-gemeinschaft Forschungs-zentren Rachada-pisek} We congratulate our colleagues in the CERN accelerator departments for the excellent performance of the LHC and thank the technical and administrative staffs at CERN and at other CMS institutes for their contributions to the success of the CMS effort. In addition, we gratefully acknowledge the computing centres and personnel of the Worldwide LHC Computing Grid for delivering so effectively the computing infrastructure essential to our analyses. Finally, we acknowledge the enduring support for the construction and operation of the LHC and the CMS detector provided by the following funding agencies: the Austrian Federal Ministry of Science, Research and Economy and the Austrian Science Fund; the Belgian Fonds de la Recherche Scientifique, and Fonds voor Wetenschappelijk Onderzoek; the Brazilian Funding Agencies (CNPq, CAPES, FAPERJ, and FAPESP); the Bulgarian Ministry of Education and Science; CERN; the Chinese Academy of Sciences, Ministry of Science and Technology, and National Natural Science Foundation of China; the Colombian Funding Agency (COLCIENCIAS); the Croatian Ministry of Science, Education and Sport, and the Croatian Science Foundation; the Research Promotion Foundation, Cyprus; the Secretariat for Higher Education, Science, Technology and Innovation, Ecuador; the Ministry of Education and Research, Estonian Research Council via IUT23-4 and IUT23-6 and European Regional Development Fund, Estonia; the Academy of Finland, Finnish Ministry of Education and Culture, and Helsinki Institute of Physics; the Institut National de Physique Nucl\'eaire et de Physique des Particules~/~CNRS, and Commissariat \`a l'\'Energie Atomique et aux \'Energies Alternatives~/~CEA, France; the Bundesministerium f\"ur Bildung und Forschung, Deutsche Forschungsgemeinschaft, and Helmholtz-Gemeinschaft Deutscher Forschungszentren, Germany; the General Secretariat for Research and Technology, Greece; the National Scientific Research Foundation, and National Innovation Office, Hungary; the Department of Atomic Energy and the Department of Science and Technology, India; the Institute for Studies in Theoretical Physics and Mathematics, Iran; the Science Foundation, Ireland; the Istituto Nazionale di Fisica Nucleare, Italy; the Ministry of Science, ICT and Future Planning, and National Research Foundation (NRF), Republic of Korea; the Lithuanian Academy of Sciences; the Ministry of Education, and University of Malaya (Malaysia); the Mexican Funding Agencies (BUAP, CINVESTAV, CONACYT, LNS, SEP, and UASLP-FAI); the Ministry of Business, Innovation and Employment, New Zealand; the Pakistan Atomic Energy Commission; the Ministry of Science and Higher Education and the National Science Centre, Poland; the Funda\c{c}\~ao para a Ci\^encia e a Tecnologia, Portugal; JINR, Dubna; the Ministry of Education and Science of the Russian Federation, the Federal Agency of Atomic Energy of the Russian Federation, Russian Academy of Sciences, and the Russian Foundation for Basic Research; the Ministry of Education, Science and Technological Development of Serbia; the Secretar\'{\i}a de Estado de Investigaci\'on, Desarrollo e Innovaci\'on and Programa Consolider-Ingenio 2010, Spain; the Swiss Funding Agencies (ETH Board, ETH Zurich, PSI, SNF, UniZH, Canton Zurich, and SER); the Ministry of Science and Technology, Taipei; the Thailand Center of Excellence in Physics, the Institute for the Promotion of Teaching Science and Technology of Thailand, Special Task Force for Activating Research and the National Science and Technology Development Agency of Thailand; the Scientific and Technical Research Council of Turkey, and Turkish Atomic Energy Authority; the National Academy of Sciences of Ukraine, and State Fund for Fundamental Researches, Ukraine; the Science and Technology Facilities Council, UK; the US Department of Energy, and the US National Science Foundation.

Individuals have received support from the Marie-Curie programme and the European Research Council and EPLANET (European Union); the Leventis Foundation; the A. P. Sloan Foundation; the Alexander von Humboldt Foundation; the Belgian Federal Science Policy Office; the Fonds pour la Formation \`a la Recherche dans l'Industrie et dans l'Agriculture (FRIA-Belgium); the Agentschap voor Innovatie door Wetenschap en Technologie (IWT-Belgium); the Ministry of Education, Youth and Sports (MEYS) of the Czech Republic; the Council of Science and Industrial Research, India; the HOMING PLUS programme of the Foundation for Polish Science, cofinanced from European Union, Regional Development Fund, the Mobility Plus programme of the Ministry of Science and Higher Education, the National Science Center (Poland), contracts Harmonia 2014/14/M/ST2/00428, Opus 2013/11/B/ST2/04202, 2014/13/B/ST2/02543 and 2014/15/B/ST2/03998, Sonata-bis 2012/07/E/ST2/01406; the Thalis and Aristeia programmes cofinanced by EU-ESF and the Greek NSRF; the National Priorities Research Program by Qatar National Research Fund; the Programa Clar\'in-COFUND del Principado de Asturias; the Rachadapisek Sompot Fund for Postdoctoral Fellowship, Chulalongkorn University and the Chulalongkorn Academic into Its 2nd Century Project Advancement Project (Thailand); and the Welch Foundation, contract C-1845.

\end{acknowledgments}

\clearpage

\bibliography{auto_generated}

\providecommand{\href}[2]{#2}\begingroup\raggedright\begin{thebibliography}{10}%
\makeatletter
\providecommand{\hrefCMSnoop }[0]{\@secondoftwo}%
\makeatother
\providecommand{\doi}{\texttt{doi:}\begingroup \urlstyle{tt}\Url}

\bibitem{Beneke:2000hk}
M.~Beneke\hrefCMSnoop {}{ {et~al.}, ``{Top quark physics}'',} (2000).
\href{http://www.arXiv.org/abs/hep-ph/0003033}{\texttt{arXiv:hep-ph/0003033}}.

\bibitem{Willenbrock:1986cr}
\hrefCMSnoop {}{S.~S.~D. Willenbrock and D.~A. Dicus, ``{Production of heavy
  quarks from W gluon fusion}'',} \textit{ Phys. Rev. D} \textbf{ 34} (1986)
  155,
\href{http://dx.doi.org/10.1103/PhysRevD.34.155}{\doi{10.1103/PhysRevD.34.155}}.

\bibitem{Glashow:1970gm}
\hrefCMSnoop {}{S.~L. Glashow, J.~Iliopoulos, and L.~Maiani, ``{Weak
  interactions with lepton-hadron symmetry}'',} \textit{ Phys. Rev. D} \textbf{
  2} (1970) 1285,
\href{http://dx.doi.org/10.1103/PhysRevD.2.1285}{\doi{10.1103/PhysRevD.2.1285}}.

\bibitem{Agashe:2014kda}
\hrefCMSnoop {}{{Particle Data Group}, K.~A. Olive {et~al.}, ``{Review of
  Particle Physics}'',} \textit{ Chin. Phys. C} \textbf{ 38} (2014) 090001,
\href{http://dx.doi.org/10.1088/1674-1137/38/9/090001}{\doi{10.1088/1674-1137/38/9/090001}}.

\bibitem{Eilam:1990zc}
\hrefCMSnoop {}{G.~Eilam, J.~L. Hewett, and A.~Soni, ``{Rare decays of the top
  quark in the standard and two Higgs doublet models}'',} \textit{ Phys. Rev.
  D} \textbf{ 44} (1991) 1473,
  \href{http://dx.doi.org/10.1103/PhysRevD.44.1473}{\doi{10.1103/PhysRevD.44.1473}}.
[Erratum: \DOI{10.1103/PhysRevD.59.039901}.

\bibitem{Atwood:1995ud}
\hrefCMSnoop {}{D.~Atwood, L.~Reina, and A.~Soni, ``{Probing flavor changing
  top-charm-scalar interactions in ${\rm e^{+} e^{-}}$ collisions}'',} \textit{
  Phys. Rev. D} \textbf{ 53} (1996) 1199,
  \href{http://dx.doi.org/10.1103/PhysRevD.53.1199}{\doi{10.1103/PhysRevD.53.1199}},
\href{http://www.arXiv.org/abs/hep-ph/9506243}{\texttt{arXiv:hep-ph/9506243}}.

\bibitem{Yang:1997dk}
\hrefCMSnoop {}{J.~M. Yang, B.-L. Young, and X.~Zhang, ``{Flavor-changing top
  quark decays in R-parity-violating SUSY}'',} \textit{ Phys. Rev. D} \textbf{
  58} (1998) 055001,
  \href{http://dx.doi.org/10.1103/PhysRevD.58.055001}{\doi{10.1103/PhysRevD.58.055001}},
\href{http://www.arXiv.org/abs/hep-ph/9705341}{\texttt{arXiv:hep-ph/9705341}}.

\bibitem{Grossman:1994jb}
\hrefCMSnoop {}{Y.~Grossman, ``{Phenomenology of models with more than two
  Higgs doublets}'',} \textit{ Nucl. Phys. B} \textbf{ 426} (1994) 355,
  \href{http://dx.doi.org/10.1016/0550-3213(94)90316-6}{\doi{10.1016/0550-3213(94)90316-6}},
\href{http://www.arXiv.org/abs/hep-ph/9401311}{\texttt{arXiv:hep-ph/9401311}}.

\bibitem{Pich:2009sp}
\hrefCMSnoop {}{A.~Pich and P.~Tuzon, ``{Yukawa alignment in the
  two-Higgs-doublet model}'',} \textit{ Phys. Rev. D} \textbf{ 80} (2009)
  091702,
  \href{http://dx.doi.org/10.1103/PhysRevD.80.091702}{\doi{10.1103/PhysRevD.80.091702}},
\href{http://www.arXiv.org/abs/0908.1554}{\texttt{arXiv:0908.1554}}.

\bibitem{Keus:2013hya}
\hrefCMSnoop {}{V.~Keus, S.~F. King, and S.~Moretti, ``{Three-Higgs-doublet
  models: symmetries, potentials and Higgs boson masses}'',} \textit{ JHEP}
  \textbf{ 01} (2014) 052,
  \href{http://dx.doi.org/10.1007/JHEP01(2014)052}{\doi{10.1007/JHEP01(2014)052}},
\href{http://www.arXiv.org/abs/1310.8253}{\texttt{arXiv:1310.8253}}.

\bibitem{DiazCruz:1989ub}
\hrefCMSnoop {}{J.~L. Diaz-Cruz, R.~Martinez, M.~A. Perez, and A.~Rosado,
  ``{Flavor changing radiative decay of the $t$ quark}'',} \textit{ Phys. Rev.
  D} \textbf{ 41} (1990) 891,
\href{http://dx.doi.org/10.1103/PhysRevD.41.891}{\doi{10.1103/PhysRevD.41.891}}.

\bibitem{Arhrib:2006pm}
\hrefCMSnoop {}{A.~Arhrib and W.-S. Hou, ``{Flavor changing neutral currents
  involving heavy quarks with four generations}'',} \textit{ JHEP} \textbf{ 07}
  (2006) 009,
  \href{http://dx.doi.org/10.1088/1126-6708/2006/07/009}{\doi{10.1088/1126-6708/2006/07/009}},
\href{http://www.arXiv.org/abs/hep-ph/0602035}{\texttt{arXiv:hep-ph/0602035}}.

\bibitem{Branco:2013tda}
\href {http://pos.sissa.it/archive/conferences/177/024/Corfu2012_024.pdf}{G.~C.
  Branco and M.~N. Rebelo, ``{New physics in the flavour sector in the presence
  of flavour changing neutral currents}'',} in \textit{ {Proceedings of the
  Corfu Summer Institute 2012}}, p.~024.
\newblock 2013.
\newblock
\href{http://www.arXiv.org/abs/1308.4639}{\texttt{arXiv:1308.4639}}.
\newblock

\bibitem{Georgi:1994ha}
\hrefCMSnoop {}{H.~Georgi, L.~Kaplan, D.~Morin, and A.~Schenk, ``{Effects of
  top compositeness}'',} \textit{ Phys. Rev. D} \textbf{ 51} (1995) 3888,
  \href{http://dx.doi.org/10.1103/PhysRevD.51.3888}{\doi{10.1103/PhysRevD.51.3888}},
\href{http://www.arXiv.org/abs/hep-ph/9410307}{\texttt{arXiv:hep-ph/9410307}}.

\bibitem{Giudice:2007fh}
\hrefCMSnoop {}{G.~F. Giudice, C.~Grojean, A.~Pomarol, and R.~Rattazzi, ``{The
  strongly-interacting light Higgs}'',} \textit{ JHEP} \textbf{ 06} (2007) 045,
  \href{http://dx.doi.org/10.1088/1126-6708/2007/06/045}{\doi{10.1088/1126-6708/2007/06/045}},
\href{http://www.arXiv.org/abs/hep-ph/0703164}{\texttt{arXiv:hep-ph/0703164}}.

\bibitem{Konig:2014iqa}
\hrefCMSnoop {}{M.~K{\"o}nig, M.~Neubert, and D.~M. Straub, ``{Dipole operator
  constraints on composite Higgs models}'',} \textit{ Eur. Phys. J. C} \textbf{
  74} (2014) 2945,
  \href{http://dx.doi.org/10.1140/epjc/s10052-014-2945-9}{\doi{10.1140/epjc/s10052-014-2945-9}},
\href{http://www.arXiv.org/abs/1403.2756}{\texttt{arXiv:1403.2756}}.

\bibitem{Agashe:2014jca}
K.~Agashe\hrefCMSnoop {}{ {et~al.}, ``{Warped dipole completed, with a tower of
  Higgs bosons}'',} \textit{ JHEP} \textbf{ 06} (2015) 196,
  \href{http://dx.doi.org/10.1007/JHEP06(2015)196}{\doi{10.1007/JHEP06(2015)196}},
\href{http://www.arXiv.org/abs/1412.6468}{\texttt{arXiv:1412.6468}}.

\bibitem{Contino:2003ve}
\hrefCMSnoop {}{R.~Contino, Y.~Nomura, and A.~Pomarol, ``{Higgs as a
  holographic pseudo-Goldstone boson}'',} \textit{ Nucl. Phys. B} \textbf{ 671}
  (2003) 148,
  \href{http://dx.doi.org/10.1016/j.nuclphysb.2003.08.027}{\doi{10.1016/j.nuclphysb.2003.08.027}},
\href{http://www.arXiv.org/abs/hep-ph/0306259}{\texttt{arXiv:hep-ph/0306259}}.

\bibitem{AguilarSaavedra:2008zc}
\hrefCMSnoop {}{J.~A. Aguilar-Saavedra, ``{A minimal set of top anomalous
  couplings}'',} \textit{ Nucl. Phys. B} \textbf{ 812} (2009) 181,
  \href{http://dx.doi.org/10.1016/j.nuclphysb.2008.12.012}{\doi{10.1016/j.nuclphysb.2008.12.012}},
\href{http://www.arXiv.org/abs/0811.3842}{\texttt{arXiv:0811.3842}}.

\bibitem{Willenbrock:2014bja}
\hrefCMSnoop {}{S.~Willenbrock and C.~Zhang, ``{Effective field theory beyond
  the Standard Model}'',} \textit{ Ann. Rev. Nucl. Part. Sci.} \textbf{ 64}
  (2014) 83,
  \href{http://dx.doi.org/10.1146/annurev-nucl-102313-025623}{\doi{10.1146/annurev-nucl-102313-025623}},
\href{http://www.arXiv.org/abs/1401.0470}{\texttt{arXiv:1401.0470}}.

\bibitem{Aaltonen:2008qr}
\hrefCMSnoop {}{{CDF} Collaboration, ``{Search for top-quark production via
  flavor-changing neutral currents in $\mathrm{W}$+1 jet events at CDF}'',}
  \textit{ Phys. Rev. Lett.} \textbf{ 102} (2009) 151801,
  \href{http://dx.doi.org/10.1103/PhysRevLett.102.151801}{\doi{10.1103/PhysRevLett.102.151801}},
\href{http://www.arXiv.org/abs/0812.3400}{\texttt{arXiv:0812.3400}}.

\bibitem{Abazov:2010qk}
\hrefCMSnoop {}{{D0} Collaboration, ``{Search for flavor changing neutral
  currents via quark-gluon couplings in single top quark production using
  2.3\fbinv of $\rm {p\bar{p}}$ collisions}'',} \textit{ Phys. Lett. B}
  \textbf{ 693} (2010) 81,
  \href{http://dx.doi.org/10.1016/j.physletb.2010.08.011}{\doi{10.1016/j.physletb.2010.08.011}},
\href{http://www.arXiv.org/abs/1006.3575}{\texttt{arXiv:1006.3575}}.

\bibitem{Aad:2015gea}
\hrefCMSnoop {}{{ATLAS Collaboration}, ``{Search for single top-quark
  production via flavour-changing neutral currents at 8\TeV with the ATLAS
  detector}'',} \textit{ Eur. Phys. J. C} \textbf{ 76} (2016) 55,
  \href{http://dx.doi.org/10.1140/epjc/s10052-016-3876-4}{\doi{10.1140/epjc/s10052-016-3876-4}},
\href{http://www.arXiv.org/abs/1509.00294}{\texttt{arXiv:1509.00294}}.

\bibitem{Khachatryan:2014iya}
\hrefCMSnoop {}{{CMS Collaboration}, ``{Measurement of the $t$-channel
  single-top-quark production cross section and of the $|V_{\mathrm{tb}}|$ CKM
  matrix element in pp collisions at $\sqrt{s} =$ 8\TeV}'',} \textit{ JHEP}
  \textbf{ 06} (2014) 090,
  \href{http://dx.doi.org/10.1007/JHEP06(2014)090}{\doi{10.1007/JHEP06(2014)090}},
\href{http://www.arXiv.org/abs/1403.7366}{\texttt{arXiv:1403.7366}}.

\bibitem{Khachatryan:2015dzz}
\hrefCMSnoop {}{{CMS Collaboration}, ``{Measurement of top quark polarisation
  in $t$-channel single top quark production}'',} \textit{ JHEP} \textbf{ 04}
  (2016) 073,
  \href{http://dx.doi.org/10.1007/JHEP04(2016)073}{\doi{10.1007/JHEP04(2016)073}},
\href{http://www.arXiv.org/abs/1511.02138}{\texttt{arXiv:1511.02138}}.

\bibitem{Aad:2014fwa}
\hrefCMSnoop {}{{ATLAS Collaboration}, ``{Comprehensive measurements of
  $t$-channel single top-quark production cross sections at $\sqrt{s} =$ 7\TeV
  with the ATLAS detector}'',} \textit{ Phys. Rev. D} \textbf{ 90} (2014)
  112006,
  \href{http://dx.doi.org/10.1103/PhysRevD.90.112006}{\doi{10.1103/PhysRevD.90.112006}},
\href{http://www.arXiv.org/abs/1406.7844}{\texttt{arXiv:1406.7844}}.

\bibitem{Boos:2004kh}
\hrefCMSnoop {}{E.~Boos {et~al.}, ``{CompHEP 4.4: automatic computations from
  Lagrangians to events}'',} \textit{ Nucl. Instrum. Meth. A} \textbf{ 534}
  (2004) 250,
  \href{http://dx.doi.org/10.1016/j.nima.2004.07.096}{\doi{10.1016/j.nima.2004.07.096}},
\href{http://www.arXiv.org/abs/hep-ph/0403113}{\texttt{arXiv:hep-ph/0403113}}.

\bibitem{FBMBook}
\href {http://www.cs.utoronto.ca/~radford/bnn.book.html}{R.~M. N., ``Bayesian
  learning for neural networks'',} Technical Report ISBN 0-387-94724-8, Dept.
  of Statistics and Dept. of Computer Science, University of Toronto, 1994.

\bibitem{Bhat:2005hq}
\hrefCMSnoop {}{P.~C. Bhat and H.~B. Prosper, ``{Bayesian neural networks}'',}
  in \textit{ Statistical Problems in Particle Physics, Astrophysics and
  Cosmology (PHYSTAT05)}, p.~151.
\newblock Oxford, England, 2005.
\newblock
Conf. Proc. C050912.

\bibitem{FBMPackage}
\href {http://www.cs.toronto.edu/~radford/fbm.software.html}{R.~M. Neal,
  ``Software for flexible Bayesian modeling and Markov chain sampling'',}
  Technical Report 2004-11-10, Dept. of Statistics and Dept. of Computer
  Science, University of Toronto, 2004.

\bibitem{JINST}
\hrefCMSnoop {}{{CMS Collaboration}, ``The {CMS} experiment at the {CERN}
  {LHC}'',} \textit{ JINST} \textbf{ 3} (2008) S08004,
\href{http://dx.doi.org/10.1088/1748-0221/3/08/S08004}{\doi{10.1088/1748-0221/3/08/S08004}}.

\bibitem{CMS-PAS-PFT-09-001}
\href {http://cdsweb.cern.ch/record/1194487}{{CMS Collaboration},
  ``Particle--flow event reconstruction in {CMS} and performance for jets,
  taus, and {\MET}'',} CMS Physics Analysis Summary CMS-PAS-PFT-09-001, 2009.

\bibitem{CMS-PAS-PFT-10-001}
\href {http://cdsweb.cern.ch/record/1247373}{{CMS Collaboration},
  ``Commissioning of the particle--flow event reconstruction with the first
  {LHC} collisions recorded in the {CMS} detector'',} CMS Physics Analysis
  Summary CMS-PAS-PFT-10-001, 2010.

\bibitem{Cacciari:2008gp}
\hrefCMSnoop {}{M.~Cacciari, G.~P. Salam, and G.~Soyez, ``{The anti-$k_t$ jet
  clustering algorithm}'',} \textit{ JHEP} \textbf{ 04} (2008) 063,
  \href{http://dx.doi.org/10.1088/1126-6708/2008/04/063}{\doi{10.1088/1126-6708/2008/04/063}},
\href{http://www.arXiv.org/abs/0802.1189}{\texttt{arXiv:0802.1189}}.

\bibitem{Cacciari:2011ma}
\hrefCMSnoop {}{M.~Cacciari, G.~P. Salam, and G.~Soyez, ``{FastJet user
  manual}'',} \textit{ Eur. Phys. J. C} \textbf{ 72} (2012) 1896,
  \href{http://dx.doi.org/10.1140/epjc/s10052-012-1896-2}{\doi{10.1140/epjc/s10052-012-1896-2}},
\href{http://www.arXiv.org/abs/1111.6097}{\texttt{arXiv:1111.6097}}.

\bibitem{Boos:2006af}
E.~E. Boos\hrefCMSnoop {}{ {et~al.}, ``{Method for simulating electroweak
  top-quark production events in the NLO approximation: SingleTop event
  generator}'',} \textit{ Phys. Atom. Nucl.} \textbf{ 69} (2006) 1317,
\href{http://dx.doi.org/10.1134/S1063778806080084}{\doi{10.1134/S1063778806080084}}.

\bibitem{Kidonakis:tch}
\hrefCMSnoop {}{N.~Kidonakis, ``{Next-to-next-to-leading-order collinear and
  soft gluon corrections for $t$-channel single top quark production}'',}
  \textit{ Phys. Rev. D} \textbf{ 83} (2011) 091503,
  \href{http://dx.doi.org/10.1103/PhysRevD.83.091503}{\doi{10.1103/PhysRevD.83.091503}},
\href{http://www.arXiv.org/abs/1103.2792}{\texttt{arXiv:1103.2792}}.

\bibitem{Hathor1}
\hrefCMSnoop {}{U.~L. M.~Aliev, H.~Lacker {et~al.}, ``{HATHOR: a HAdronic Top
  and Heavy quarks crOss section calculatoR}'',} \textit{ Comput. Phys.
  Commun.} \textbf{ 182} (2011) 1034,
  \href{http://dx.doi.org/10.1016/j.cpc.2010.12.040}{\doi{10.1016/j.cpc.2010.12.040}},
\href{http://www.arXiv.org/abs/1007.1327}{\texttt{arXiv:1007.1327}}.

\bibitem{Hathor2}
P.~Kant\hrefCMSnoop {}{ {et~al.}, ``{HatHor for single top-quark production:
  Updated predictions and uncertainty estimates for single top-quark production
  in hadronic collisions}'',} \textit{ Comput. Phys. Commun.} \textbf{ 191}
  (2015) 74,
  \href{http://dx.doi.org/10.1016/j.cpc.2015.02.001}{\doi{10.1016/j.cpc.2015.02.001}},
\href{http://www.arXiv.org/abs/1406.4403}{\texttt{arXiv:1406.4403}}.

\bibitem{Alioli:2010xd}
\hrefCMSnoop {}{S.~Alioli, P.~Nason, C.~Oleari, and E.~Re, ``{A general
  framework for implementing NLO calculations in shower Monte Carlo programs:
  the POWHEG BOX}'',} \textit{ JHEP} \textbf{ 06} (2010) 043,
  \href{http://dx.doi.org/10.1007/JHEP06(2010)043}{\doi{10.1007/JHEP06(2010)043}},
\href{http://www.arXiv.org/abs/1002.2581}{\texttt{arXiv:1002.2581}}.

\bibitem{Alwall:2011uj}
J.~Alwall\hrefCMSnoop {}{ {et~al.}, ``{MadGraph 5: going beyond}'',} \textit{
  JHEP} \textbf{ 06} (2011) 128,
  \href{http://dx.doi.org/10.1007/JHEP06(2011)128}{\doi{10.1007/JHEP06(2011)128}},
\href{http://www.arXiv.org/abs/1106.0522}{\texttt{arXiv:1106.0522}}.

\bibitem{Czakon:2013goa}
\hrefCMSnoop {}{M.~Czakon, P.~Fiedler, and A.~Mitov, ``{Total top-quark
  pair-production cross section at hadron colliders through
  $\mathcal{O}(\alpha^4_S)$}'',} \textit{ Phys. Rev. Lett.} \textbf{ 110}
  (2013) 252004,
  \href{http://dx.doi.org/10.1103/PhysRevLett.110.252004}{\doi{10.1103/PhysRevLett.110.252004}},
\href{http://www.arXiv.org/abs/1303.6254}{\texttt{arXiv:1303.6254}}.

\bibitem{Czakon:ttbar}
\hrefCMSnoop {}{M.~Czakon and A.~Mitov, ``{Top++: a program for the calculation
  of the top-pair cross-section at hadron colliders}'',} \textit{ Comput. Phys.
  Commun.} \textbf{ 185} (2014) 2930,
  \href{http://dx.doi.org/10.1016/j.cpc.2014.06.021}{\doi{10.1016/j.cpc.2014.06.021}},
\href{http://www.arXiv.org/abs/1112.5675}{\texttt{arXiv:1112.5675}}.

\bibitem{Gavin:2010az}
\hrefCMSnoop {}{R.~Gavin, Y.~Li, F.~Petriello, and S.~Quackenbush, ``{FEWZ 2.0:
  A code for hadronic Z production at next-to-next-to-leading order}'',}
  \textit{ Comput. Phys. Commun.} \textbf{ 182} (2011) 2388,
  \href{http://dx.doi.org/10.1016/j.cpc.2011.06.008}{\doi{10.1016/j.cpc.2011.06.008}},
\href{http://www.arXiv.org/abs/1011.3540}{\texttt{arXiv:1011.3540}}.

\bibitem{Campbell:2010ff}
\hrefCMSnoop {}{J.~M. Campbell and R.~K. Ellis, ``{MCFM for the Tevatron and
  the LHC}'',} \textit{ Nucl. Phys. Proc. Suppl.} \textbf{ 205-206} (2010) 10,
  \href{http://dx.doi.org/10.1016/j.nuclphysbps.2010.08.011}{\doi{10.1016/j.nuclphysbps.2010.08.011}},
\href{http://www.arXiv.org/abs/1007.3492}{\texttt{arXiv:1007.3492}}.

\bibitem{Sjostrand:2006za}
\hrefCMSnoop {}{T.~{Sj{\"o}strand}, S.~Mrenna, and P.~Z. Skands, ``{PYTHIA 6.4
  physics and manual}'',} \textit{ JHEP} \textbf{ 05} (2006) 026,
  \href{http://dx.doi.org/10.1088/1126-6708/2006/05/026}{\doi{10.1088/1126-6708/2006/05/026}},
\href{http://www.arXiv.org/abs/hep-ph/0603175}{\texttt{arXiv:hep-ph/0603175}}.

\bibitem{Kidonakis:twsch_new}
\hrefCMSnoop {}{N.~Kidonakis, ``{Top quark production}'',} (2013).
\href{http://www.arXiv.org/abs/1311.0283}{\texttt{arXiv:1311.0283}}.

\bibitem{Alekhin:2011sk}
\hrefCMSnoop {}{S.~Alekhin {et~al.}, ``{The PDF4LHC Working Group interim
  report}'',} (2011).
\href{http://www.arXiv.org/abs/1101.0536}{\texttt{arXiv:1101.0536}}.

\bibitem{Gao:2013xoa}
J.~Gao\hrefCMSnoop {}{ {et~al.}, ``{CT10 next-to-next-to-leading order global
  analysis of QCD}'',} \textit{ Phys. Rev. D} \textbf{ 89} (2014) 033009,
  \href{http://dx.doi.org/10.1103/PhysRevD.89.033009}{\doi{10.1103/PhysRevD.89.033009}},
\href{http://www.arXiv.org/abs/1302.6246}{\texttt{arXiv:1302.6246}}.

\bibitem{Chatrchyan:2013gfi}
\hrefCMSnoop {}{{CMS Collaboration}, ``{Study of the underlying event at
  forward rapidity in pp collisions at $\sqrt{s}$ = 0.9, 2.76, and 7 TeV}'',}
  \textit{ JHEP} \textbf{ 04} (2013) 072,
  \href{http://dx.doi.org/10.1007/JHEP04(2013)072}{\doi{10.1007/JHEP04(2013)072}},
\href{http://www.arXiv.org/abs/1302.2394}{\texttt{arXiv:1302.2394}}.

\bibitem{Khachatryan:2015pea}
\hrefCMSnoop {}{{CMS Collaboration}, ``{Event generator tunes obtained from
  underlying event and multiparton scattering measurements}'',} \textit{ Eur.
  Phys. J. C} \textbf{ 76} (2016), no.~3, 155,
  \href{http://dx.doi.org/10.1140/epjc/s10052-016-3988-x}{\doi{10.1140/epjc/s10052-016-3988-x}},
\href{http://www.arXiv.org/abs/1512.00815}{\texttt{arXiv:1512.00815}}.

\bibitem{Chatrchyan:2012xi}
\hrefCMSnoop {}{{CMS Collaboration}, ``{Performance of CMS muon reconstruction
  in pp collision events at $\sqrt{s}= $ 7\TeV}'',} \textit{ JINST} \textbf{ 7}
  (2012) P10002,
  \href{http://dx.doi.org/10.1088/1748-0221/7/10/P10002}{\doi{10.1088/1748-0221/7/10/P10002}},
\href{http://www.arXiv.org/abs/1206.4071}{\texttt{arXiv:1206.4071}}.

\bibitem{CMS-DP-2013-009}
\href {http://cds.cern.ch/record/1536406}{{CMS Collaboration}, ``Single muon
  efficiencies in 2012 Data'',} {CMS} Detector Performance Note
  CMS-DP-2013-009, 2013.

\bibitem{BTag2011}
\hrefCMSnoop {}{{CMS Collaboration}, ``{Identification of b-quark jets with the
  CMS experiment}'',} \textit{ JINST} \textbf{ 8} (2013) P04013,
  \href{http://dx.doi.org/10.1088/1748-0221/8/04/P04013}{\doi{10.1088/1748-0221/8/04/P04013}},
\href{http://www.arXiv.org/abs/1211.4462}{\texttt{arXiv:1211.4462}}.

\bibitem{Mahlon:1996pn}
\hrefCMSnoop {}{G.~Mahlon and S.~J. Parke, ``{Improved spin basis for angular
  correlation studies in single top quark production at the Tevatron}'',}
  \textit{ Phys. Rev. D} \textbf{ 55} (1997) 7249,
  \href{http://dx.doi.org/10.1103/PhysRevD.55.7249}{\doi{10.1103/PhysRevD.55.7249}},
\href{http://www.arXiv.org/abs/hep-ph/9611367}{\texttt{arXiv:hep-ph/9611367}}.

\bibitem{Savedra:cos}
\hrefCMSnoop {}{J.~A. Aguilar-Saavedra and R.~V. Herrero-Hahn,
  ``{Model-independent measurement of the top quark polarisation}'',} \textit{
  Phys. Lett. B} \textbf{ 718} (2012) 983,
  \href{http://dx.doi.org/10.1016/j.physletb.2012.11.031}{\doi{10.1016/j.physletb.2012.11.031}},
\href{http://www.arXiv.org/abs/1208.6006}{\texttt{arXiv:1208.6006}}.

\bibitem{Barger:1987nn}
\hrefCMSnoop {}{V.~D. Barger and R.~J.~N. Phillips, ``{Collider physics}'',}
  \textit{ Redwood City, USA: Addison-Wesley (1987) (Frontiers In Physics, 71)}
(1987).

\bibitem{Chatrchyan:2011vp}
\hrefCMSnoop {}{{CMS Collaboration}, ``{Measurement of the $t$-channel single
  top quark production cross section in pp collisions at $\sqrt{s} =$
  7\TeV}'',} \textit{ Phys. Rev. Lett.} \textbf{ 107} (2011) 091802,
  \href{http://dx.doi.org/10.1103/PhysRevLett.107.091802}{\doi{10.1103/PhysRevLett.107.091802}},
\href{http://www.arXiv.org/abs/1106.3052}{\texttt{arXiv:1106.3052}}.

\bibitem{Chatrchyan:2012ep}
\hrefCMSnoop {}{{CMS Collaboration}, ``{Measurement of the single-top-quark
  $t$-channel cross section in pp collisions at $\sqrt{s} =$ 7\TeV}'',}
  \textit{ JHEP} \textbf{ 12} (2012) 035,
  \href{http://dx.doi.org/10.1007/JHEP12(2012)035}{\doi{10.1007/JHEP12(2012)035}},
\href{http://www.arXiv.org/abs/1209.4533}{\texttt{arXiv:1209.4533}}.

\bibitem{JEC2010}
\hrefCMSnoop {}{{CMS Collaboration}, ``{Determination of jet energy calibration
  and transverse momentum resolution in CMS}'',} \textit{ JINST} \textbf{ 6}
  (2011) P11002,
  \href{http://dx.doi.org/10.1088/1748-0221/6/11/P11002}{\doi{10.1088/1748-0221/6/11/P11002}},
\href{http://www.arXiv.org/abs/1107.4277}{\texttt{arXiv:1107.4277}}.

\bibitem{JER2010}
\href {https://cds.cern.ch/record/1339945}{{{CMS}} Collaboration, ``Jet energy
  resolution in {CMS} at $\sqrt{s} =$ 7\TeV'',} {CMS} Physics Analysis Summary
  CMS-PAS-JME-10-014, 2011.

\bibitem{Lumi7TeV}
\href {http://cds.cern.ch/record/1434360}{{{CMS}} Collaboration, ``Absolute
  calibration of the luminosity measurement at {CMS}: winter 2012 update'',}
  {CMS} Physics Analysis Summary CMS-PAS-SMP-12-008, 2012.

\bibitem{Lumi8TeV}
\href {http://cdsweb.cern.ch/record/1598864}{{CMS Collaboration}, ``CMS
  Luminosity Based on Pixel Cluster Counting - Summer 2013 Update'',} CMS
  Physics Analysis Summary CMS-PAS-LUM-13-001, 2013.

\bibitem{Chatrchyan:2012nj}
\hrefCMSnoop {}{{CMS Collaboration}, ``{Measurement of the inelastic
  proton-proton cross section at $\sqrt{s} =$ 7\TeV}'',} \textit{ Phys. Lett.
  B} \textbf{ 722} (2013) 5,
  \href{http://dx.doi.org/10.1016/j.physletb.2013.03.024}{\doi{10.1016/j.physletb.2013.03.024}},
\href{http://www.arXiv.org/abs/1210.6718}{\texttt{arXiv:1210.6718}}.

\bibitem{Alwall:2008qv}
\hrefCMSnoop {}{J.~Alwall, S.~de~Visscher, and F.~Maltoni, ``{QCD radiation in
  the production of heavy colored particles at the LHC}'',} \textit{ JHEP}
  \textbf{ 02} (2009) 017,
  \href{http://dx.doi.org/10.1088/1126-6708/2009/02/017}{\doi{10.1088/1126-6708/2009/02/017}},
\href{http://www.arXiv.org/abs/0810.5350}{\texttt{arXiv:0810.5350}}.

\bibitem{Chatrchyan:2012saa}
\hrefCMSnoop {}{{CMS Collaboration}, ``{Measurement of differential top-quark
  pair production cross sections in pp collisions at $\sqrt{s} =$ 7\TeV}'',}
  \textit{ Eur. Phys. J. C} \textbf{ 73} (2013) 2339,
  \href{http://dx.doi.org/10.1140/epjc/s10052-013-2339-4}{\doi{10.1140/epjc/s10052-013-2339-4}},
\href{http://www.arXiv.org/abs/1211.2220}{\texttt{arXiv:1211.2220}}.

\bibitem{Khachatryan:2015oqa}
\hrefCMSnoop {}{{CMS Collaboration}, ``{Measurement of the differential cross
  section for top quark pair production in pp collisions at $\sqrt{s} =$
  8\TeV}'',} \textit{ Eur. Phys. J. C} \textbf{ 75} (2015) 542,
  \href{http://dx.doi.org/10.1140/epjc/s10052-015-3709-x}{\doi{10.1140/epjc/s10052-015-3709-x}},
\href{http://www.arXiv.org/abs/1505.04480}{\texttt{arXiv:1505.04480}}.

\bibitem{Barlow:1993dm}
\hrefCMSnoop {}{R.~J. Barlow and C.~Beeston, ``{Fitting using finite Monte
  Carlo samples}'',} \textit{ Comput. Phys. Commun.} \textbf{ 77} (1993) 219,
\href{http://dx.doi.org/10.1016/0010-4655(93)90005-W}{\doi{10.1016/0010-4655(93)90005-W}}.

\bibitem{Buchmuller:1985jz}
\hrefCMSnoop {}{W.~Buchm{\"u}ller and D.~Wyler, ``{Effective lagrangian
  analysis of new interactions and flavor conservation}'',} \textit{ Nucl.
  Phys. B} \textbf{ 268} (1986) 621,
\href{http://dx.doi.org/10.1016/0550-3213(86)90262-2}{\doi{10.1016/0550-3213(86)90262-2}}.

\bibitem{Kane:1991bg}
\hrefCMSnoop {}{G.~L. Kane, G.~A. Ladinsky, and C.~P. Yuan, ``{Using the top
  quark for testing Standard Model polarization and CP predictions}'',}
  \textit{ Phys. Rev. D} \textbf{ 45} (1992) 124,
\href{http://dx.doi.org/10.1103/PhysRevD.45.124}{\doi{10.1103/PhysRevD.45.124}}.

\bibitem{Abazov:2008sz}
\hrefCMSnoop {}{{D0} Collaboration, ``{Search for anomalous Wtb couplings in
  single top quark production}'',} \textit{ Phys. Rev. Lett.} \textbf{ 101}
  (2008) 221801,
  \href{http://dx.doi.org/10.1103/PhysRevLett.101.221801}{\doi{10.1103/PhysRevLett.101.221801}},
\href{http://www.arXiv.org/abs/0807.1692}{\texttt{arXiv:0807.1692}}.

\bibitem{Abazov:2011pm}
\hrefCMSnoop {}{{D0} Collaboration, ``{Search for anomalous Wtb couplings in
  single top quark production in $\rm {p\bar{p}}$ collisions at $\sqrt{s} = $
  1.96\TeV}'',} \textit{ Phys. Lett. B} \textbf{ 708} (2012) 21,
  \href{http://dx.doi.org/10.1016/j.physletb.2012.01.014}{\doi{10.1016/j.physletb.2012.01.014}},
\href{http://www.arXiv.org/abs/1110.4592}{\texttt{arXiv:1110.4592}}.

\bibitem{Boos:2016zmp}
\hrefCMSnoop {}{E.~Boos, V.~Bunichev, L.~Dudko, and M.~Perfilov, ``{Modeling of
  anomalous Wtb interactions using subsidiary fields}'',} \textit{ Int. J. Mod.
  Phys. A} \textbf{ 32} (2016) 1750008,
  \href{http://dx.doi.org/10.1142/S0217751X17500087}{\doi{10.1142/S0217751X17500087}},
\href{http://www.arXiv.org/abs/1607.00505}{\texttt{arXiv:1607.00505}}.

\bibitem{Aad:2012ky}
\hrefCMSnoop {}{{ATLAS Collaboration}, ``{Measurement of the W boson
  polarization in top quark decays with the ATLAS detector}'',} \textit{ JHEP}
  \textbf{ 06} (2012) 088,
  \href{http://dx.doi.org/10.1007/JHEP06(2012)088}{\doi{10.1007/JHEP06(2012)088}},
\href{http://www.arXiv.org/abs/1205.2484}{\texttt{arXiv:1205.2484}}.

\bibitem{Chatrchyan:2013jna}
\hrefCMSnoop {}{{CMS Collaboration}, ``{Measurement of the W-boson helicity in
  top-quark decays from $t\bar{t}$ production in lepton+jets events in pp
  collisions at $\sqrt{s} =$ 7\TeV}'',} \textit{ JHEP} \textbf{ 10} (2013) 167,
  \href{http://dx.doi.org/10.1007/JHEP10(2013)167}{\doi{10.1007/JHEP10(2013)167}},
\href{http://www.arXiv.org/abs/1308.3879}{\texttt{arXiv:1308.3879}}.

\bibitem{Khachatryan:2014vma}
\hrefCMSnoop {}{{CMS Collaboration}, ``{Measurement of the W boson helicity in
  events with a single reconstructed top quark in pp collisions at $ \sqrt{s}=
  $ 8\TeV}'',} \textit{ JHEP} \textbf{ 01} (2015) 053,
  \href{http://dx.doi.org/10.1007/JHEP01(2015)053}{\doi{10.1007/JHEP01(2015)053}},
\href{http://www.arXiv.org/abs/1410.1154}{\texttt{arXiv:1410.1154}}.

\bibitem{Liu:2005dp}
\hrefCMSnoop {}{J.~J. Liu, C.~S. Li, L.~L. Yang, and L.~G. Jin,
  ``{Next-to-leading order QCD corrections to the direct top quark production
  via model-independent FCNC couplings at hadron colliders}'',} \textit{ Phys.
  Rev. D} \textbf{ 72} (2005) 074018,
  \href{http://dx.doi.org/10.1103/PhysRevD.72.074018}{\doi{10.1103/PhysRevD.72.074018}},
\href{http://www.arXiv.org/abs/hep-ph/0508016}{\texttt{arXiv:hep-ph/0508016}}.

\bibitem{Zhang:2008yn}
J.~J. Zhang\hrefCMSnoop {}{ {et~al.}, ``{Next-to-leading order QCD corrections
  to the top quark decay via model-independent FCNC couplings}'',} \textit{
  Phys. Rev. Lett.} \textbf{ 102} (2009) 072001,
  \href{http://dx.doi.org/10.1103/PhysRevLett.102.072001}{\doi{10.1103/PhysRevLett.102.072001}},
\href{http://www.arXiv.org/abs/0810.3889}{\texttt{arXiv:0810.3889}}.

\end{thebibliography}\endgroup
\cleardoublepage \appendix\section{The CMS Collaboration \label{app:collab}}\begin{sloppypar}\hyphenpenalty=5000\widowpenalty=500\clubpenalty=5000\textbf{Yerevan Physics Institute,  Yerevan,  Armenia}\\*[0pt]
V.~Khachatryan, A.M.~Sirunyan, A.~Tumasyan
\vskip\cmsinstskip
\textbf{Institut f\"{u}r Hochenergiephysik,  Wien,  Austria}\\*[0pt]
W.~Adam, E.~Asilar, T.~Bergauer, J.~Brandstetter, E.~Brondolin, M.~Dragicevic, J.~Er\"{o}, M.~Flechl, M.~Friedl, R.~Fr\"{u}hwirth\cmsAuthorMark{1}, V.M.~Ghete, C.~Hartl, N.~H\"{o}rmann, J.~Hrubec, M.~Jeitler\cmsAuthorMark{1}, A.~K\"{o}nig, I.~Kr\"{a}tschmer, D.~Liko, T.~Matsushita, I.~Mikulec, D.~Rabady, N.~Rad, B.~Rahbaran, H.~Rohringer, J.~Schieck\cmsAuthorMark{1}, J.~Strauss, W.~Treberer-Treberspurg, W.~Waltenberger, C.-E.~Wulz\cmsAuthorMark{1}
% ...
\vskip\cmsinstskip
% ...
1:~~Also at Vienna University of Technology, Vienna, Austria\\
% ...

\end{sloppypar}
\end{document}